# Understanding Phonon Transport Properties Using Classical Molecular Dynamics Simulations: A Review

Murali Gopal Muraleedharan,[a] Kiarash Gordiz,[b] Andrew Rohskopf,[b] Spencer Thomas Wyant,[b] Zhe Cheng,[c] Samuel Graham,[d] and Asegun Henry[b,*]

[a]Department of Mechanical Engineering, The Pennsylvania State University

[b]Department of Mechanical Engineering, Massachusetts Institute of Technology

[c]Deparment of Materials Science and Engineering, University of Illinois at Urbana Champaign

[d]George W. Woodruff School of Mechanical Engineering, Georgia Institute of Technology

*Corresponding author: ase@mit.edu

**Abstract**

Predictive modeling of the phonon/thermal transport properties of materials is vital to rational design for a diverse spectrum of engineering applications. Classical Molecular Dynamics (MD) simulations serve as a tool to simulate the time evolution of the atomic level system dynamics and enable calculation of thermal transport properties for a wide range of materials, from perfect periodic crystals to systems with strong structural and compositional disorder, as well as their interfaces. Although MD does not intrinsically rely on a plane wave-like phonon description, when coupled with lattice dynamics calculations, it can give insights to the vibrational mode level contributions to thermal transport, which includes plane-wave like modes as well as others, rendering the approach versatile and powerful. On the other hand, several deficiencies including the lack of vibrationally accurate interatomic potentials and the inability to rigorously include the quantum nature of phonons prohibit the widespread applicability and reliability of Molecular Dynamics simulations. This article provides a comprehensive review of classical Molecular Dynamics based formalisms for extracting thermal transport properties: thermal conductivity and thermal interfacial conductance and the effects of various structural, compositional, and chemical parameters on these properties. Here, we highlight unusual property predictions, and emphasize on the needs and strategies for developing accurate interatomic potentials and rigorous quantum correction schemes.

**Table of contents**

## 1. Introduction

Understanding thermal transport in solids is important to design and engineer materials with desired thermal conductivity (TC) and thermal interfacial conductance (TIC) for myriad applications.[1,2] Examples include thermal management materials for maintaining the operating temperatures of high power electronic devices,[3–5] batteries,[6] and photovoltaics,[7,8] variable thermal conductance building envelopes,[9,10] thermal energy storage materials,[11,12] materials for long-distance thermal energy transmission,[13,14] as well as thermoelectric materials for power generation[15] and solid-state refrigeration applications.[16] Below ~1000 $^0$C, the two major carriers of thermal energy in solids are phonons and electrons.[17] In crystalline solids, atoms located at lattice sites oscillate about their equilibrium positions, creating a displacement field that comprises its thermal energy. From a quantum mechanical perspective, this atomic displacement field can be described by quantized oscillators, termed phonons.[18,19] They are the major heat carriers in semiconductors and insulators, where conduction band electron density is low. Phonons are bosons; their distribution in thermal equilibrium can therefore be described by Bose-Einstein distribution. They interact with each other, electrons, impurities, boundaries, and other imperfections through what are most often termed scattering events.

For crystalline solids, phonon transport is well-understood and can be theoretically described by the phonon gas model (PGM) wherein phonons are treated as particles with momentum proportional to its wave vector, i.e., $\hbar \mathbf{q}$, while the energy is a multiple of $\hbar \omega$, where $\hbar$ is the reduced Planck constant, $\omega$ is the angular frequency of the phonon, $\mathbf{q}$ is its wave vector, i.e., $2\pi/\lambda$, and $\lambda$ is its wavelength.[20–25] For crystalline materials, when there is high symmetry, the idea of treating phonons as particles seems well founded, because the normal modes of vibration will consist of displacement fields that are shaped like waves (i.e., the *"mode shape"*). If waves of similar direction and wavelength are super imposed it results in a wave packet that translates in time at the group velocity, and thus resembles a particle translating at a constant speed. Each phonon has a polarization vector field which describes the magnitude and direction of the vibration of atoms in a phonon (i.e., the mode shape). However, for materials with broken symmetry such as crystalline solids with varying levels of disorder: weak structural and compositional defects to strongly disordered amorphous solids and alloys, as well as interfaces, the normal modes can deviate dramatically from the base case of waves in a crystalline solid. As a result, the idea of forming wave packets becomes questionable, and the PGM is not readily applicable because the plane wave nature of phonons becomes questionable. Experimental and theoretical understanding of thermal transport in such disordered materials are summarized in recent dedicated review articles.[26,27]

Nonetheless, predictive modeling of thermal transport is crucial to rational materials design. Several computational methods have been proposed and applied to predict phonon TC (denoted by $k$ in equations) and TIC (denoted by $G$ in equations) of materials. Among them, the most commonly used are anharmonic lattice dynamics (ALD) based on Boltzmann transport equation (BTE)[28,29] and classical Molecular Dynamics (MD)[30,31]. ALD approach evaluates TC by solving the BTE, based on the assumption that the collective vibrations of atoms in a crystal lattice can be thought of as phonons, wherein the problem reduces to calculating the phonon lifetimes from the anharmonic phonon interactions, or mathematically, the anharmonic interatomic force constants (IFC). Similarly, lattice-dynamics (LD) based approaches for calculating TIC are based on Landauer formalism,[32] where energy transfer at the interface is described by harmonic[33–38] and anharmonic[39] interactions of propagating phonons at the interface and evaluating the degree of energy transfer by each phonon to the other side (transmission coefficient).[40] Hence, for both TC and TIC, anharmonic IFC's act as perturbations to the harmonic phonon Hamiltonian in quantum perturbation theory, the lowest order of which i.e. the third order derivative of interatomic potential describes the three phonon interactions.[29,41] Therefore, in estimating both the phonon lifetimes and interfacial transmission coefficients, an accurate description of interatomic interaction is important. Various descriptions of interatomic interactions, using empirical potentials,[42–44] bond-charge models[45] and more recently, density functional theory (DFT)[41] have been reported in the literature. DFT-BTE approach has lately been widely adopted for crystalline materials for their high accuracy without the requirement of any empirical fitting parameters.[41,46–49] However, there are a few disadvantages with DFT-BTE. Firstly, as the method relies on the phonon quasiparticle description, it fails when phonons are not well-defined, for example in the case of crystals with strong intrinsic disorder, highly anharmonic systems, or at high temperatures where the underlying perturbation theory may fail.[29]

MD-based methods, on the other hand, are "numerical experiments" wherein the heat flow is calculated from tracking the atomic motion and appropriate post-processing techniques are performed to evaluate TC. MD has several unique advantages over ALD based methods. Firstly, it includes anharmonicity to full order and hence computationally expensive higher order anharmonic interactions need not be explicitly modeled as in the case of DFT-BTE. Secondly, since it does not rely on a plane wave definition of phonons, it can be reliably applied not only to perfect crystals but also to disordered materials such as amorphous solids, alloys, lower dimensional materials, nanostructures, or even fluids. But when combined with lattice dynamics (LD), MD simulations can be used to sample the vibrational phase space of atoms in terms of the vibrational modes (i.e. phonons in the case of perfect crystals). Most importantly, large length scales (several nm) and long timescales (several ns) which are cost prohibitive with DFT at the length scales needed (≥ 1 nm → $10^2$-$10^4$ atoms), can be achieved in MD simulations. However, there are also several deficiencies with MD-based methods. Firstly, the accuracy of results depends on MD statistics, which hinges on the efficacy of interatomic potential in describing vibrational modes accurately. And secondly, being based on classical mechanics, it does not include quantum effects such as phonon energy level quantization per se, leading to reduced accuracy at temperatures below Debye temperature. Hence the results are often required to be treated with semiempirical quantum correction strategies for ensuring accuracy.[50,51]

Insofar as the systems considered are materials with broken symmetry, classical MD is more effective than ALD based methods. Although *ab initio* MD has also been recently explored to compute TC and seems promising,[52,53] the formalism has not been widely adopted due to the computational costs associated with using large system sizes and long simulation time required to extract relevant statistics. Therefore, in this article, we review the works in the literature that use classical MD simulations to extract phonon properties. We focus on giving a perspective of how MD trajectory and statistics can be used to extract the physics phonon thermal transport, mainly TC and TIC. This review article is structured as follows: In section 2, we briefly summarize the theory of MD, LD, and interatomic potentials. In section 3, different MD-based formalisms for evaluating TC are discussed, followed by the application of LD for understanding vibrational modes and their contributions. Then we comprehensively review MD based formalisms for calculating TC, where we summarize bulk (system-level) and mode-level knowledge gained from MD. In section 4, we review general and modal formalisms for TIC followed by a synopsis of prior works for understanding TIC of different material interfaces. In section 5, we compare the capabilities of different MD-based modal analysis techniques in predicting TC and TIC. In section 6, the different experimental techniques for validating and benchmarking MD results are briefly presented and finally, in section 7, we direct the attention of readers to open questions and challenges in this field. At the foundation of methods discussed in this paper, lies the general theory of MD simulations and phonons.

## 2. Theory of Molecular Dynamics Simulations and Phonon Transport

Statistical mechanics has been well-developed to describe macroscopic observables in terms of the microscopic behaviors of atoms or other energy carriers. For example, kinetic theory describes the thermodynamics of dilute gases well, and it is often a common practice to approximate quasiparticles like phonons as a system of dilute gas particles by solving the BTE.[54] This works well for crystals,[55] but when dealing with real systems, such as systems with interfaces and broken structural symmetry in solids, such as amorphous structures, defects, and alloys, this physical picture becomes questionable. In such systems, MD simulations serve as numerical experiments to some extent, since they allow the system dynamics to evolve temporally and one can extract insights about the thermal transport physics from the statistical data. Combined with lattice dynamics (LD), MD simulations may be used to sample the vibrational phase space of atoms in terms of the vibrational modes (i.e., phonons).

### 2.1. Molecular Dynamics

MD simulations proceed in time by solving for the motion of atoms and sampling dynamical properties in regions of phase space, from which this data is then input to calculate thermodynamic and statistical mechanical quantities. Dynamics are determined by solving Newton's equations of motions numerically for a system of interacting particles, where the interactions between particles are approximated as mathematical functions known as interatomic potentials, forcefields, or are often just referred to as "potentials". Obtaining the dynamics first involves calculating the force on every atom. In classical MD, the force, $\mathbf{F}_i$ on an atom $i$ is given by the classical mechanical definition as the negative gradient of the system potential energy with respect to the atom position $\mathbf{r}_i$. When the system potential energy is an analytical function that approximates the potential energy surface (PES), it is often known as an interatomic potential (IAP).

By knowing the force on each atom from the potential, the acceleration $\ddot{\mathbf{r}}_i$ of each atom is obtained via Newton's second law, $\mathbf{F}_i = m_i \ddot{\mathbf{r}}_i$, given the mass $m_i$ of atom $i$. These classical equations of motion have been shown to approximate the Schrodinger equation reasonably well[56], provided $m_i$ is not too small and the temperature is high enough to neglect the gap in quantum energy levels of the systems vibrational modes. If the temperature is sufficiently below the Debye temperature of the material, the motion of nuclei is quantum in nature[57], and classical mechanics alone will not accurately represent the true dynamics. Recent advances in path integral MD tackle this issue[58], but in principle the basic idea of MD remains the same – to solve the equations of motion throughout time.

The equations of motion can then be integrated in time using a variety of numerical integration schemes. For systems of 2$^{nd}$ order ODEs such as the equations of motion for a system of atoms, the Verlet algorithms are a common choice and position-Verlet has been shown to possess better numerical stability compared to velocity-Verlet[59]. Although much research has been dedicated to the creation of more advanced integrators with less truncation error[60], the effects on thermal properties calculated in MD have not been studied as far as we know. The commonly used Verlet algorithms are examples of symplectic integrators[61]; this implies that the integration conserves the total energy of the system (if the potential is conservative), which is critical for sampling phase space in the microcanonical ($NVE$) ensemble with constant number of atoms $N$, system volume $V$, and total energy $E$. An ensemble in statistical mechanics is a collection of points in phase space and is of crucial importance when analyzing the statistics of different dynamics associated with different thermodynamic states. Different ensembles associated with different thermodynamic constraints (e.g. energy, pressure, or temperature) often occupy different regions of phase space, and these different ensembles refer to the phase space trajectories that would exist under such thermodynamic constraints.

### 2.1.1. Phonon Transport Applications of Thermostats and Barostats

It is often desirable, for example, to study a system at constant temperature $T$ in the canonical ($NVT$) ensemble or at constant pressure and pressure ($NPT$) so as to replicate a complementing experiment in a controlled environment. These thermostats and/or barostats are imposed by either rescaling the atom velocities or system pressure during the simulation via the Andersen and Berendsen methods, or modifying the equations of motion with Nose-Hoover dynamics[62]; such methods are discussed in books and other reviews. Langevin dynamics[63] is another example of a thermostat, and involves modifying the equations of motion for each atom to include a dissipative term along with a random force; this models the motion of ions interacting with some external bath. Aside from equilibrating the system at a given temperature or pressure, these methods of enforcing certain temperatures and pressures in MD simulations are useful for studying thermal transport properties in extreme conditions where measuring thermal transport properties is either difficult or impossible. Measurements of thermal transport properties in these scenarios is challenging, however, so MD simulations are an easier alternative. As an example, Jones and Ward recently incorporated barostats and thermostats to equilibrate LiF at high temperatures (1000-4000

K) and pressures (100-400 GPa), and then calculated thermal conductivity at these extreme conditions using MD. Thermostat/barostat modifications to the equations of motion are not only useful for thermalizing and pressurizing systems to study thermal transport at extreme conditions; they may also help model the dynamics of ions/phonons due to electronic degrees of freedom.

To realize the utility of thermostats in modeling electron-phonon interactions, consider typical MD simulations where the electronic degrees of freedom are eliminated since they respond adiabatically to the motion of ions. This means that the electrons typically remain in the ground state, thus allowing the use of the Born-Oppenheimer approximation in which the ions interact only with each other via an interatomic potential. The adiabatic approximation is problematic, however, in situations where electrons are excited by external fields (e.g. lasers) and transfer their energy to phonons[64], or when fast moving ions excite electrons (e.g. radiation damage[65]). Such situations result in coupling of the electron and phonon degrees of freedom and have been modeled by the electron-phonon two temperature model (TTM)[66] in which the electrons and lattice have two separate temperatures, and couple with each other via an effective electron-phonon coupling constant. This two-temperature model may be combined with Langevin dynamics in MD by introducing an electron-phonon friction coefficient[67] as the dissipative force for ions. This method is not without its issues, however, as Tamm *et al*. recently showed that the Langevin model of an electronic bath results in the same lifetime for all phonons[68], which places a serious limitation on studying electron-phonon equilibration. Their recent modifications of the method alleviates this issue[69] and offers a promising method of simulating electron-phonon interactions in metals via classical MD.  Recent advances[70] in  machine learning the electron charge density as a function of atomic positions may also be used to study electron-phonon interactions, since one may study the effect of changing band structure (and hence electron transport properties) due to the change in ionic positions via vibrational modes; such a scheme would not require artificial thermostats.

## 2.2. Lattice Dynamics for MD

To study phonon contributions to dynamical phenomena such as heat flow, one must first obtain the normal modes for the system under study. The behavior of modes/phonons depends on the interatomic potential and masses of a structure in dynamical equilibrium (i.e., zero force on all atoms). Specifically, the vibrational frequency and mode shape depend on the 2nd spatial derivatives of the potential with respect to atomic displacements, and the interactions between modes which allow them to exchange energy with each other depend on higher (anharmonic) derivatives of the potential. To clearly elucidate what is meant by harmonic and anharmonic parts of the potential, it is instructive to Taylor expand the potential energy surface in terms of atomic displacements $u_i^\alpha$ for each atom $i$ in the $\alpha$ Cartesian direction, resulting in the Taylor expansion potential (TEP):

$$U = \frac{1}{2}\sum_{ij,\alpha\beta} \Phi_{ij}^{\alpha\beta} u_i^\alpha u_j^\beta + \frac{1}{3!}\sum_{ijk,\alpha\beta\gamma} \Psi_{ijk}^{\alpha\beta\gamma} u_i^\alpha u_j^\beta u_k^\gamma + \cdots \qquad (1)$$

where $\Phi_{ij}^{\alpha\beta} = \partial^2 U / \partial u_i^\alpha \partial u_j^\beta$ and $\Psi_{ijk}^{\alpha\beta\gamma} = \partial^3 U / \partial u_i^\alpha \partial u_j^\beta \partial u_k^\gamma$ are the 2nd order interatomic force constants (IFC2s) and 3rd order interatomic force constants (IFC3s), respectively.

To determine the frequency and mode shape of individual modes (i.e., no anharmonic interaction with other modes), consider the equations of motion for a system of atoms interacting via a harmonic TEP, and rewrite these harmonic equations of motion in the matrix form of an eigenvalue problem[71]

$$\omega^2(\mathbf{k},n) \cdot \mathbf{e}(\mathbf{k},n) = \mathbf{D}(\mathbf{k}) \cdot \mathbf{e}(\mathbf{k},n) \qquad (2)$$

which gives the frequencies $\omega(\mathbf{k},n)$ as the square root of the eigenvalues, and the eigenvectors $\mathbf{e}(\mathbf{k},n)$ describing the shape of a mode with wave vector $\mathbf{k}$ in branch $n$ (there may be multiple modes with wave vector $\mathbf{k}$ on a dispersion curve, in different branches). The dynamical matrix $\mathbf{D}(\mathbf{k})$ houses information on masses and stiffness between interactions in the system, and is an $N \times N$ array of smaller $3 \times 3$ submatrices, each having Cartesian components determined by

$$D_{\alpha\beta}(jj',\mathbf{k}) = \frac{1}{\sqrt{m_j m_{j'}}} \sum_{l'} \Phi_{\alpha\beta}\binom{jj'}{ll'} \exp\left(i\mathbf{k} \cdot [\mathbf{r}(j'l') - \mathbf{r}(jl)]\right) \qquad (3)$$

where atom $j$ is in unit cell $l$ (often designated as $l = 0$ for the reference cell in a periodic system), and atom $j'$ is in unit cell $l'$. In this notation, the IFC2s are represented as $\Phi_{\alpha\beta}\binom{jj'}{ll'}$ to denote the force constant between atom $j$ in unit cell $l$ and atom $j'$ in unit cell $l'$. In periodic systems, only the atoms within an individual unit cell $l$ are distinguishable, while all others are repetitions that have the same solutions to the equation of motion modulated by different plane waves whose wavelength and direction are described by the corresponding wave vector $\mathbf{k}$.

### 2.3. Supercell Lattice Dynamics to Study Systems with Broken Symmetry

Due to the assumption of crystalline periodicity in the structure of atoms described by Equation 3, such an expression is unsuitable for describing vibrational modes in systems that lack crystalline symmetry. To extract the modes of any arbitrary supercell of atoms, without the assumption of symmetry, one may consider only the $\mathbf{k} = 0$ modes, whereby the entire supercell is treated as the basis of a cubic lattice. The eigenvalue problem of Equation 2 remains the same, but the dynamical matrix takes a much simpler form given by the IFC2s divided by the masses of atoms $i$ and $j$: $D_{ij}^{\alpha\beta} = \frac{1}{\sqrt{m_i m_j}} \Phi_{ij}^{\alpha\beta}$. This approach is termed super-cell LD (SCLD) and it allows the study of various types of vibrations that exist within a structure, regardless of whether the structure is crystalline or not. Performing SCLD on non-crystalline materials such as amorphous materials and alloys yields vibrational modes that differ in character compared to those found in a perfectly crystalline material. Figure 1(a)-(c) for example shows the three main types of vibrational modes found by Lv and Henry[72] in amorphous silicon, with blue vectors representing the eigenvectors of each mode.

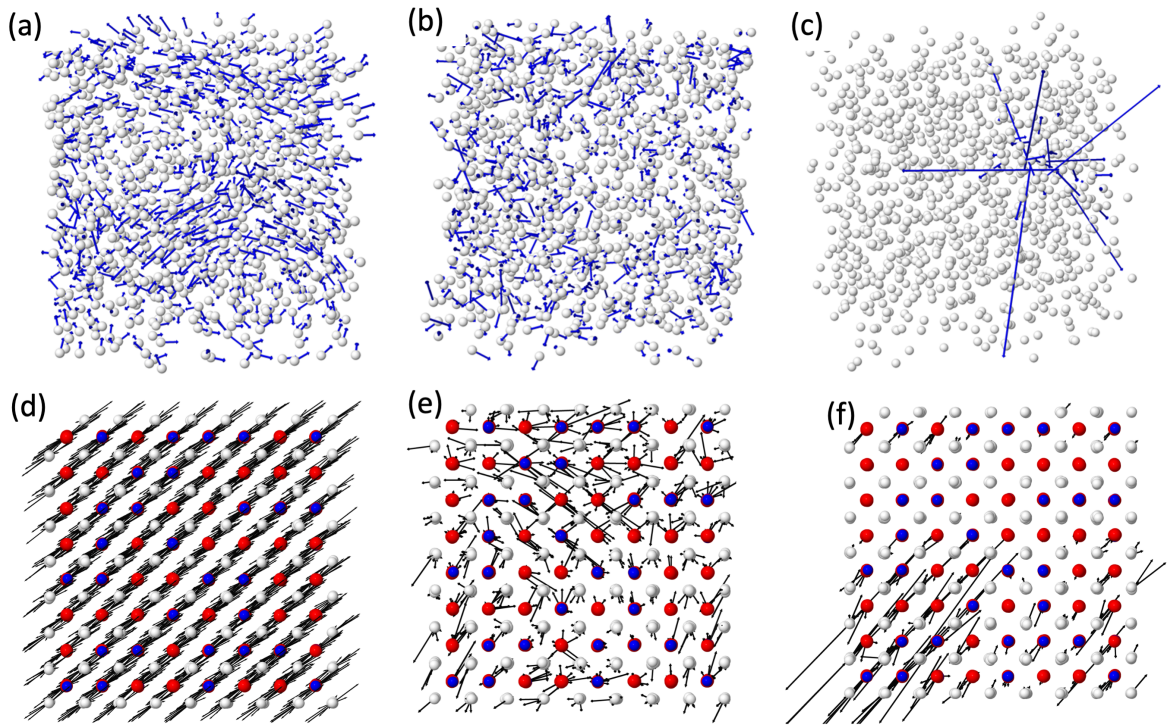

**Figure 1**. (a)-(c) Three main types of vibrational modes in non-crystalline structures (propagons, diffusons, and locons from left to right), obtained from an SCLD calculation on amorphous silicon based on Lv and Henry[73] (d)-(f) Examples of a propagon, diffuson, and locon from left to right

in a compositional alloy obtained from an SCLD calculation based on Seyf and Henry[74]. Atoms are represented as white, blue, and red spheres, and the blue vectors are eigenvectors for each atom.

Aside from amorphous materials, the same categories of modes (propagons, diffusons, and locons) exist in situations such as crystalline alloys where the masses of atoms are randomly distributed although the structure itself is crystalline[74]. Seyf *et al.*[75] devised a method to quantitatively distinguish between propagons, diffusons, and locons based on their eigenvectors, and such categorizations are useful in determining how different vibrational modes contribute to thermal transport. Non-propagating vibrational modes also exist in other broken symmetry systems such as interfaces[76].

Notwithstanding bulk materials, vibrational modes are known to exist in individual molecules and notably large molecules such as proteins. In proteins, vibrational modes drive conformational changes which alter the biological function of the protein, but identifying which vibrational modes are responsible for such changes remains a challenge[77]. For instance, Go *et al.*[78] solved the eigenvalue problem to obtain the normal modes of a protein, and the visual results for some modes are shown in Figure 2.

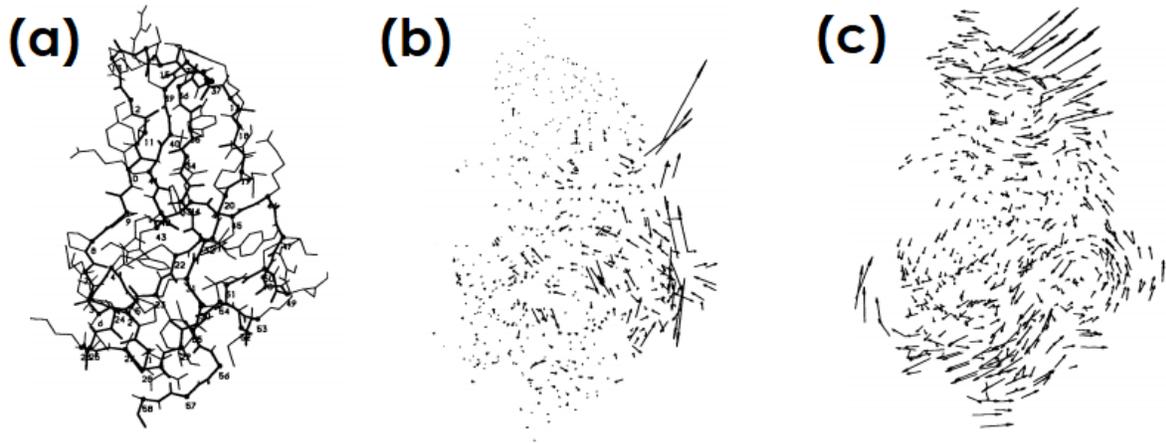

**Figure 2** – (a) The protein structure for which a SCLD calculation was performed by Go et al[78]. (b) A high-frequency localized mode, with eigenvectors shown as arrows centered at atom positions. (c) A low-frequency diffuson-like mode, with eigenvectors shown as arrows centered at atom positions. Reprinted from Go et al[78] (authors could not be reached to request permission).

More recently, statistical similarity analyses[79] have also been performed to compare the eigenvectors of protein vibrational modes to the known displacement field of the conformational change, as a means of identifying which vibrational modes contribute to the conformational change. Despite these efforts, the study of thermal transport in biomolecules such as proteins remains largely unexplored by the phonon transport community, although showing how the modes which drive conformational changes in proteins obtain their energy via thermal transport would provide more insight into the nature of conformational changes.

To study the behavior of vibrational modes in a MD simulation, one may use the Fourier series representation of normal mode coordinates in terms of atomic displacements and eigenvectors, given by

$$X_n = \sum_i \sqrt{m_i}\, \mathbf{e}_i \cdot \mathbf{u}_i \qquad (4)$$

where $X_n$ is the mode amplitude of mode $n$. Similarly, for mode velocities, one obtains

$$\dot{X}_n = \sum_i \sqrt{m_i}\, \mathbf{e}_i \cdot \mathbf{v}_i \tag{5}$$

Equations 4 and 5 may be calculated during a MD simulation to observe the dynamics of vibrational modes. The inverse transformations also yield

$$\mathbf{u}_i = \frac{1}{\sqrt{m_i}} \sum_n X_n \mathbf{e}_i \tag{6}$$

and

$$\mathbf{v}_i = \frac{1}{\sqrt{m_i}} \sum_n \dot{X}_n \mathbf{e}_i \tag{7}$$

for the atomic displacements and velocities in terms of normal mode coordinates; these expressions may be substituted into other dynamical quantities such as the heat flux which depend on their values, thus allowing one to study mode contributions to quantities like heat flow. Performing such analyses accurately, however, depends on the accuracy of the IAP.

## 2.4. Interatomic Potentials

Since individual mode properties and anharmonic interactions between modes depend on the IFCs described in the TEP of Equation 1, it is of interest to use potentials that accurately model these IFCs. Obtaining the IFCs by fitting the TEP to DFT has been known to result in a TEP that accurately predicts forces,[80] harmonic properties such as mode frequencies,[81] and anharmonic properties arising from mode-mode interactions.[82] Thus, the TEP functional form is the IAP that in principle best suited for studying phonons in classical MD. However, there is an important problem that arises when attempting to use the TEP to run a MD simulation, namely stability.

Here, in using the term "stability", what is meant is the ability for the energy in a MD simulation to be conserved, and for the dynamics to give rise to stable vibrations of the atoms around their equilibrium sites when simulating at a temperature below the material's melting/sublimation/decomposition temperature. When IAPs such as the TEP exhibit instability, it is often the case that the energy is not properly conserved in a microcanonical ensemble, and the temperature quickly rises causing the atoms to deviate far from their equilibrium sites, often flying apart in space – effectively turning into a gas. For example, Murakami *et al.* showed that the anharmonic TEP yields accurate TC in MD simulations, but also noted that dynamical instability of the TEP prohibited simulations at high temperatures.[83] This issue of instability when using the TEP to perform MD simulations has been noted in several reports in the literature.[81,83,84] Although it is not clear why this stability problem arises, many suspect that the lack of higher order anharmonic terms may be a cause, so researchers have attempted to improve stability and better model anharmonicity by incorporating higher order force constants. A number of methods have been proposed to circumvent the immense computational cost of obtaining ever higher orders of anharmonicity, including the use of compressed sensing methods[85] and regularized least squares[86]. Instead of using anharmonic IFCs, Rohskopf *et al.* recently used a translationally invariant version of the harmonic TEP to exactly model harmonic forces and phonon dispersion, while a separate potential models the anharmonic force components[87]; this potential exactly predicts phonon dispersion curves but suffers from the lack of transferability to different structures due to the fact that the IFC2s are defined for a particular structure. More transferrable than the TEP, however, are traditional IAPs which include so-called "empirical potentials" that describe the system geometry in terms of interatomic distances, angles, or dihedrals, with adjustable parameters to help the potential accurately model interatomic forces. Since the transferability and ease-of-use of empirical potentials makes them popular and easy to use, a discussion of their accuracy in describing phonon transport is warranted.

There have been several surveys on the predictive capacity of different empirical potentials, mainly concerning silicon as a model system.[88–90] Abs da Cruz *et al.* examined the broader accuracy of different potentials relative to dispersion and thermal expansion for bulk and silicon nanowires, concluding that a later parametrization of the Tersoff potential had the best performance for isotopically pure silicon, and that an earlier parameterization of Tersoff and 1NN MEAM should be avoided.[88] However, Howell *et al.* highlighted that both the Stillinger-Weber potential or the Tersoff potential overestimate the thermal conductivity relative to experiment, and have somewhat different temperature dependences; they also differ with respect to each other, with the Tersoff potential predicting twice the thermal expansion compared to Stillinger-Weber.[89] Other surveys have explored the accuracy of different interatomic potentials for 2D materials like graphene[91,92] and $MoS_2$[93]. Zou *et al.* determined that a 2010 parameterization of Tersoff, optimized to phonon dispersion, significantly outperformed all other classical potentials, demonstrating the value of using potentials meant to optimize thermal properties.[91]

Due to their convenience and computational speed, empirical potentials have remained popular in literature despite their issues with accuracy in describing phonons. Most strikingly, however, is the inability of empirical potentials to capture phonon dispersion in crystalline materials. To realize this downside of empirical potentials, consider the phonon dispersion for crystalline silicon calculated using different cutoffs of a harmonic TEP, with force constants obtained from DFT. This is shown in Figure 3.

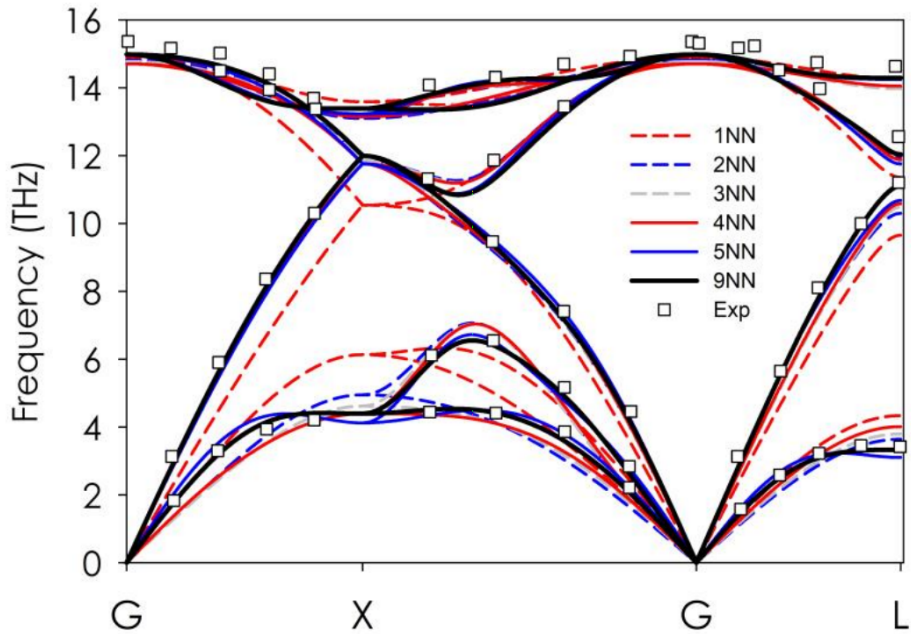

**Figure 3**. Convergence of the phonon dispersion for crystalline silicon as the cutoff of force constant interactions are increased from $1^{st}$ nearest neighbors (NN) to $9^{th}$ NN, calculated with DFT force constants. This shows the importance of cutoff in the interatomic potential for converging the phonon dispersion, as seen with the $9^{th}$ NN results converging to experimental values from literature[45].

From Figure 3, it is apparent that the phonon dispersion for crystalline silicon is not correct unless interatomic interactions extend out to $8^{th}$ nearest neighbors, although typical empirical potentials such as Tersoff or MEAM only include $1^{st}$ or $2^{nd}$ neighbors. To alleviate this issue, Rohskopf *et al.* attempted exhaustive genetic algorithm fits of empirical potentials with added long-range radial terms such as Coulomb or van der Waals, but the phonon dispersion could still not be exactly matched. Similar procedure was also followed for developing a phonon-optimized potential for aluminum metal[94] using a functional form that used 2 and 3-body terms, which although did not reproduce phonon dispersion relations accurately, predicted DFT-BTE evaluated phonon TC and aluminum oxide[95] using a bond-order based functional form which reproduced phonon dispersion relations with minor errors. Rohskopf *et al.*[87] later showed that radial potentials simply lack the geometrical flexibility to describe IFCs in all three Cartesian directions, so a long-range non-radial potential is required for this purpose. The more recently popular machine learned interatomic potentials (MLIPs) incorporate long-range non-radial interactions through their non-radial descriptors[96], such as the Gaussian approximation potential (GAP) which extends out to $5^{th}$ neighbors for silicon[97]; these long-range non-radial interactions in MLIPs are attractive for describing phonons since Rohskopf *et al*[87] showed that such interactions are crucial to exactly describe phonon dispersion curves, thus warranting further attention to MLIPs for describing phonons.

MLIPs, especially Gaussian approximation potential (GAP) and neural network potentials (NNP),[98] have been successful in modelling thermal transport. Zhang and Sun used a GAP to study the thermal conductivity of silicene, matching well corresponding DFT-BTE results [99]. Another study used GAP to study the TC of crystalline and amorphous Si and obtained reasonable agreement with experiment.[100] Their potential was trained by obtaining stochastically sampled snapshots by perturbing the dynamical matrix, an efficient way of generating fitting data. Gu and Zhao developed a SNAP potential to model the thermal conductivity of $MoS_{2(1-x)}Se_x$ 2D alloys.[101] Babaei *et al.* developed a GAP potential to study crystalline Si with vacancies, appreciably outperforming empirical potentials like Stillinger-Weber and Tersoff.[102] Li *et al.* used a NNP to model the TC in three different phases of Si, namely crystalline, amorphous and liquid Si [103]. One study from Korotaev *et al.* trained a moment tensor potential to investigate a complex material, the skutterudite $CoSb_3$, matching *ab initio* and experimental phonon dispersion and thermal conductivity well [104]. Sosso *et al.* used neural network potentials to study the thermal transport behavior of the amorphous phase of GeTe, a technologically relevant phase change material.[105] Similarly, Campi *et al.* used a similar NNP to investigate the TIC of a crystalline/amorphous GeTe boundary.[106]

The lack of accurate interatomic potentials is the main bottleneck in studying phonon transport via classical MD, thus motivating current research in the creation of accurate potentials; MLIPs are a promising step forward. If one has an accurate representation of the potential energy surface, one can accurately simulate the dynamics of atoms. This allows for an accurate sampling of phase space, which in turn can be used with further statistical mechanical theories such as Green-Kubo to extract bulk thermal properties in terms of the underlying dynamical components; an accurate interatomic potential is therefore paramount to accurately calculating properties like TC in an MD simulation, since thermal transport depends on the dynamics.

## 3. Thermal Conductivity

### 3.1. Different MD methods to calculate Thermal Conductivity

Dynamical quantities relating to thermal transport, such as the heat flux of Equation 10. or the system temperature as an average of atomic kinetic energies, are readily calculated on the fly in MD simulations, and extracting thermal properties from this microscopic information is the main motivation of using classical MD for phonon transport. One such property that may be extracted is the TC of a material. Three main approaches for calculating TC in MD simulations are being widely used: Equilibrium MD (EMD), Non-equilibrium MD (NEMD), and Approach to Equilibrium MD (AEMD). In EMD, Green-Kubo (GK) relations, which are based on the fluctuation-dissipation theorem can be used to calculate the TC from the fluctuations of the heat current correlation function during an equilibrium simulation, in the microcanonical (*NVE*) ensemble. In the NEMD, TC can be computed directly from the response of the system to a perturbation, which can be a temperature gradient resulting in a heat flux[107] or in the reverse form, a constant heat flux resulting in a temperature profile[108]. In AEMD method, TC is computed from the time response of the system to a perturbation, which is usually in the form of a square or sinusoidal temperature profile.[109,110] In the succeeding sections, we summarize the theoretical formulation of these three approaches.

### 3.1.1 Equilibrium Molecular Dynamics

Since the time average of instantaneously calculated properties corresponds to the real macroscopic observable (a numerical experiment can be thought of a sample of a small time), observable properties can therefore be calculated by time averaging the instantaneous MD properties. Statistical mechanical theories, however, relate *ensemble averages* to observable properties. The application of such statistical mechanical theories therefore requires that the time average of a property in MD be equal to the ensemble average, which is true if the system is ergodic. Most systems studied in MD are ergodic, but care should be taken that this is not always the case as in some glasses, metastable phases, or highly harmonic systems.[111] Given an ergodic system, one can apply statistical mechanical theories such as the GK formula to calculate transport coefficients in using time averages in MD simulations, which correspond to ensemble averages in statistical mechanics. For TC, the GK formula takes the form:

$$\kappa^{\alpha\beta} = \frac{V}{k_B T^2} \int_0^\infty \langle Q^\alpha(t+t')Q^\beta(t) \rangle dt' \qquad (8)$$

where $\kappa^{\alpha\beta}$ is the TC tensor and $Q^{\alpha}$ is the heat flux vector component in the $\alpha$ Cartesian direction. The quantity $\langle Q^{\alpha}(t+t')Q^{\beta}(t)\rangle$ is the heat flux autocorrelation function at time $t$, which describes correlations between equilibrium fluctuations in the heat flux.

For the application of Equation 8 in an MD simulation, one must calculate the heat flux as a function of time. Recent derivations[112] of the atomistic heat flux vector $\mathbf{Q}$ in terms of the dynamics of ions begin with the expression

$$\mathbf{Q} = \frac{d}{dt}\sum_i \mathbf{r}_i E_i \tag{9}$$

where the sum is over all atoms $i$ with position vector $\mathbf{r}_i$ and energy $E_i$, but the authors are unaware of a rigorous basis for such an expression. These modern derivations agree, however, with Hardy's result for the atomistic heat flux derived rigorously from quantum mechanical arguments:

$$\mathbf{Q} = \frac{1}{V}\sum_i \left[ E_i \mathbf{v}_i + \sum_{j\neq i}\left(\frac{\partial U_j}{\partial \mathbf{r}_{ij}}\bullet \mathbf{v}_i\right)\mathbf{r}_{ij}\right] \tag{10}$$

where $V$ is the system volume, $U_j$ is the potential energy of atom $U_j$, and $\mathbf{v}_i$ are the atomic velocities. Equation 10. has many similar forms, especially those which replace the virial term with an atomic stress tensor for two-body interactions, as used by the popular MD code LAMMPS; Boone *et al.* recently showed that this atomic stress approximation of the heat flux is incorrect for many-body potentials, although it is appropriate for 2-body potentials[113]. To avoid this issue while still retaining the convenient use of the stress tensor in calculating heat flux, Surblys *et al.* used a more general stress tensor appropriate for many-body interactions in evaluating the heat flux.[114] This recent controversy surrounding the peculiarities of Equation 10 applied to many-body potentials[113] stemmed from the popular assumption that $\mathbf{F}_{ij} = \frac{\partial U_j}{\partial \mathbf{r}_{ij}}$, which may not be true for potential containing more than 2-body interactions, although it is true for potentials like Tersoff, which have definable $\mathbf{F}_{ij}$ terms as sums over 3-body terms.[115] Such attention is warranted, as it has been shown that TC calculations are sensitive to the form of heat flux used,[116] and may result in drastically miscalculating thermal conductivity if the 2-body heat flux expression is used with many-body potentials[115].

Regardless of the heat flux formalism used, the instantaneous heat flux is readily calculated on the fly during MD simulations, and its time averaged behavior is related to thermal conductivity (TC) via the GK formula of Equation 8. Equation 8 is therefore used to calculate phonon TC in EMD simulations, where the system is first thermodynamically equilibrated at a temperature and pressure, and no energy or temperature gradients exist except for equilibrium fluctuations. The time average of Equation 8 is equivalent to an ensemble average for an ergodic system, so Gordiz and Henry recently showed that simulation time can be decreased by taking many ensemble averages of shorter simulations instead.[117] The EMD method is, however, not without its difficulties; at long times, the signal-to-noise ratio can be unfavorable for time-correlation functions.[31] It can therefore be difficult to converge the autocorrelation function with time.

### 3.1.2 Nonequilibrium Molecular Dynamics

If one chooses not to equilibrate the system, a nonequilibrium state can be maintained by introducing variables that control system temperature, pressure, energy, heat flux, etc. Using such thermostats, barostats, ergostats, etc. at system boundaries to maintain a nonequilibrium constitutes NEMD, and the nonequilibrium states are often held in a steady-state situation for simplicity. Some examples of nonequilibrium steady states include Couette flow[118] with two different velocity boundary conditions resulting in a linear velocity profile, or a slab with two different constant temperature boundary conditions, which yields a linear temperature profile[108]. These boundary conditions in MD simulations are enforced using thermostats or artificial kinetic energy swapping which control atomic velocities in desired regions. For these conditions to result in a nonequilibrium steady state, however, the time-averaged values of local dynamical quantities such as velocity and temperature must vary along the system.[119] It is from these gradients in instantaneous dynamical quantities, such as velocity and temperature, that transport properties like viscosity or TC may be obtained.

To make use of nonequilibrium in a simulation for thermal transport, one may set up an analogy to macroscopic experiments where a temperature gradient is imposed on the system of atoms, and then calculate the time averaged heat flux through the system, from which the TC is obtained via Fourier's law. One would find that this is difficult, however, because the instantaneous heat flux of Equation 10 experiences large fluctuations and therefore has a slowly converging average. Furthermore, large temperature gradients are required to distinguish heat flow from the noise of these fluctuations. To overcome these issues, Muller-Plathe devised a method, often called "reverse" NEMD (RNEMD), that makes use of the reverse analysis; impose a known heat flux (i.e., a known difference between energies at hot and cold reservoirs) on the system and measure the resulting temperature gradient.[108] The advantage of the Muller-Plathe method is that the highly fluctuating and slowly converging heat flux is known *a priori*, and thus need not be calculated during a simulation, although the temperature and its gradient must also be calculated as averages. From the ensemble average of the temperature gradient, and the difference in kinetic energy between the hot and cold reservoir, the TC is readily calculated.[108] The geometry and boundary conditions of a NEMD simulation are of importance, and most researchers simply use the scheme in Muller-Plathe's original paper[108], where a slab of material has a hot reservoir in the center and two cold reservoirs at ¼ and ¾ of the length of the slab. Such a configuration results in the steady-state temperature profile of Figure 4 (a), and allows continuous temperature distributions at the periodic boundary conditions.

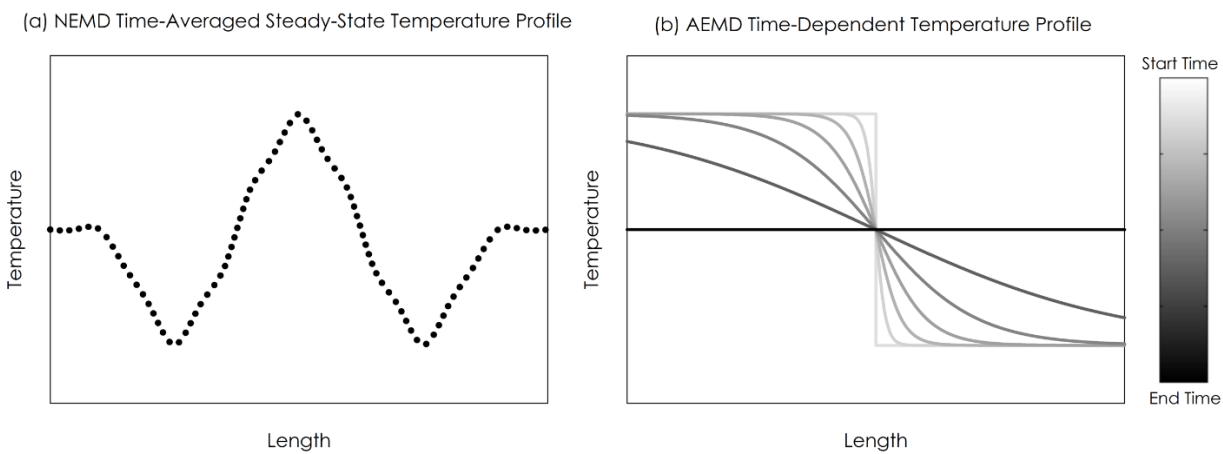

**Figure 4**. (a) Steady-state temperature profile of a NEMD simulation using the energy distribution scheme in Muller-Plathe's original paper[108], where two cold reservoirs are at ¼ and ¾ the box length, and a hot reservoir is located at the middle. (b) Time-dependent temperature profile in the AEMD method, showing the evolution of the temperature profile from the start of the simulation (light grey) until equilibrium is reached (black line).

One subtlety of the RNEMD is that the kinetic energy from the cold reservoir is artificially transferred to the hot reservoir to conserve total energy, and this exchange of velocities may disturb the stability of numerical integration of the equations of motion, resulting in a drift in total energy. Care must therefore be taken to use smaller timesteps than usual to ensure numerical stability.[120] Extensions of the RNEMD method by Kuang and Gezelter have also sought to fix this issue via "velocity scaling and shearing" on all molecules in the system.[121]

The NEMD method of Jund *et al.*, on the other hand, uses thermostats to control the temperature of the two reservoirs, instead of artificially swapping velocities between the two reservoirs.[122] This method, often called "direct" NEMD, has been shown to give slightly different temperature profiles than RNEMD, although both are within the range of experimental measurements for simple systems like single layer graphene.[120] In the case of graphene, both methods resulted in identical results, although the direct method of Jund *et al.* may be more convenient due to avoid numerical stability issues with swapping kinetic energy between atoms. Direct method has been used extensively, with various studies such as correctly simulating the high TC of carbon nanotubes[123], providing insight into the nature of low TC of layered $WeS_2$ crystals[124], and showing equivalence to calculations performed via EMD for various material structures.[125] In addition to its wide applicability, NEMD improves the signal-to-noise ratio in a measured response by inducing a much larger-than-equilibrium response artificially and measuring the steady state response to such a perturbation, compared to EMD. While EMD and NEMD are perhaps the most widely known and available to researchers, a third method has recently emerged with notable benefits; that is the "approach-to-equilibrium" MD (AEMD) method developed by Lampin *et al.*[109]

### 3.1.3 Approach-to-Equilibrium Molecular Dynamics

In AEMD, the system is initially driven out-of-equilibrium by declaring part of the system heated to a different temperature than the rest. The system is then allowed to relax, i.e. approach equilibrium, and the time evolution of the temperature difference between the two parts is monitored, resulting in the time-dependent temperature profile of Figure 4(b). Specifically, both sides of a material are heated to different temperatures, and their time to approach equilibrium is fitted to a bi-exponential curve as described by Lampin *et al.*[109] This time to approach equilibrium is then input to an analytical model derived by from the heat conduction equation, from which TC is readily obtained.

This method offers three notable advantages: (1) The time to reach equilibrium is on the order of 0.1-1000 ps depending on the material and size, thus greatly reducing computational cost of computing an autocorrelation function in EMD, or establishing a steady state current in NEMD, (2) The absolute energy flux of Equation 10 need not be known, thereby reducing the need to deal with a numerically noisy quantity, and (3) The calculation of the average temperature over extended portions of the system corresponds to more realistic temperature measurements, as opposed to local (< 1 nm) definitions of temperature. The latter of these advantages is most notable, since this peculiarity of AEMD allows clear correspondence between simulation and experiment, and indeed the AEMD method was inspired by the experimental set up of the laser-flash method[126] to measure macroscopic thermal diffusivity[109]. Furthermore, the reduced computational requirements allow the study of large systems, which may contain realistic structural features such as grain boundaries that can be incorporated in the MD simulation; Melis et al recently took advantage of this aspect of AEMD to compute TC of nanocrystalline and amorphous silicon, yielding excellent agreement with experiment.[127]

In addition to the various techniques and methods to set up or analyze EMD, NEMD, or AEMD to calculate the total or bulk TC, it is also important to compute the modal contributions to TC. The next discussion discusses the computational aspects of modal analysis of TC.

### 3.2. Modal Analysis for TC

The equilibrium and nonequilibrium methods discussed in Sections 3.1.1 and 3.1.2 to estimate thermal conductivity rely on expressions like the atomistic heat flux of Equation 10 or the atomistic temperature defined as an average of all atomic kinetic energies, which are useful for extracting the overall behavior of heat conduction or bulk TC. In terms of insights into phonon contributions to heat transfer, however, expressions like the atomistic heat flux or temperature, depend only on atomic dynamical variables (displacements, energies, velocities, etc.), and therefore do not provide direct insight into normal mode contributions. This is the topic of more current research, where atomistic quantities like temperature[128] or heat flux[129] are decomposed into contributions from individual modes at any instantaneous time in a simulation. Furthermore, the use of MD simulations to sample phase space allows one to extract anharmonic phonon quantities such as relaxation time.[130] Overall, most MD-based approaches to study phonon transport fall into two main categories: (1) approaches that extract the phonon relaxation times from anharmonic MD simulations and utilize the BTE formulations to calculate the phonon contributions to TC, and (2) approaches that determine the spectral or modal contributions to heat flux in the system and then use the EMD or NEMD formulation of TC to extract the spectral or modal contributions to TC.

### 3.2.1 Modal Analysis Based on Relaxation Time Calculations: Normal Mode Analysis

Normal mode analysis (NMA) involves analyzing the behavior of normal modes that ultimately lead to transport properties. One way of doing this involves observing the time correlation of individual mode amplitudes, which provides insight into how that mode decays due to anharmonicity in the interatomic potential. First, one must obtain the normal modes of the system using the descriptions in section 2.3. Once the modes are identified, their amplitudes and velocities during a MD simulation are obtained using Equations 4 and 5, respectively. It is important to note here that these equations are gamma point ($\mathbf{k}=0$) relations, which can be applied to any structure regardless of symmetry, although one may also utilize the symmetry of crystals to deal with modes other than just the gamma point ($\mathbf{k} \neq 0$); doing so leads to an expression for the amplitude/coordinate of mode $n$ as a plane-wave modulated version of Equation 4,

$$X(\mathbf{k},n) = \frac{1}{\sqrt{N}} \sum_{jl} \sqrt{m_j} \exp[-i\mathbf{k}\cdot\mathbf{r}(jl)]\mathbf{e}(j,\mathbf{k},n)\cdot\mathbf{u}(jl) \tag{11}$$

Where $N$ is the number of primitive cells in the periodic crystalline structure (e.g. for an SCLD calculation, we treat the entire supercell as a single primitive cell so that $N=1$), and the combination of indicies $jl$ denotes the basis atom $j$ of primitive cell $l$.

During an MD simulation, the mode amplitudes of Equation 4 (or Equation 11 if one desires to incorporate crystalline symmetry and $\mathbf{k} \neq 0$ modes) may be analyzed to study contributions to thermal transport; the oldest of such methods is referred to as "time-domain normal mode analysis" (TDNMA) [129,131–136] and involves estimating the relaxation time from MD simulations, which are then input to kinetic models of TC. Ladd *et al*. first applied a TDNMA method by noting that the amplitude of a normal mode decays according to terms related to its anharmonic interactions with other modes. In the context of EMD, this leads to an expression for the phonon relaxation time $\tau(\mathbf{k},n)$ of a mode as the autocorrelation of that mode's deviation $\delta X$ from its mean amplitude[131]

$$\tau(\mathbf{k},n) = \frac{\int \langle \delta X(\mathbf{k},n,t+t')\cdot\delta X(\mathbf{k},n,t)\rangle dt'}{\langle \delta X(\mathbf{k},n,t)^2 \rangle} \tag{12}$$

Where $X(\mathbf{k},n,t)$ is the coordinate/amplitude of mode branch $n$ with wave-vector $\mathbf{k}$ at time $t$, given during a MD simulation by Equation 4. The idea here is that once one knows $\tau(\mathbf{k},n)$, then this mode's contribution to TC is readily obtained via the phonon gas model of TC. Similarly, spectral energy density (SED) techniques involve projecting atomic displacements and velocities onto the normal modes, which are then input to a spectral energy density parameter; this parameter then fit to a Lorentzian function to get the half-width at half-maximum $\Gamma$, from which $\tau \propto \Gamma^{-1}$.[137–140] Similar to the TDNMA method, the utility of the SED method is obtaining mode relaxation times, which can then be used to study contributions to the kinetic model of TC. The relaxation times provided by TDNMA and SED methods have been used in literature for studying phonon transport in mostly crystalline materials, such as solid argon,[132] silica,[141] silicon,[129] and carbon nanotubes.[138] By calculating the structure factors of the disordered modes from a supercell gamma point lattice dynamics calculation, Larkin and McGaughey [142] could obtain an effective dispersion curve and hence the group velocities belonging to the propagating modes of vibration in amorphous solids. This then allowed them to extend the SED approach to study phonon transport in amorphous silica and amorphous silicon.[142] Although the effective dispersion technique has been utilized in other experimental [143–146] and numerical [134,147–150] studies as well, it can only capture the effect of propagating modes under the SED framework. Therefore, having access to a modal analysis approach that is independent of the definition of group velocity particularly when dealing with broken-symmetry systems allows for determining the contribution by all classes of vibrational modes to the transfer of heat in the system. The group velocity-independent modal analysis approaches function by directly decomposing the heat flux in the system and are explained in the following subsection.

### 3.2.2 Modal analysis based on direct decomposition of heat current

Two modal analysis methods have been proposed by Zhou *et al*.,[151,152] both of which are based on NEMD implementation. In their first method,[151] called time domain direct decomposition (TDDD) method, the modal contributions to thermal conductivity is obtained by directly replacing the modal contributions to velocity, total energy and second-order stress tensor of each atom into the definition of heat flux in Equation 10. In the second formalism proposed by Zhou *et al.*,[152] called frequency domain direct decomposition (FDDD) method, the authors calculate the spectral contributions to the exchanged heat flow between two atoms in the system and replace them in the definition of heat flow for TC, which results in the spectral contributions to TC. The authors' approach to find the spectral contributions to is based on a previous study by Chalopin and Volz[153] where such a methodology was developed to determine the spectral contributions to the interfacial heat flow, which will be more talked about in Section 4.4. The applicability of TDDD method to all families of solids seems to be limited since the approach is based on well-defined wave-vectors, which only exist in crystalline solids. On the other hand, the FDDD method can technically be applied to all families of materials, however by being based on spectral decomposition, as the authors themselves point out,[152] the formalism cannot determine the contributions by individual modes of vibration to the thermal conductivity and all the contributions coming from similar frequency vibrations are lumped together.[151]

GKMA is another modal analysis approach proposed by Lv and Henry[154] that utilizes modes of vibration obtained via SCLD. Knowing the eigenvectors from a SCLD calculation allows one to extract modal velocities during a MD simulation via Equation 5, from which modal contributions to atomic velocities can be obtained using Equation 7. As proposed by Lv and Henry, [154] by directly replacing the atomic velocities in the heat flux of Equation 10 by their modal contributions, the modal contributions to heat flux $\left(\mathbf{Q}_n\right)$ are obtained,

$$\mathbf{Q} = \sum_n^{3N} \mathbf{Q}_n = \sum_n^{3N} \frac{1}{V} \sum_i \left[ E_i \mathbf{v}_i(n) + \sum_{j \neq i} \left( \frac{\partial U_j}{\partial \mathbf{r}_{ij}} \bullet \mathbf{v}_i(n) \right) \mathbf{r}_{ij} \right] \qquad (13)$$

where $\mathbf{v}_i(n)$ is the contribution of mode $n$ to the velocity of atom $i$. Substituting this expression into the GK formula of Equation 8, therefore, also decomposes TC into mode contributions, and constitutes the GKMA method.[154]

It should be noted that there are many ways to go about decomposing the heat current, but only decomposing based on velocity can be implemented in general. One could modally decompose the force, but this leads to issues with periodic images, since one cannot distinguish the difference between contributions from atoms within the super-cell vs. the same atoms in periodic images of the super-cell.[73] Others have tried decomposing the displacements, e.g. Sun and Allen have used a power series of the relative displacements to decompose heat current.[155] Decomposing the displacement term in the heat flux is problematic if one simply considers the initial configuration of an MD simulation, which often has zero displacements, but nonzero heat flux due to finite velocities.[73] Decomposing the displacements, however, cannot resolve the distinct mode contributions in this unique scenario, so it is not correct in general. In conclusion, the decomposition based on velocity is both rigorous and correct. For this reason, the GKMA method is only based on the velocity decomposition approach to modally analyze the Hardy's heat flux expression. By finding the modal contributions to TC under the correlation picture, GKMA can explain several previously unexplained experimental results compared to phonon gas model methods. For example, the study of Seyf *et al.* showed that the virtual crystal approximation for alloys cannot match experimental TC values for InGaAs alloys.[74] Another example showed, a quantum correction may be made to individual modes which then helps explain low temperature TC behavior in amorphous materials.[156]

While time domain methods such as TDNMA, TDDD and FDDD have been applied to NEMD[151] simulations, general modal analysis of modes determined by SCLD for disordered systems has yet to be applied to NEMD. Although one could easily obtain mode contributions to the system temperature by substituting mode contributions to atomic velocities (Equation 7) into the atomistic definition of temperature as an average of all atom kinetic energies. From mode contributions to temperature, one could easily study mode contributions to TC via the NEMD approach. This would be particularly useful for the study of interfaces, which is the topic of Section 4.4. Aside from interfaces, however, there are other ways to break crystalline symmetry in a material such as defects, alloying, doping, etc. These changes to a crystalline structure influence thermal transport property, and such effects may be studied using MD, which is the topic of the next section.

## 3.3. Effects of system parameters on TC

In this section, we survey how MD simulations have been utilized by various researchers to gain insights on how various intrinsic and extrinsic physicochemical parameters (system size, structural and compositional disorder, and isotopes) affect TC of phonon-dominated materials.

### 3.3.1. Size effects

For many nanoscale device applications, reducing the system dimensions to a few nm is important. While system sizes do not usually impact the specific heat and group velocities above cryogenic temperatures, the MFPs of phonons are significantly affected by system sizes due to increased phonon-boundary scattering events.[1,2] Although experiments have shown that systems smaller than a few micrometers (μm) often exhibit size effects,[157] in MD simulations, however, accessing μm-scale systems is difficult and usually system sizes of a few nm are used. Consequently, MD simulations often seem to underpredict TC, primarily because calculated TC is a function of number of atoms in the MD cell, converging in the infinite limit to the bulk value. This numerical artifact is called 'finite system-size effects' which should not be confused with the physical phenomena of size-effects on TC.[141,158–162] Both EMD and NEMD are affected by finite-size effects. Schelling *et al.*[160] assessed the effects of finite simulation cell size on TC computation and found that TC was dependent on the simulation cell size if there are insufficient phonon modes to accurately reproduce the phonon-phonon scattering in the associated bulk material, because the Brillouin-zone (BZ) resolution is too coarse. This effect can, however, be alleviated by increasing the simulation cell size until the TC reaches a 'size-independent' value.[162–164] Other authors have attributed finite-size effects to "memory" effects.[165,166] In this explanation, due to the PBC's, a phonon may pass the same point in space several times without scattering. Since the system may retain some dynamical information during the passage of the phonon, artificial correlations may exist in the autocorrelation function. In this case, the correlation function may contain artificial components.[158]

In NEMD, finite system-size effects are found to be more severe due to an additional mechanism - phonon scattering or perturbation at the sample/reservoir boundaries.[160,162] This effect is pronounced when the system size is smaller than the bulk phonon MFP. For instance, Chantrenne *et al.* observed higher size effect for PBC than for free BC's which they attributed to numerical errors from a phonon mode re-entering the simulation box and undergoing more scattering.[165] But in combination with wave vector analysis, they show that, qualitatively, the size dependence of TC is due to the discretization of the wave vectors for small systems and the influence of the phonon scattering at the boundary surfaces. Subsequently, finite-size effects usually affect long phonon-wavelength systems more than short wavelength systems.

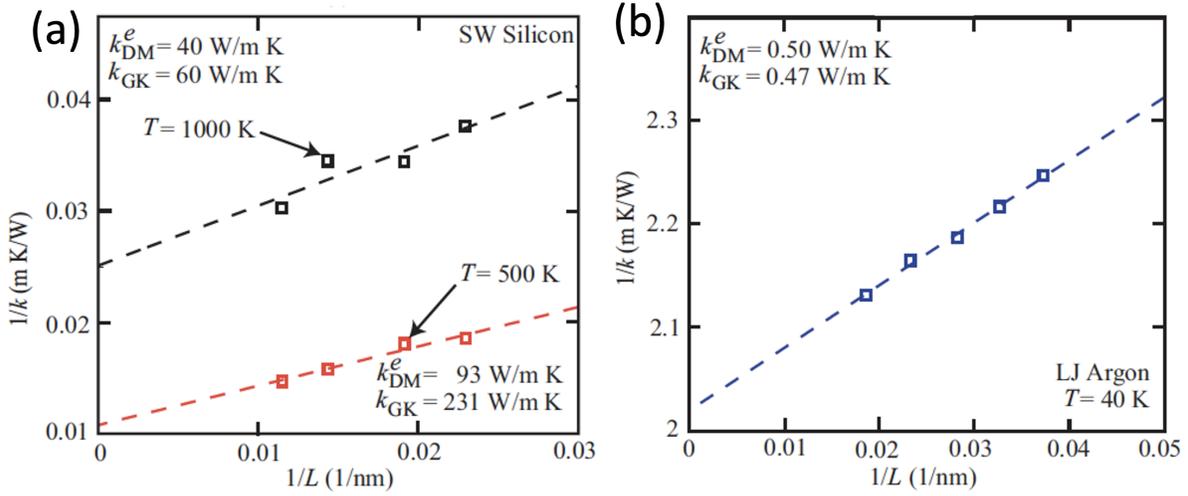

**Figure 5.** TC calculated using NEMD and EMD methods for (a) Si using Stillinger-Weber potential and (b) Argon using Lennard-Jones potential (reprinted from Sellan *et al.*[162], copyright 2010, with permission from the American Physical Society).

Sellan *et al.* recommend protocols to evaluate TC reliably and accurately using MD.[162] If system sizes are smaller than the largest bulk mean-free paths that dominate the TC are considered, a linear relationship between *1/k* and *1/L* may be incorrectly inferred and the TC can be severely underestimated. They also recommend when to appropriately use GK method and when to use NEMD. They recommend using GK method every time unless a converged HCACF is not obtained in which case, NEMD may be used. If NEMD TC in the infinite length limit does not agree with the experimental TC, then it cannot be used to study size effects. In such a case, a LD-based method needs to be used. Therefore, generally speaking, sufficiently large systems with EMD-GK with PBC in all directions is enough to evaluate bulk TC, whereas for NEMD, an extrapolation method based on *1/k* v/s *1/L* relation may need to be adopted.[165,167,168] Figure 5 (a) and (b) shows *1/k* v/s *1/L* applied for Si and Ar systems respectively by Sellan *et al.*[162] Following these protocols, MD simulations have been used to evaluate TC of size-affected systems such as thin films,[169–173] nanowires,[165,174,175] nanoribbons[164,176–178], and superlattices.[179–182]

Size effects on the TC of semiconductors have been widely studied using EMD and NEMD simulations.[129,165,175] One of the earliest studies involves elucidating spectral contributions to TC of Si by Henry and Chen[129] who used EMD-GK in combination with LD to calculate TC as a function of MFP and frequency. Essentially, each eigen state was assigned an MFP and their contribution to TC was calculated, which gives the actual effect of size on TC in terms of the MFP of phonons. Likewise, Wang *et al.* studied the dependence of diameter on TC of GaN nanowires in longitudinal and transverse directions using NEMD.[175] They observed reduced TC for smaller diameters and attributed it to: (i) the change of phonon spectrum in one dimensional structures, which modifies the phonon group velocity and the scattering mechanisms, and (ii) the boundary inelastic scattering, which increases diffuse reflections on the surfaces. As the diameter of nanowires increases, so does the TC, mainly because the boundary scatting rate decreases. The significant reduction of TC was attributed to the high surface to volume ratios of the nanowires. Specifically, the relatively large fractions of surface atoms enhance surface scattering of phonons and decrease the phonon MFP, resulting in lower TC that is proportional to the MFP. Similar observations were also observed by Chantrenne *et al.*[165]

With growing interest in graphene for various thermal management applications, graphene nanoribbons and sheets have been considered for various MD studies.[120,176,178,183–185] Cao *et al.* used NEMD to predict size effects on TC of GNR and found that the reduction in TC can be attributed to the resonance of out-of-plane modes.[183] Similar observations were also made by others[176] for a multilayer graphene system, where an increase in number of layers reduced TC due to interlayer van der Waal interactions that constraints the in-plane phonon transport.[185] Xu *et al.* also make important observations on the divergent TC of graphene, obtained from RNEMD calculations.[120] They attributed the length dependent TC to multiple mechanisms. Firstly, to the ballistic propagation of extremely long-wavelength, low-frequency acoustic phonons. As sample size increases, more low-frequency acoustic phonons were excited which contributed to thermal conduction, resulting in a length-dependent behavior. Secondly, to the selection rules for three-phonon scattering, the phase space of which was found to be strongly restricted by the reduced dimensionality. Upon comparing the phonon populations in MD simulations at equilibrium and non-equilibrium conditions, Xu *et al.*[120] found that in the latter case, the population of out-of-plane modes is augmented, whereas in-plane modes with polarization in the direction of the heat flux propagation was slightly depopulated.

In addition to graphene, 2D materials like $MoS_2$,[186,187] phosphorene,[173] and carbon allotropes[188] have also been systems of MD investigations. Jiang *et al.* investigated $MoS_2$ phonons using NEMD and predicted the in-plane TC accurately (~6 W/mK) for both zigzag and armchair configurations.[186] Moreover, in $MoS_2$ nanosheet and nanoribbon, strong size-effects have been observed using NEMD calculations.[187] TC of monolayer $MoS_2$ was computed to be 1.35 W/mK, which is three orders of magnitude lower than that of graphene while the MFP is 5.2 nm, which is two orders of magnitude lower than that of graphene. On the other hand, TC of monolayer $MoS_2$ NRs was found to be insensitive to width and edge-type, indicating that the Umklapp scattering process dominated the thermal transport in $MoS_2$ NRs.[187] For phosphorene systems, Zhang *et al.* reported the in-plane TC computed using NEMD to show strong dependence on the sample length and mechanical strain, which they credited to the competition between the boundary scattering and phonon-phonon scattering.[173] On the other hand, EMD and RNEMD studies on carbon allotropes using a modified Tersoff potential have shown that TC does not depend on the sample width for a fixed length of 50 nm but increases monotonically when length is increased from 50 to 1000 nm, indicative of the strong contributions from acoustic phonons with longer wave-length.[188]

Thermal transport in polymers has also been studied with MD. Using EMD-GK simulations Henry *et al.*[189] delineated the divergent TC behavior of polymers. Their studies show a transition from 1D-3D transition of phonon heat conduction in polyethylene, wherein a single chain's TC was found to be large but when two chains were allowed to interact, TC decreased by ~40% signifying a sharp 1D-2D transition. Upon allowing 2D structures to interact and form a bulk crystal, TC decreased by another 10% indicating a less sharp 2D-3D transition. They ascribed these observations to the competition between heat conduction contribution and increased phonon-phonon scattering resulting from the addition of new phonon modes. Similar divergent behavior was also observed in NEMD simulations of Wang *et al.*[190] in 1D polymeric chains.

In summary, EMD, NEMD, and RNEMD simulations have been effectively implemented to qualitatively and quantitatively study phonon transport in size-affected systems yielding valuable insights on the boundary scattering mechanisms. Another physical parameter that can be tuned to control phonon transport is the structural disorder. In the next section, we look at how strong structural disorders, as in the case of amorphous solids and polymers affect TC.

### 3.3.2. Strong structural disorder (amorphous solids and polymers)

As described in section 2.3, for amorphous solids, rather than adopting the traditional PGM physical picture, it is more appropriate to distinguish three vibrational quasiparticles: propagons, diffusons, and locons. Based on the SCLD calculations on amorphous silicon (a-Si), Allen and Feldman (A-F)[191,192] first observed this different natural categorization of vibrational modes.[191] Propagons are spatially delocalized low frequency modes that extended through the entire system and exhibited periodicity in the eigenvectors, resembling the traditional picture of phonons. They exist at low frequencies below the Ioffe-Regel (IR) crossover as shown in Figure 6. Locons were spatially localized high frequency modes involving only with a small cluster of atoms. Diffusons were in the majority and had random eigenvectors, that lack periodicity.

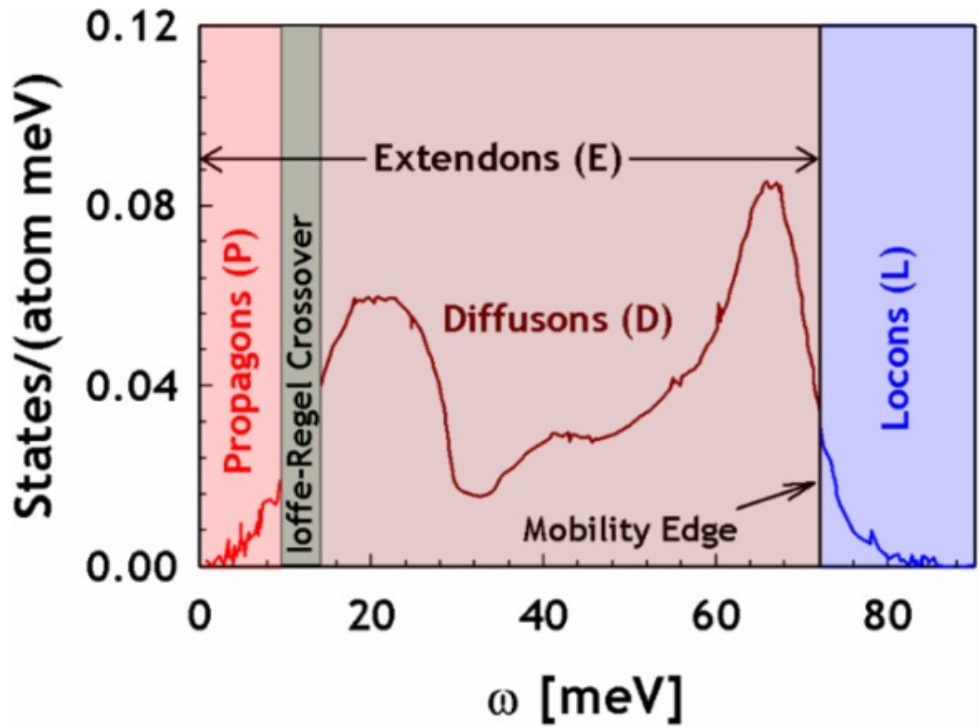

**Figure 6.** Density of vibrational states for a model of amorphous silicon and classification of vibrational modes(reprinted from Ref. [193], copyright 2017, with permission from John Wiley and Sons)

The focus of MD simulations has been on investigating how strongly structural disorder can reduce TC from the perspective of thermoelectric and insulating materials. Both NMA and GKMA have been used for this purpose to study different materials, which can be broadly classified as glasses, amorphous semiconductors, and polymers.[142,156,167,194,195] More specifically, MD has been used to gain insights on the contributions from propagons, diffusons, and locons to thermal transport. Larkin and McGaughey used NMA and GK approaches to study the contributions of propagating and non-propagating modes to TC of a-Si and a-SiO$_2$.[142] Their studies show that propagons contribution to a-SiO$_2$ is very low. It was shown that only 6% of modes in a-SiO$_2$ are propagating, whereas 35% in a-Si are propagons. Their calculations also agreed with that of He *et al.* with respect to propagon contribution.[195] The substantially smaller percentage of propagons in a-SiO$_2$ was attributed to weak bonding between SiO$_4$ tetrahedra in a-SiO$_2$, whereas in a-Si, a network of strong tetrahedral bonds ensured the existence of propagons at higher frequencies. They observed a substantial difference in propagating and non-propagating mode characters owing to bonding behavior (weak tetrahedra in a-SiO$_2$ vs strong in a-Si). Another attempt in elucidating TC of amorphous solids was conducted by Lv and Henry using GKMA.[156,194] For a-SiO$_2$ they found that locons contributed to more than 10% of TC,[156] whereas for a-C, propagons were found to contribute only 13% to TC and the majority was from diffusons.[194] Co-existence of propagons and diffusons have also been observed in MD simulations.[167] In an early work, Oligschleger *et al.* used EMD-GK and NEMD to study TC of Se and SiO$_2$ glasses and observed the co-existence of propagons and diffusons in the low frequency regime.[167]

Another class of amorphous materials that have been investigated are polymers.[196–201] Polymers consist of long-chain molecules that are entangled when in their amorphous state, are usually melted and molded or cured at moderate temperatures, and used as alternatives for metal-based heat exchangers in automobiles, data-centers, and hand-held electronics.[202] The earliest works on evaluating TC of polymers were reported by Lussetti *et al.*[197] and Terrao *et al.*[196]

who used Muller-Plathe NEMD method to evaluate TC of LJ polymer systems. Their studies gave a TC value of 0.15-0.45 W/m-K. Amorphous polymers have thermal conductivities on the order of 0.1 Wm$^{-1}$ K$^{-1}$ and are generally considered as thermal insulators. The general consensus is that the TC is positively correlated with the degree of cross-linking.[198,199,203,204] However, Luo *et al.*[200] report contrasting observations. They show using NEMD simulations that not only is the TC of a single chain limited by conformational disorder, but even in bulk amorphous structures, structural disorder within collections of chains can limit TC. From a standpoint of modal contributions, Shenogin *et al.* report that the TC is dominated by very low-frequency propagons while most modes were localized and high frequency, and had negligible contributions. Contrastingly, Hsieh *et al.* used NEMD simulations to show that diffusons contribute the most to polymer TC.[205]

Shenogin *et al.* also compared the results of harmonic approximation of Kubo linear response theory of Allen and Feldman and NEMD results for inorganic and polymeric glasses.[201] In general, they found that the TC of a material was almost entirely dictated by low frequency propagating modes, while most high frequency modes (>5 THz) were found to be localized and negligibly contributed to TC. In addition, for a-Si, LJ glass, and bead-spring glass, A-F theory was found to accurately predict TC whereas for a-SiO$_2$ and polystyrene, it underpredicted TC. They observed that for a-Si and LJ-glass, harmonic theory was good enough to predict TC accurately, whereas for a-SiO$_2$, MD results were 30% higher than A-F theory. They attributed it to anharmonic energy transfer between locons or between locons and extendons.

How TC and vibrational properties of individual materials affect the TIC of the interfaces they form has also been studied by several researchers.[206–208] For instance, Giri *et al.* studied how structural disorder in bulk materials influences the TIC of a-superlattices (SL) using NEMD.[206]

They found that increasing the mass-mismatch in amorphous SLs yields lower TIC, indicating that a mismatch in vibrational spectra is responsible for the thermal resistance. They also found that thermal transport in SL's were dominated by diffusons for amorphous Si/Ge and Si/heavy-Si SLs.

Combined MD-LD approach has been effective in studying phonon transport of systems with strong structural defects and inspecting under what conditions they converge to the A-F theory predictions. For systems with weak structural defects, however, this taxonomy of vibrational modes is not essential because most modes would follow a propagating nature. In the following section, we examine the role of weak structural defects on phonon transport.

### 3.3.3. Weak structural defects

All real materials have structural defects that can reduce TC relative to ideal structures. Whereas TC of materials with strong structural disorder is better modeled by the A-F theory, materials with weak structural disorder can be dealt with analytic models that provide a solution to the BTE under various simplifying assumptions. Specifically, the Klemens-Callaway (K-C) model has been widely used to model the effect of structural defects on TC.[209–212] At the core of the model is Callaway's expression for TC, which adopts the relaxation time approximation (RTA), assumes a Debye spectrum, and separately treats normal and Umklapp processes.[209] The effect of defects is fed into the K-C model through their effect on scattering rates in which, under the RTA, Matthiessen's rule applies to different scattering mechanisms. The scattering rates of point defects and dislocations are then modeled using simple analytic expression like those derived by K-C based on first-order perturbation theory.[210–212] In the more expensive DFT-BTE approach, defect scattering is incorporated in a rigorous way via a T-matrix formalism, which goes beyond first order perturbation theory and incorporates both mass and force constant variance.[213] Researchers have adopted this methodology to study point defects in diamond,[214] BAs,[49] InN,[215] and graphene,[216] among other systems. Such a formalism has also been applied to study of edge dislocations in silicon[217] and GaN,[218] though the limited supercell size used in DFT calculations necessitated the use of small dislocation dipole models, and in both studies the BTE was solved with the RTA. These methods used experimentally or computationally derived material properties to parameterize the BTE, or simplified analytic versions of it, which are then solved to obtain TC.

On the other hand, in MD, calculations are typically repeated at different concentrations of point defects and at different temperatures in order to directly assess the concentration and temperature dependency of TC reduction. Point defects in many bulk systems have been investigated in this way using various MD formalisms: Si, [219–221] SiC, [222–224] $UO_2$,[225] PbTe,[139,226] $Bi_2Te_3$,[227] $In_4Se_3$,[228] and $CoSb_3$,[229] among many others. Likewise, the effects of point defects in many 2D and nanostructure materials have also been investigated in this way, including: carbon nanotubes,[230,231] ideal and nanostructured graphene,[232–236] ideal and nanostructured $MoS_2$,[237,238] silicene,[239] and Si nanowires[240].

MD-based studies of dislocations and their effects of on TC take on a somewhat different character, as they tend to focus on the effect of a single dislocation or dislocation dipole within the simulation cell. One approach is to perform an NEMD simulation with a single well separated dislocation dipole.[241–243] Note that the dislocation dipoles used in these simulations are much farther apart than those used in DFT calculations, ensuring there is no direct dislocation-dislocation interaction. Dislocation density can be varied by varying the size of the periodic system, with larger periodic systems corresponding to lower dislocation densities, albeit simulating an unrealistic ordered array of dislocations. This approach has been adopted to study edge dislocation in Fe[241], $UO_2$[242] and PbTe[243]. A similar approach was used to study edge and screw dislocation in GaN, but under an EMD framework and using dislocation quadrapoles[244], demonstrating that screw dislocations have a stronger effect on thermal conductivity reduction than edge dislocations. An alternative approach is to study an isolated dislocation within a nanowire geometry, typically using an EMD framework. In this approach, the primary emphasis is on assessing the effects of different types of dislocations and/or the Burgers-vector-dependence of TC reduction. Studies utilizing this strategy include those investigating screw dislocation in PbSe and SiGe nanowires [245], screw dislocations in SiC [246], and screw/edge dislocations in GaN [247]

It is illustrative to consider how the results of MD-based studies compare to the predictions of the K-C model. For example, studies of point defects generally exhibit concentration dependencies consistent with that predicted by K-C models, i.e. a linear or square root dependence below or above a critical concentration.[223,229,248] A closer look at those comparisons will, however, reveal noticeable discrepancies between the specific prediction of the analytic model (sometimes fit to MD data) and the direct MD results. Likewise, Deng *et al.* compared the accuracy of a K-C model for dislocations in $UO_2$ to direct MD results.[242] While the MD data exhibited the same dislocation and temperature dependence predicted by the K-C model, the analytic model analytic model systemically overpredicted TC.[242] One interesting example to consider is that Liu et al, who used a K-C model to extend the range of validity of their MD-derived data. In this work, MD calculated TC of $UO_2$ were used to fit a K-C model that incorporated an experimentally parameterized phonon-spin scattering term. In this way, the MD data was fit to a model that was valid at low temperatures, where spin and quantum statistics effect dominate and are unaccounted for in classical MD simulations..[225]

Unsurprisingly, the simple analytic models like that of K-C have limitations that can be overcome using direct MD simulations. Ni *et al.* demonstrated this by investigating the thermal transport along a screw dislocation in SiC, a limiting case which the K-C model treats as contributing zero phonon scattering. They demonstrated a Burgers-vector dependent reduction in TC that stems from significantly increased anharmonic phonon scattering in the highly distorted core region.[246] Likewise, Yao *et al.* used MD to explore the range in which K-C model is valid for point defects, and when it begins to break down at higher point defect concentrations and for high frequency modes.[249] Sun *et al.* highlighted the discrepancy between the smoothly varying dependence of mean free path on assumed by K-C models compared to a much more varied MFPs determined from MD simulations, often deviating significantly from the theoretical prediction in certain frequency ranges.[243] Other studies using EMD of defective silicon showed a significant coupling between phonon-phonon scattering and phonon-defect scattering, resulting in an overestimation in phonon relaxation times when using Matthiessen's rule.[220]

MD has also been used for investigating more complex point defects and dislocations that require larger simulation cells, including defect complexes,[222] defects at interfaces,[250] dopant-decorated dislocations,[251] and screw dislocations in superlattices spanning a large range of periodicities[252]. In a study performed by Jones *et al.*, the combined effect of vacancies and dislocations in highly defective samples of LiF were modeled using a statistical model fit to MD data.[253] Their study involved sampling different concentrations of vacancies alone, dislocations alone, and vacancies and dislocations together. The EMD-GK predictions of each sample point, along with their respective uncertainty, were then used to fit a Bayesian model that could predict

– with quantifiable uncertainty – the TC of a sample with arbitrary concentrations of vacancies and dislocations. Figure alpha shows the prediction and uncertainty of thermal conductivity at different concentrations of vacancies and dislocations, as predicted by their model.

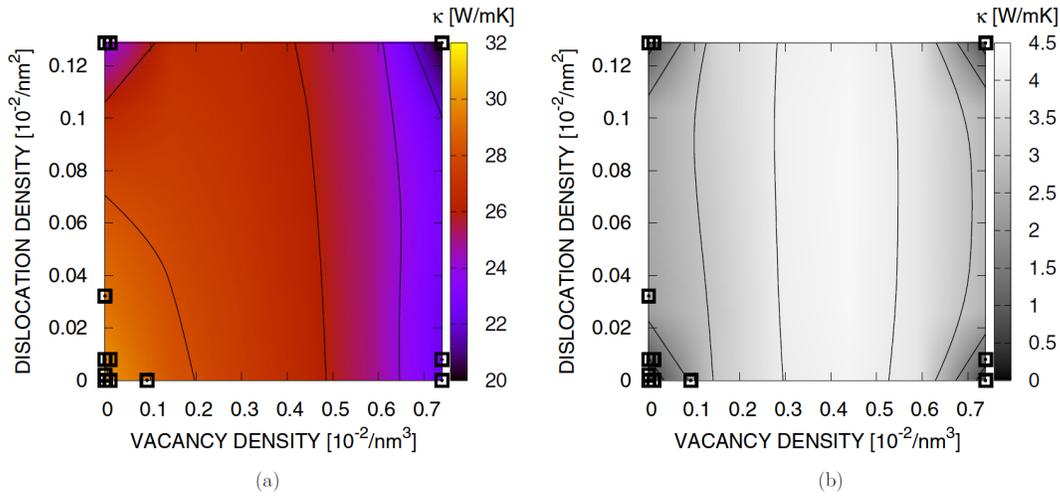

**Figure 7**. (a) Prediction of thermal conductivity perpendicular to dislocations as a function of vacancy and dislocation density. (b) Uncertainty in prediction. Square points indicate GK sample locations (reprinted from Jones *et al.* [253], copyright 2018, with permission from the American Physical Society).

Other kinds of structural defects are well beyond what can be investigated using *ab initio* methods and necessitate the use of MD. Such structural defects include pores and voids within a variety of 2D and 3D materials[219,254–260]. Likewise, large-scale structural defects in nanomaterials have been studied, including wrinkles and folds in 2D materials [261–265] and kinks/modulations in 1D material [266–268]. At the largest scale, MD has also been used to directly simulate nanocrystalline samples and predict their TC. Typically, such nanocrystals are instantiated using Voronoi tessellation, and TC is assessed at different grain sizes. Bulk nanocrystals that have been studied in this way include silicon,[269–271] PbTe,[226] alloyed

SiGe,[272] and a diamond-SiC nanocomposite[273]. Similar investigations have been made into 2D materials, primarily graphene,[184,274–277] but also h-BN[278].

While it is useful to directly simulate the effects of structural defects on thermal transport, there is a need to understand and explain such effects. In many MD studies, thus far, such analysis has been provided by simple kinetic theory arguments, looking at vibrational density of states (VDOS), group velocities, and relaxation times or mean free path of phonons. This kind of analysis is comparable to those of studies using the BTE or simpler analytic models, which themselves are based on a kinetic theory of heat transport, although the relevant data is obtained in different ways. In some cases, the VDOS and group velocities are obtained from lattice dynamics calculations using the same potentials used in their simulations.[255,279] More commonly, Fourier transformed velocity autocorrelation function to obtain the VDOS[184,239,246,264] and/or spectral energy density methods to obtain group velocities and relaxation times[139,237,246,254] have been adopted.

For the most part, these kinds of analyses are still rooted in the PGM picture. However, one of the main advantages of MD-based methods is their ability to study thermal transport beyond the assumptions of PGM. For example, the work of Estreicher, Bebek, *et al.* have challenged the picture of phonons simply scattering off of point defects, and instead identified phonon trapping in localized defect modes as an important mechanism of TC reduction.[280,281] While those studies were performed using *ab initio* MD, future studies could investigate a similar phenomenon using classical MD, exploring longer time scales, more realistic defect concentrations, and larger structural defects. In fact, it should be noted that some authors have made a connection between TC reduction and localized defect modes, utilizing either a participation ratio analysis.[238,264,282] or in some cases visualizing localized modes[238,266]. In this regard, the modal analysis developed under the GKMA framework could be a useful tool for understanding the correlation between localized defect modes and heat carrying propagating modes. To date, no such work has been attempted that the authors are aware of.

Broadly speaking, MD-based studies of structural defects take different approaches, the most straightforward of which provide direct quantitative predictions and analysis of how defects affect thermal transport in materials, as described above. However, MD-based studies are also well-suited for exploratory investigations that can provide insights into more complex defect related phenomena, helping to clarify and tease out possible mechanisms of TC reduction. For example, researchers have investigated how radiation-induced voids can affect TC in nuclear materials.[256–258] In particular, the papers by Lee *et al.* [257] and Chen *et al.* [256], who respectively studied He and Xe bubbles in $UO_2$, demonstrate the value of such exploratory MD studies. Both these works reported unexpected decrease in TC with an increase of pressure within the voids. While in the He case, this was attributed to the diffusion of He into the surrounding void interface, creating a disordered region that enhanced phonon scattering, the Xe atoms were too large to significantly diffuse into the surrounding regions, and hence was found to induce distortions in the void shape, thereby increasing scattering by increased strain.

Some researchers have adopted a survey-like approach when using MD-based method to study structural defects and their effects on thermal transport. Essentially, they scanned across the effects of many different kinds of structural defects, leveraging the computational inexpensiveness of classical MD relative to *ab initio* methods. This approach has been adopted to study a different kinds of point defects in SiC,[222,224] different interstitials in Si,[279] and ten different kinds of substitutional and intrinsic defects in $UO_2$[225]. Similarly a 2019 study by de Sousa Oliveira and Neophytou extensively scanned of different pore geometries in nanoporous silicon [255]. Their work revealed that pore density and surface area were better predictors of TC reduction than porosity and pore size and that the most important geometrical feature was the "line of sight", i.e. the extent to which direct paths of phonon transport were interrupted by pore boundaries, inducing scattering. Such kinds of studies would be difficult to perform in a purely *ab initio* context, due to both length scale constraints and the large number of calculations that are necessary when investigating many kinds of defects.

Ultimately, the accuracy and validity of MD-based studies is rooted in the accuracy of the underlying interatomic potential used. Regarding studies of structural defects and thermal transport, a few papers have used more than one potential and compared the results between them. Abs da cruz *et al.*'s study of nanocrstyalline SiGe presented similar TC predictions between a Stillinger-Weber and Tersoff potential for a given nanograin size, though the predicted values were twice as small as the experimentally observed value [272]. In Liu *et al.*'s study of point defects in UO2, two separate potentials were used, a pairwise Busker potential and a many-body potential developed by Cooper, Rushton, and Grimes; these two potentials predicted similar reductions in TC, though the many-body potential predicted a stronger conductivity reduction for certain point defects [225]. Nevertheless, caution is necessary, as interatomic potentials may not correctly capture certain behaviors, even at a qualitative level. For example, Kuryliuk *et al.* explored the effects of hydrostatic strain in silicon and observed vastly different behavior with different potentials: out of four different potentials, only the Tersoff potential correctly exhibited the right behavior of increasing TC with compressive strain and decreasing TC with tensile strain[283]. It is expected, nonetheless, that future work will show a wider adoption of potentials that can approximate *ab initio* accuracy, imbuing classical MD with high fidelity predictive power.

Evidently, classical MD has been extensively used to study the effects of weak structural defects on TC and examine the general applicability of K-C model. In addition to structural, compositional disorder can also affect phonon transport. In the following section, we consider systems with strong compositional disorder such as semiconductor alloys and look at how MD simulations have provided insights to phonon transport behavior.

### 3.3.4. Strong compositional disorder (alloys)

Semiconductor alloys find application in thermoelectric materials, where alloying has been widely employed to reduce the lattice TC to improve the material figure of merit.[284,285] Alloying gives rise to local mass and bond strength heterogeneity, both of which have been explored to understand how alloying can be used to reduce TC.[83,135,286] The majority of work in this area has looked at the mass-difference phonon scattering, based on Tamura model which describes the mass-difference phonon scattering rate based on perturbation theory.[287] The standard approximation in LD calculations is Virtual Crystal Approximation (VCA) where the alloy is replaced with a perfect, single-species crystal with properties (e.g., density and cohesive energy) equivalent to an average composition (e.g., atomic mass and/or bond strength). For predictive modeling, VCA is coupled with phonon scattering from mass disorder and used in first principles approaches. In this sense, dissimilar elements in an alloy lattice are treated as 'scattering centers' for the phonon gas, and the expressions used to model this effect can be derived. However, BTE based on VCA has been found to be insufficient in predicting TC accurately. [74,75]

On this front, MD simulations have been used to predict TC and also to verify the applicability of VCA and Tamura model.[74,134,288] Larkin and McGaughey investigated the utility of VCA for predicting the mode properties and TC of LJ argon and SW silicon alloys by a detailed comparison of the VC-NMD, VC-ALD, and GK methods (Figure8(b)-(e)).[134] In general, they found that as the compositional disorder increased, the VCA accuracy decreased. They also observed that the transport in ordered and disordered lattices could be separated into low-frequency dominated and full-spectrum materials. Low-frequency dominated materials tend to have high TC that are significantly higher than the high-scatter limit due to the large group velocities and long lifetimes of low-frequency modes. They also calculated the mass-difference phonon scattering rates for various alloy concentrations and found that the Tamura model fails in the high-frequency regime even with the lowest alloy concentration (5%), which was attributed to the lack of higher-order terms in the mass-difference perturbation. Another work by Shiga *et al.* examined the effect of mass contrast on alloy phonon scattering in mass-substituted LJ alloy crystals using EMD and NMD for different mass ratios.[288] Their studies suggest that TC prediction may be significantly underestimated if the critical frequency of mass-mismatch scattering is much lower than maximum mode frequency in the system. They also identified a critical phonon frequency, above which the mass-difference scattering rate predicted by the Tamura model deviates from that calculated directly by MD. The critical phonon frequency was also found to decrease as the mass ratio increases or decreases from 1.0, which further reduces the applicable frequency regime of the Tamura model. Studies by Mei and Knezevic on the TC of III-V semiconductor alloys using EMD also questions the VCA.[289] They stated that the larger the mass difference between the elements, the larger the deviation from VCA.

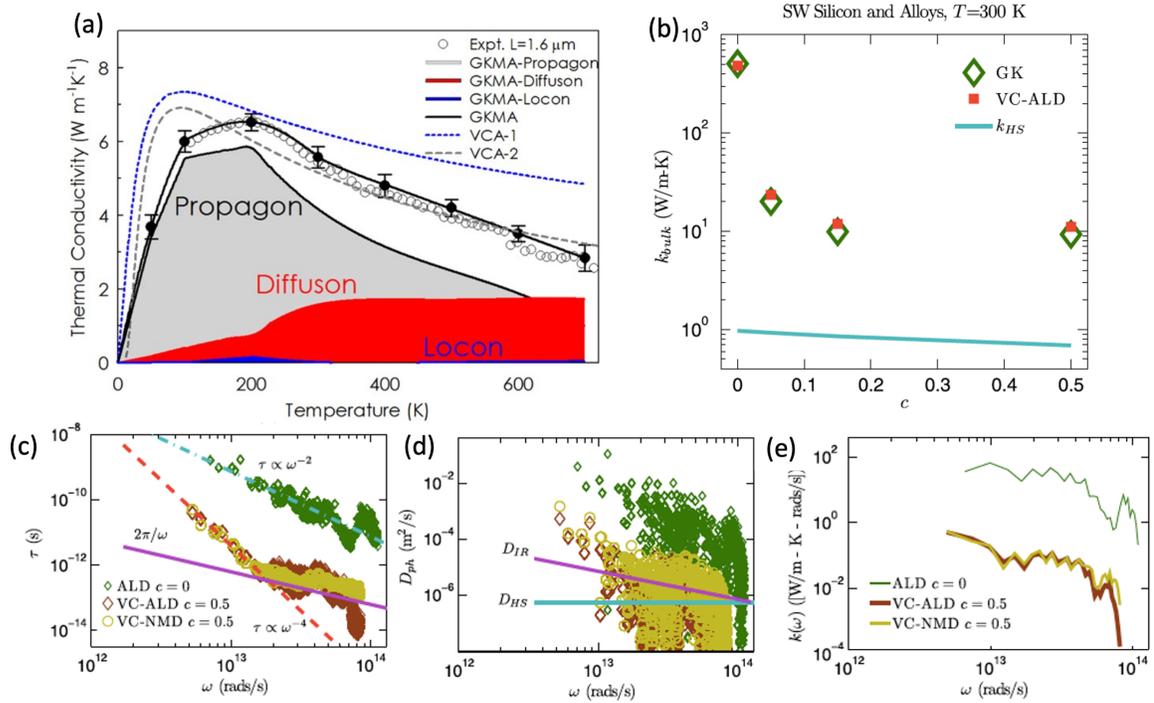

**Figure 8**. (a) Thermal conductivity of $In_{0.53}Ga_{0.47}As$ alloy and increasing contributions by diffusons and locons with increase in temperature (reprinted from Seyf et al.[74], copyright 2017, under Creative Commons Attribution 4.0 International License), (b) TC of SW-Si and alloys calculated using GK and VC-ALD methods (c) Lifetimes predicted using VC-ALD and VC-NMD, (d) Mode diffusivities compared with high-scatter limit and IR limit, and (e) Spectral TC, which show TC peaking at low frequency((b)-(e) reprinted from Larkin et al.[134], copyright 2013, with permission from AIP publishing).

Another revisit to VCA approximation was made by Seyf et al. who used GKMA to evaluate TC of InGaAs alloys (Figure8(a)).[74] Not only did they show that VCA fails, but they also provided a different perspective, arguing that alloys tend to behave more like amorphous materials. This is based on the observations that the mode character changes dramatically with composition and a steep drop in TC between 0–2% impurity corresponds with a decrease in propagating plane wave character. Furthermore, when 15–85% of the lattice sites are occupied by dissimilar atoms, the diffusons were found to dominate TC, whose effects were found to intensify at high temperatures (Figure 7(a)). Several other researchers have also observed mode localization in alloys.[171,290,291] Zhang *et al.* observed phonon localization in compositionally graded alloys which reflects a large part of the wave packet in Si/Ge alloys and thereby reduces alloy TC.[290] Giri *et al.* also studied the localization of vibrational modes in Si/Ge alloys using NEMD simulations. They attributed the reduction in TC of superlattice structures, i.e. by comparison to their disordered alloy counterparts, to mode localization around the cutoff frequency of the a-Ge layer.[291]

Effects of alloying and other approaches like nanostructuring and amorphization have been comparatively studied using MD simulations.[292] A consensus has been achieved that structural disorder tends to have higher effect than compositional disorder on TC reduction. For instance, Xiong *et al.* used EMD-GK to study TC of SiGe alloy-based nanophononic metamaterials to understand the role of phonon resonance in reducing TC.[292] They quantified the role of alloying and resonators on low frequency modal contributions: the combined effect of alloy scattering and resonance contributed to 22-fold decrease of Si TC. In another investigation of TC of $Si_{1-x}Ge_x$ alloys using AEMD, it was found that the TC is reduced due to scattering at the grain boundaries rather than by mass differentials.[293] Norouzzadeh *et al.* also studied the TC of amorphous SiGe alloys using EMD-GK and found that for the amorphous alloy, TC was nearly an order of magnitude lower than the lowest TC of $Si_xGe_{1-x}$ and compositional disorder has much lower effect than structural disorder.[294]

Like materials with weak structural disorder, weak compositional disorder can also affect TC. In the following section, we inspect how MD simulations have been utilized to predict TC of materials with weak intrinsic compositional disorder and the effect of doping.

### 3.3.5. Weak compositional disorder and doping

Intrinsic weak compositional disorder/defects and sometimes doping are both examples of break in symmetry to a periodic crystalline lattice that affect phonon transport in solids. Compositional defects are known to change the normal modes of vibration, and the effect on normal modes have been studied computationally[74] and verified experimentally by the occurrence of peaks in the Raman spectra that depend on mixture composition[295]. This effect on the normal modes, therefore, directly impacts thermal transport. Lyver *et al.* studied this effect in binary crystals by using EMD-GK simulations, and showed that the typical trend of lattice TC scaling inversely with temperature fails when the crystal contains disordered atoms of differing diameters.[296] It was not until the past decade that researchers started investigating the underlying mechanisms of such phenomena in terms of the normal modes. Skye *et al.* showed that thermal transport is greatly decreased due to mass disorder, and that point-defect scattering models based on the Debye spectrum cannot fit experimental results, thus suggesting that high frequency modes are important (and not just the low frequency traveling modes as assumed by the bulk of literature).[297] Schemes for more elaborate studies on disorder and compositional defects, especially in alloys, are highlighted in recent tutorial reviews,[27] which incorporate normal mode analysis to study the effect of normal modes on thermal transport properties.

Regarding the accuracy of MD studies of compositional defects, recent work by Somayajulu *et al*. studied the effect of oxide composition on nuclear energy materials, where they showed the agreement of EMD-GK TC calculations with experimental results.[298] While such studies exhibit the accuracy of MD for thermal transport in defect/disordered materials, a major advance in computational materials science would be utilizing MD to design or tailor thermal transport properties; indeed, recent work by Dettori *et al*. used NEMD simulations to show thermal rectification in bulk semiconductors via structural defect engineering.[299] The effects of compositional defects in nanomaterials have also been recently studied using MD simulations, with some authors noting up to a 15% decrease in TC with random defect patterns.[300] Zhang *et al*. also used NEMD simulations to study the effect of nitrogen doping on bilayer graphene TC, with guidelines for tuning TC for device applications.[301] MD simulations were also used to study the effect of nitrogen doping on the thermal insulation properties of aerogels.[302] Such works in the atomistic design of thermal transport paves the way for rationally reducing lattice TC, thus benefiting energy applications such as heat retainment and thermoelectrics.

The quest for more efficient thermoelectrics is directly benefited by computational studies of the effect of doping on thermal transport, since doping is readily achieved and one of the easiest ways to tune phonon transport properties. Navid *et al.* showed used MD simulations to study the reduction in TC of nanoribbons with silicon and carbon doping.[303] Wei *et al*. also used MD simulations to show that Cd impurities can reduce lattice TC in $CuInTe_2$, thus boosting its thermoelectric performance.[304] Aside from thermal transport calculations, some researchers studied temperature dependent force constants, and thus study the effect of dopants on thermoelectric materials via the phonon density of states.[305] Regardless of the method used, MD simulations are increasingly becoming more of a valuable tool to study the effect of defects and disorder on lattice thermal transport.

Another special case of weak compositional disorder is the presence of isotopes. Although chemically similar to the atoms that constitute the crystal lattice, the major difference lies is in the atomic mass of defect atoms. The next section will summarize MD studies on how presence of isotopes affects phonon transport behavior.

### 3.3.6. Isotopic effects

TC of solids can also be influenced by the presence of isotopes. This effect is only pronounced when other factors such as boundary and anharmonic scattering are not strong.[306] This is usually the case in single crystal, chemically pure high TC compounds at moderate to low temperatures.[307–309] Nevertheless, a study has even shown that Si isotopes affect the TC of amorphous Si with its known low TC.[310] In general, all MD,[177,306,310–314] DFT-BTE,[47,48,315] and experimental[307–309,316,317] studies show that the existence of isotopes in a structure decrease the TC compared to the value of its isotopically pure form. In this regard, the majority of the literature utilizes this opportunity to either decrease the TC by introducing more isotopes to the structure,[177,306,310–314] or increase the TC by synthesizing isotopically pure compounds – both of which are consistent with the intuition based on the PGM.[46,307–309]

Reducing TC is particularly interesting for thermoelectric applications, since isotopes only affect the mass of atoms, which allows for reduction of TC without sacrificing electronic performance.[311] Many MD applications have studied the reduction of TC caused by isotopic effects. In this review, we only focus on the simulations that have explicitly included different mass isotopes in their simulated atomic structures, as these are the only simulations that can directly capture the modal interactions that are caused by different mass atoms in the system.

Using EMD simulations Zhang *et al.*[311] showed that the TC of graphene decreases by 80% when isotope concentrations of $^{13}$C as low as 25% are present in the structure. Using NEMD approach, a similar decrease in TC have also been reported for single-walled carbon nanotube[312,313] (Figure 1) and graphene nano-ribbon.[177,313] By studying the phonon participation ratio, the authors attributed the observed reduction in TC of SWNTs[312,313] and GNRs[5] to the localized modes that appear in the structure after isotopes are introduced to the system, which mitigates the efficient flow of heat in the lattice. One approach to understand such a reduction in TC is by using the prescribed formulas[318] under the BTE framework that has been successfully used in multiple studies to explain the isotope scattering. [47,48,315] By having access to the actual dynamics, MD-based modal analysis approaches[151,152,154,156] have the potential to more accurately describe the isotope-based reduction in TC and explain specifically how the induced localized modes by isotopes interact with the rest of the vibrational mods in the lattice and decrease the TC.

Based on NEMD simulations, Park *et al.*[310] observed a similar reduction in TC for Si and amorphous Si. They explained the observed reduction in the TC of crystalline Si using the traditional BTE based approaches. However, since these BTE based approaches are not applicable to amorphous solids, the mechanism behind the TC reduction in amorphous Si is still unclear, but in theory could be analyzed using a MD-based modal analysis technique. In another study, Pei *et al.*[314] showed that variations in Si mass lower the TC of Silicene (2D monolayer of silicon atoms arranged in honeycomb lattice) as well. The authors[314] also showed that ordered positioning of the Si isotopes in the structure in form of a superlattice leads to a much larger reduction in TC than random positioning of the isotope atoms in the system (Figure 9). The higher effectiveness of superlattice structures in decreasing the TC has also been demonstrated in other studies,[177,314] where such a reduction is generally attributed to the additional interfaces that are introduced to system through the superlattice[252,319] and the consequent interface scattering that is usually augmented by the localized phonons across these interfaces.[319] Follow-up studies based on MD-based modal analysis can shed more light on the interaction of these localized modes with the other modes in the system that lead to the reduction of TC in superlattice structures.

The importance of having MD-based approaches independent of any PGM formulations can be further understood by noting the study by Wu *et al.,*[306] where using NEMD, they studied the isotopic effect on thermal transport in single layer molybdenum disulfide ($MoS_2$). The authors[306] showed that the phonon relaxation times that are obtained from MD simulations combined with phonon modal spectral energy density (SED)[137,139] analysis are drastically different from the ones that are obtained from scattering formulations[318] commonly used in the BTE approach, where Tamura model[318] is combined with Matthiessen's rule to include the isotope effects. The observed differences become much more pronounced at higher isotope concentrations. The authors[306] also found that Mo mass variation has more impact on TC, which is directly related to the heavier mass of Mo compared to S. Higher mass variations for the heavier atom in a compound have been shown to have a larger impact on the TC by different theoretical[47] and experimental[307–309] studies. These studies have shown that the behavior of heat-carrying acoustic phonons is dominated by the vibrations of heavy atom, which is usually justified by the isotope scattering term[47,318] used in the BTE analysis of phonons. Exploring these BTE based observations by MD-based modal analysis approaches are interesting topics for future studies.

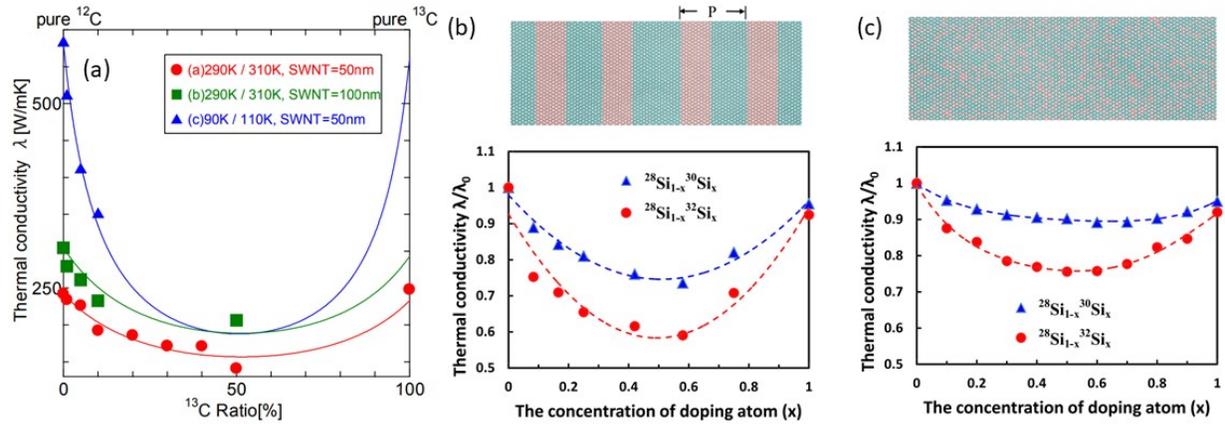

**Figure 9.** Reduction of TC by adding (a) C isotopes to single walled carbon nanotube structure at 300K and 100K (reprinted from Ref [312], Copyright 2006, with permission from the Japan Society of Mechanical Engineers), and (b-c) Si isotopes in (b) orderly (superlattice) and (c) random fashions to silicene structure (reprinted from Ref [314], Copyright 2013, with permission from AIP Publishing). $\lambda_0$ shown in panels (b) and (c) is equal to ~45W/m-K.

Synthesizing isotopically pure materials can increase their TC. Particularly, as mentioned before, if (*i*) the boundary scattering is weak (e.g., by having a single crystal), and (*ii*) the anharmonic interactions are small (e.g., by having a simple binary compound with large mass ratio[320] between the comprising elements and bunching of the acoustic branches[320]) the isotopic effects can dominate the TC. This has been clearly shown in recent experimental measurements of boron arsenide (BAs),[16] where theoretical predictions predicted a high TC (>1000W/m-K),[46] however this value could not be achieved[321] until the single crystal BAs was synthesized in an isotopically pure form, particularly for As, which is the heavier element.[307–309] Although DFT-BTE calculations have shown good agreement with experimental measurements, using MD-based modal analysis approaches[151,152,154] might provide additional insight into the mechanisms of heat conduction in the high TC compounds such as BAs, and how a small degree of isotopic effect can drastically decrease their TC.

In sections 3.3.1-3.3.6, we summarized how MD studies have been utilized to study the effects of intrinsic and extrinsic physicochemical parameters on the TC of materials and how MD-LD methods have been implemented to shed light on the modal contributions to TC. In many practical applications, however, systems are not monolithic, i.e. they consist of numerous material interfaces. When the system sizes approach the nano dimensions, the ratio of surface area to volume of materials increases, and consequently the role of interfaces starts getting more pronounced. Hence, interfacial thermal transport grows in importance and demands an independent theoretical treatment. Therefore, in section 4, we discuss the theory, computational approaches and prior studies using MD simulations to understand and quantify TIC of various material combinations.

## 4. Thermal Interfacial Conductance

When heat flows across the interface of two different materials, a temperature discontinuity forms at the interface (see Figure 10). The interfacial heat flow $(Q)$, can be written as the product of the TIC (denoted by $G$), the contact area $(A)$, and the temperature difference across the interface $(\Delta T)$ (i.e., $Q = GA\Delta T$). Interfaces play a key role in the thermal behavior of nanostructures.[1,2] Due to tremendous advances in nanostructuring in recent years,[322,323] exquisite structures with characteristic lengths on the order of nanometers can be fabricated for applications in nanoelectronics[324] and nanoscale energy conversion.[2] In these small scales, interfaces can become the dominant resistance to heat transfer, which on one side impedes the progress towards achieving improved performance in nano-electronics,[325] nano-optoelectronics,[326] nano-energetics,[327] or energy conversion devices such as multi-junction solar cells,[328,329] and on the other side sets interface engineering as a promising path to reach higher ZT thermoelectric materials,[330–336] or lower-TC thermal barrier coatings[337–339] through reducing TC (e.g., by making grain boundaries[340] or superlattices[340]). Having access to powerful computational and theoretical techniques to study interface heat transfer not only allows us to predict and assess the effect of different factors on interfacial heat transfer, but also helps us gain valuable insight which can be used to expedite our progress towards better performing devices for multitude of applications.

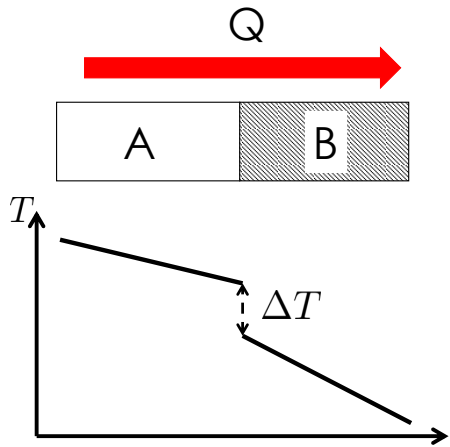

**Figure 10.** Heat flow $(Q)$ causes a temperature jump $(\Delta T)$ at the interface of two solid materials labeled A and B with contact area $A$. $Q$ is proportional to $\Delta T$, and the constant of proportionality is the thermal interface conductance $G$ (i.e., $Q = GA\Delta T$).

Since the first experimental observations of thermal interface resistance across the interface of Cu and liquid He,[38,341] different theoretical models have been proposed to explain the transfer of heat across interfaces and to predict TIC.[33–38] Several of these methods are based on static

LD calculations and do not require any simulation in time domain to capture the dynamics of the system. The acoustic mismatch model (AMM),[33,34] the diffuse mismatch model (DMM),[35,342,343] the atomistic Green's function (AGF) approach,[36,344–347] the harmonic LD based approaches,[348–350] and frequency domain perfectly matched layer (FD-PML) method[37,351] are notable examples of these models. Although each of these techniques is devised for different sets of operating conditions (*e.g.*, temperature range, interface quality, etc.), they are all based on the PGM and usually invoke some version of the Landauer formalism.[55,352,353] Many improvements have been made to these methods,[342,354–358] but still capturing roughness, interatomic diffusion, stress, imperfections etc. with full inclusion of anharmonicity is challenging and a limiting issue for most of these methods. It has been argued in many reports that anharmonic interactions have notable contributions to TIC at high temperatures.[153,342,359–362] English *et al.*,[363] Stevens *et al.*,[361] and Gaskins *et al.*[364] have shown that the increase in TIC at higher temperatures can be attributed to anharmonic interactions between phonons across interfaces. Thus, it is crucial that anharmonic effects are included in the analysis of interfacial heat transfer. The development of the AGF method was a major step forward, as it incorporated the atomic level details and also accounts for quantum effects.[346,365] However, most applications of the AGF method have been limited to small system sizes and harmonic interactions, due to analytical complexity and computational expense.[346] Anharmonic AGF was first formulated by Mingo and applied to simple low-dimensional structures, and it has recently been extended to three-dimensional systems by Dai and Tian.[39] Although promising, more analysis is needed to assess the applicability of this method to generalized three-dimensional interfaces. Although exceptions exist,[366] the majority of the literature using DMM has also only been able to evaluate elastic scattering interactions where the transmission of a mode's energy across the interface is purely governed by whether or not other modes with similar frequency exist on the other side of the interface.[2,347]

Several reviews exist that summarize the advances in interfacial heat transfer.[35,160,366–368] The seminal work by Swartz and Pohl[35] emphasizes on the important experimental cryogenic measurement techniques and the correct applicability of AMM and DMM in theoretical analysis of interfacial heat transfer. Hopkins[366] and Monachon *et al.*[367] have also published reviews with the emphasis on experimental measurements of thermal interface conductance. Recently Giri and Hopkins[368] also reviewed recent theoretical and experimental advancements on treating the TIC. The focus of this review is on the applicability of MD to study interfacial heat transfer and on the importance of modal analysis in investigating the conductance across solid-solid interfaces. MD simulations can be used to calculate TIC based on three main implementations: (*i*) NEMD, (*ii*) EMD, which are similar to the NEMD and EMD methods explained in previous sections for TC, and (*iii*) thermal relaxation method. These methods for capturing conductance across interfaces using MD has been thoroughly reviewed by Schelling *et al.*[160]

4.1. Different Molecular Dynamics Implementations to Calculate TIC

4.1.1. NEMD

For calculating TIC in NEMD simulations, we follow a similar approach as described in Section 3.1.2, where a heat flow is forced to flow through the structure and the interface by placing hot and cold reservoirs in different locations in the system. Depending on the position of these thermal reservoirs in the structure and the type of periodic boundary conditions used in the simulations two different implementations of NEMD can be envisioned, which are schematically shown in Figure 11. For fixed boundary conditions, the thermal reservoirs are placed at the ends of the system, where the entire energy input to the hot reservoir flows in one direction through the one interface that is usually placed in the middle of the structure. For PBC's, there will be two identical interfaces: one in the middle, and the other one at the ends of the system. In this implementation, thermal reservoirs should be placed in exactly the middle of the bulk structures at the sides of the interface, so the power input to the hot reservoir flows in equal parts in both directions (each part equal to the half of the total power input to the hot reservoir). As Figure 11 shows, in either of these implementations, at steady state, a temperature profile forms in the structure and by directly measuring the temperature difference ($\Delta T$) across the interface, which is determined by extrapolation of the temperature gradients in each respective material,[369–372] TIC can be calculated using the $G = \frac{\overline{Q}}{A \cdot \Delta T}$ formula, where $\overline{Q}$ is the steady state, time-averaged interfacial heat flow in the NEMD simulation. Due to the suppression of vibrational modes at the fixed boundaries, the fixed boundary implementation of NEMD usually requires larger structures for simulation, compared to the periodic boundary one.[207]

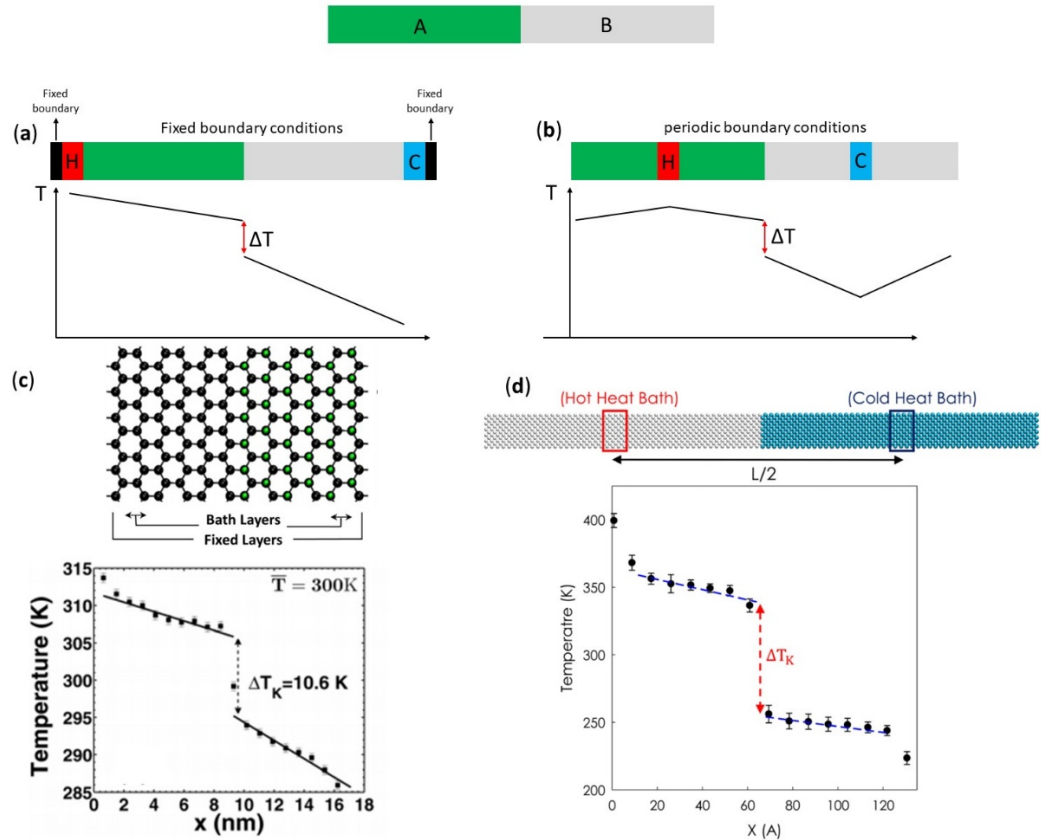

**Figure 11.** Two different implementations of NEMD simulation for evaluating TIC across the interface: (**a**) fixed boundaries, and (**b**) periodic boundaries, with their respective schematic temperature profiles shown in the bottom of the schematics. Panels (**c**) and (**d**) compare two interface structures that were analyzed in two separate studies and have used the (**c**) (reprinted from Ref [371], Copyright 2011, with permission from AIP Publishing) fixed and (**d**) (reprinted from Ref [207], Copyright 2017, with permission from AIP Publishing) periodic boundary implementation of NEMD.

### 4.1.2. EMD

TIC can also be calculated based on fluctuation-dissipation theorem as described in Section 3.1.1 but with minor additions to account for interfacial heat flux.[373] Puech *et al.*[374] as well as Domingues *et al.*[375] and Barrat *et al.*[376] have shown that the conductance is proportional to the autocorrelation of the equilibrium fluctuations of heat flow across the interface via the equilibrium definition of TIC based on the following formula,

$$G = \frac{1}{Ak_B T^2} \int_0^\infty \langle Q(t)\, Q(0) \rangle dt \qquad (14)$$

where $A$ is the interface contact area, $k_B$ is the Boltzmann constant, $T$ is the equilibrium temperature of the system, $Q$ is the instantaneous interfacial heat flux, and $\langle \cdots \rangle$ represents the autocorrelation function. The EMD formulation is not only another way of calculating TIC across interfaces in MD simulations, but it also suggests a possible alternative picture to scattering – i.e., an alternative picture based on correlation, which was discussed in the previous section and will be referred to more when reviewing the results from different studies. Two alternative formulations have been proposed for Eq. (14) in the literature. The first alternative formulation is by Chalopin *et al.*,[377] where they showed that TIC can be calculated from an EMD simulation using the following formula,

$$G = \frac{1}{Ak_B T^2} \left\langle \frac{dH_A(t)}{dt} H_B(t) \right\rangle \qquad (15)$$

Here, $H_A$ and $H_B$ are the total Hamiltonian of sides $A$ and $B$ across the interface and $\langle \cdots \rangle$ is the averaging (equal-time correlator) operator. Chalopin *et al.*[377] claim that evaluating TIC using the above formula converges faster and hence is computationally more efficient than Eq. (14). The second alternative formulation is by Rajabpour and Volz[378] and is dependent on the instantaneous temperature difference across the interface $(\Delta T)$ instead of the interfacial exchanged heat current,

$$G = \int_0^\infty \frac{\langle \Delta T(0) \Delta T(t) \rangle}{k_B \langle \Delta T(0) \rangle^2} dt \left( \frac{1}{N_1} + \frac{1}{N_2} \right) \qquad (16)$$

where $N_1$ and $N_2$ are the number of degrees of freedom at different sides across the interface (e.g., $N_1$ and $N_2$ are equal to three times the number of atoms on sides 1 and 2, respectively). Rajabpour and Volz[378] note that the calculation of temperature difference across the interface is easier than interfacial heat flux calculations. It should be noted no matter what EMD method is utilized, the values obtained for TIC from EMD are usually different from the ones from NEMD. Nonetheless, this is typically associated with finite size effects from the reservoirs, since the two values usually converge when the system size in NEMD is increased.[207,379]

### 4.1.3. Thermal Relaxation Method

Defining the temperature drop $(\Delta T)$ obtained from NEMD across interfaces of non-crystalline solids or interfaces with roughness and interatomic mixing can become difficult and ambiguous.[207,380] Therefore, relaxation method has been proposed to make the analysis more robust. The principle behind the thermal relaxation method consists of instantaneously heating one of the two solids and recording the temporal evolution of the temperature of the hot side.[381–384] The conductance $(G)$ is then obtained from the time $(\tau)$ characterizing the thermal relaxation of the hot side,

$$G = \frac{3 N_1 k_B}{4 A_0 \tau} \qquad (17)$$

where $N_1$ is the number of atoms of solid 1 (e.g., the hot side), $A_0$ is the interfacial area projected normal to the interface, and the factor 4 accounts for the presence of two interfaces if periodic boundary conditions are considered.

The above methods have been used extensively to study the effect of different parameters such as roughness, interatomic mixing and bonding strength neat the interface region on the interfacial thermal transport. These findings and the most prominent studies representing them will be examined in section 4.3. Before discussing the effect of these different parameters in section 4.3, it is worth noting the general definition of interfacial heat flow and why the interfacial region and the dynamics of atoms around the interface plays a key role in determining the TIC.

In a two-body interatomic interaction, the exchanged heat flow $(Q)$ across interface is defined by,[76,153,368,385]

$$Q = -\frac{1}{2} \sum_{i \in A}^{N_A} \sum_{j \in B}^{N_B} \mathbf{f}_{ij} \cdot (\dot{\mathbf{v}}_i + \dot{\mathbf{v}}_j) \tag{18}$$

where, $i$ and $j$ refer to atom indices at the sides of the interface $A$ and $B$, $N_A$ and $N_B$ are the number of atoms on sides $A$ and $B$, $\mathbf{v}_i$ and $\mathbf{v}_j$ are the atomic velocities and $\mathbf{f}_{ij}$ is the pairwise interaction between atoms $i$ and $j$. The above formula is generalized to higher order interactions between atoms based on the following formulation,[76,368,385–387]

$$Q = \sum_{i \in A}^{N_A} \sum_{j \in B}^{N_B} \left\{ \frac{\partial \Phi_j}{\partial \mathbf{r}_i} \cdot \mathbf{v}_i - \frac{\partial \Phi_i}{\partial \mathbf{r}_j} \cdot \mathbf{v}_j \right\} \tag{19}$$

where $\Phi_i$ is the potential energy assigned to the individual atom $i$ with the condition that $\sum_{i}^{N} \Phi_i = \Phi$ [129,388] (with $\Phi$ being the total potential energy in the system). This definition of the heat flux is free of any assumptions in contrast to PGM approaches where the definition of heat flux is merely an extension from of kinetic theory.[55] The definition in Eq. (19) shows that the atomic velocities and interactions around the interface region are the only ones that directly affect the interfacial heat flow across interfaces. Because of this, several approaches[76,153,386,387,389,390] have been proposed to alter the TIC based on modifying the position of atoms and interaction of atoms in the interface region. One of the main advantages of MD is that all these effects can be included with full interface topology and anharmonicity without any limiting assumptions. These advancements are discussed in the forthcoming sections.

### 4.2. Modal Analysis of TIC

Like TC, although NEMD, EMD, and thermal relaxation techniques can be used to capture the total TIC values by including the atomic details and full anharmonic interactions. To achieve modal insights, MD-based modal analysis methods explained in this section can be used, as knowing these contributions is the ultimate tool for a more targeted design.[391]

One of the first modal analysis approaches that was proposed to study heat transfer and phonon transport across interfaces was the wave packet (WP) method.[40,392] The WP method allows the calculation of a mode's transmissivity in the context of a MD simulation. Although founded on MD simulation, WP analysis requires that all other modes have zero amplitude.[346,359,361,365] This effectively corresponds to the $T = 0\ K$ limit and therefore is unable to examine the effects of temperature dependent anharmonicity. As a result, the WP method simply reproduces the same results as the harmonic AGF approach.[2,393] To capture the finite temperature anharmonic contributions, other MD based techniques have been proposed.[76,153,386,387,389,390] The commonality between all these techniques is that they are all based on the general definition of heat transfer across interfaces (Equations (18) and (19)),[368,385] and the difference between them is how they decompose this general definition of interfacial heat flow.

Chalopin and Volz[153] deduce the spectral contributions to heat flux $(Q(\omega))$ from the equilibrium displacements fluctuations of the contact atoms using the following formula,

$$Q(\omega) = \frac{1}{2} \sum_{\substack{i \in A, j \in B \\ \alpha, \delta \in \{x,y,z\}}} k_{i,j}^{\alpha,\delta} \left[ v_i^\alpha(\omega) u_j^{*\delta}(\omega) - u_i^{*\alpha}(\omega) v_j^\delta(\omega) \right] \quad (20)$$

where, $u_i$ refers to the atomic displacement, $v_i$ is the instantaneous velocity of atom $i$, the $\alpha$ and $\delta$ exponents refer to the $x$, $y$, or $z$ component, $k_{i,j}^{\alpha,\delta}$ is the interatomic force constant between atoms $i$ and $j$, and $u^*$ is the complex conjugate of $u$. Chalopin and Volz[153] also consider the following decomposition of atomic velocities based on different wave-vector $(\mathbf{k})$ contributions,

$$v_\mathbf{k}^\alpha(\omega) = \sum_i v_i^\alpha(\omega) e^{i\mathbf{k}\cdot r_i^0} \quad (21)$$

where $r_i^0$ is the equilibrium position of atom $i$. When Equations (20) and (21) are combined with the spectral definition of Landauer,[32]

$$Q(\omega) = \hbar\omega \left( n_a^0(\omega) - n_b^0(\omega) \right) T(\omega) \quad (22)$$

the frequency and wave-vector dependent heat flux and also the degree of transmission from phonon with frequency $\omega$ and wave-vector $\mathbf{k}$ to another phonon with frequency $\omega$ and wave-vector **k'** is found. In Equation (21) $n_{a,b}^0$ corresponds to the Bose Einstein occupation numbers of systems A and B. Although insightful, the above formalism based on the formula in Equation (21) supposes the existence of plane waves as the correct sets of vibration for the system. However, as has recently been shown[389], the correct basis set of vibrations for an interfacial system does not necessarily solely consist of plane waves, since symmetry is broken by the introduction of interface.

Based on the method developed by Chalopin and Volz,[153] Sääskilahti *et al.*[390] introduced another spectral analysis approach to capture the frequency contributions to interfacial heat transfer. In addition, by utilizing the second-order and third-order force constants and decomposing the forces into their harmonic and anharmonic components, Sääskilahti *et al.*[390] could determine the elastic and anharmonic contributions to interfacial heat flux. However, the method developed by Sääskilahti *et al.*[390] only includes a limited number of first-order anharmonic phonon scattering terms in their formulation.[386,387] Later, Zhou and Hu[387] extended the work by Sääskilahti *et al.*,[390] so it can capture the full third order inelastic terms. The anharmonic contributions calculated using the method developed by Zhou and Hu[387] will ultimately be limited by the truncation of force constants,[390] and higher inclusion of anharmonic terms will be difficult because of the computational cost associated with inclusion of higher order terms.

To go beyond the spectral calculations and determine the contribution by an individual mode of vibration, Gordiz and Henry[76,389] introduced the interface conductance modal analysis (ICMA) method. ICMA is based on the modal decomposition of atomic velocities in the general definition of interfacial heat flow (Equation (19)), which results in the following formula for individual contribution of mode $n$ to interfacial heat flow $(Q_n)$,

$$Q_n = \sum_{i \in A}^{N_A} \sum_{j \in B}^{N_B} \left\{ \left( \frac{\partial \Phi_j}{\partial \mathbf{r}_i} \right) \cdot \left( \frac{1}{(m_i)^{1/2}} \mathbf{e}_{n,i} \dot{X}_n \right) + \left( \frac{-\partial \Phi_i}{\partial \mathbf{r}_j} \right) \cdot \left( \frac{1}{(m_j)^{1/2}} \mathbf{e}_{n,j} \dot{X}_n \right) \right\} \quad (23)$$

where, $\dot{X}_n$ is the normal mode coordinate of velocity for mode $n$ (Equation 5),[18] $\mathbf{e}_{n,i}$ is the eigen vector for mode $n$ assigning the direction and displacement magnitude of atom $i$, and $m_i$ is the mass of atom $i$. Equation (23) is the modal contributions to the interfacial heat flow and can be replaced in any definition of TIC (either in NEMD or in EMD) to obtain the modal contributions to TIC $(G_n)$, where $G = \sum_n G_n$. When using the equilibrium definition of conductance (Equation 14), it will result in the following formula for $G_n$,

$$G_n = \frac{1}{A k_B T^2} \int \langle Q_n(t) Q(0) \rangle dt \quad (24)$$

Furthermore, by substituting for both heat flows in the equilibrium definition of TIC (Equation (14)), individual contributions to TIC from pairs of modes $n$ and $n'$ $(G_{n,n'})$ are found to be,

$$G_{n,n'} = \frac{1}{A k_B T^2} \int \langle Q_n(t) Q_{n'}(0) \rangle dt \quad (25)$$

$G_{n,n'}$ quantifies how two modes of vibration in the system couple to each ether to contribute to interfacial heat transfer.

The aforementioned modal analysis approaches are based on only the atomic vibrations around the interface region. Based on the spectral contributions to TC formulation proposed by Zhou *et* *al.*,[151] recently, Feng *et al.*[386] introduced a method that shows how the modes of vibration in the bulk region interact with the interface region. Their work is quite insightful, and more study is warranted, particularly if follow on studies examine the physics using the modes native to the structure.[389]

In the following sections, we examine how various physicochemical parameters that characterize the interface affect TIC. Specifically, we will be surveying the role of roughness, interfacial bond modifications, atomic mixing, structural defects such as dislocations and interfaces between amorphous materials.

## 4.3. Effects of system parameters on TIC

As the definition of interfacial heat flow in Equations. 18 and 19 shows, the heat flow between two materials is dictated by the atomic velocities and forces in the interface region. The dynamics of the atoms in the interface region can be influenced by various parameters such as dislocations, roughness, interatomic mixing and interface bonding. Since many of these parameters are naturally present at interface because of synthesis conditions, having access to powerful analysis tools to reveal the change they induce on heat transfer is crucial. This also brings opportunities to control the interfacial heat flow either by modifying the atomic positions in the interface region through topological changes such as roughness or interatomic mixing or by changing the interface bonding to manipulate the forces. Understanding the effect of each of these parameters has been active areas of research, where EMD, NEMD and relaxation methods have been successfully utilized to give us insight into the changes in TIC by modifying any of these parameters in the interface region. In this regard, the majority of the conducted analysis to understand the changes in heat transfer has been based on the vibrational DOS overlap, in which the increase in heat transfer has been shown to be correlated with the degree of overlap between the VDOS on the sides of the interface.[207,394] The goal of this section is to summarize the key findings for each of the main categories of parameters that effect the interface heat transfer and discuss possible follow-on studies that can be based on more detailed modal analysis to gain more insight into the mechanisms of heat transfer that are changed after modifying any of these parameters.

### 4.3.1. Roughness and nano-structuring

The effect of surface roughness and nano-structuring on the TIC of various material interfaces has been the subject of several MD investigations. [384,395–397] Using the thermal relaxation approach, Merabia and Termentzidis[384] (Figure 12(a)) have shown that small scale roughness at the interface of two different solids does not induce a noticeable change to the conductance compared to its value for planar interfaces. However, large roughness increases the conductance by a considerable amount, which is in agreement with other MD studies.[395–397] The increase in TIC by introduced roughness can be explained by increased net area of heat transfer at the contact area. Since the total resistance associated with an interface $\left(1/GA\right)$ scales inversely with total contact area between the two materials, it is conceivable that by increasing the total contact area, the total resistance can be reduced. In contrast, the intuition obtained from the PGM perspective suggests that more interface area would increase the probability of boundary phonon scattering and thus would reduce TIC. In the correlation paradigm (Equation (14)), however, additional surface area via increased roughness would induce the formation of new localized modes, which may offset any penalties associated with more "scattering impedance". More studies are needed to understand these effects and to what extent an increase in area can enhance or suppress the interfacial heat transfer.

Another strategy to increase TIC, which also follows the idea of increased contact area is based on adding nanostructured features to the interface region. Recent experimental measurements across multiple rough and nanostructured interfaces between Al/Si[398] and Al/GaN[399] also support the idea that increasing the contact area increases the TIC. Recently, in another insightful study, the experimental measurements by Lee and Luo[399] showed that the temperature dependent conductance at the nanostructured (pillared) and planar interfaces between two materials are similar to each other at low temperatures but different at higher temperatures (Figure 12(b)-(d)). This shows that different groups of phonons, as they get excited at different temperature ranges, will interact differently across these two interfaces. Capturing the temperature-dependent anharmonic phonon–phonon interactions that lead to such an observation can be an interesting follow-on study based on any of the MD based modal analysis techniques.[76,153,387,389,390]

In addition to the idea of increased interfacial area, Liang *et al.* [400] showed in an insightful study that interfacial roughness leads to enhanced interface conductance because of the thermalizing scattering event that it induces around the interface region. This thermalizing scattering event mitigates the transmitted phonon to travel back to the origin side after passing through the interface, which enhances the interfacial heat transfer. The example simulation by Liang *et al*. [400] on a Substrate (GaN)/thin-film (AlN) interface system showed that introduced surface roughness at the ends of the system decreased the interfacial thermal resistance (Figure 12(j)). This observation is interesting since it proposes the possibility of influencing the interfacial heat transfer by modifying the surface roughness at each sides of the system without the need to modify the interface region itself. Interestingly, when high degrees of surface roughness are introduced to the sides of the interface conductance increases almost completely independent of the length of the structure (Figure 12(k)). The theoretical analysis by Liang *et al*. [400] based on the PGM model and the AMM description of transmission probability showed a promising agreement with the obtained results from the NEMD simulations. However, it would be fruitful to examine the underlying reasons for their reported enhanced interface conductance using the proposed MD-based modal analysis methodologies.[76,153,386,387,389,390]

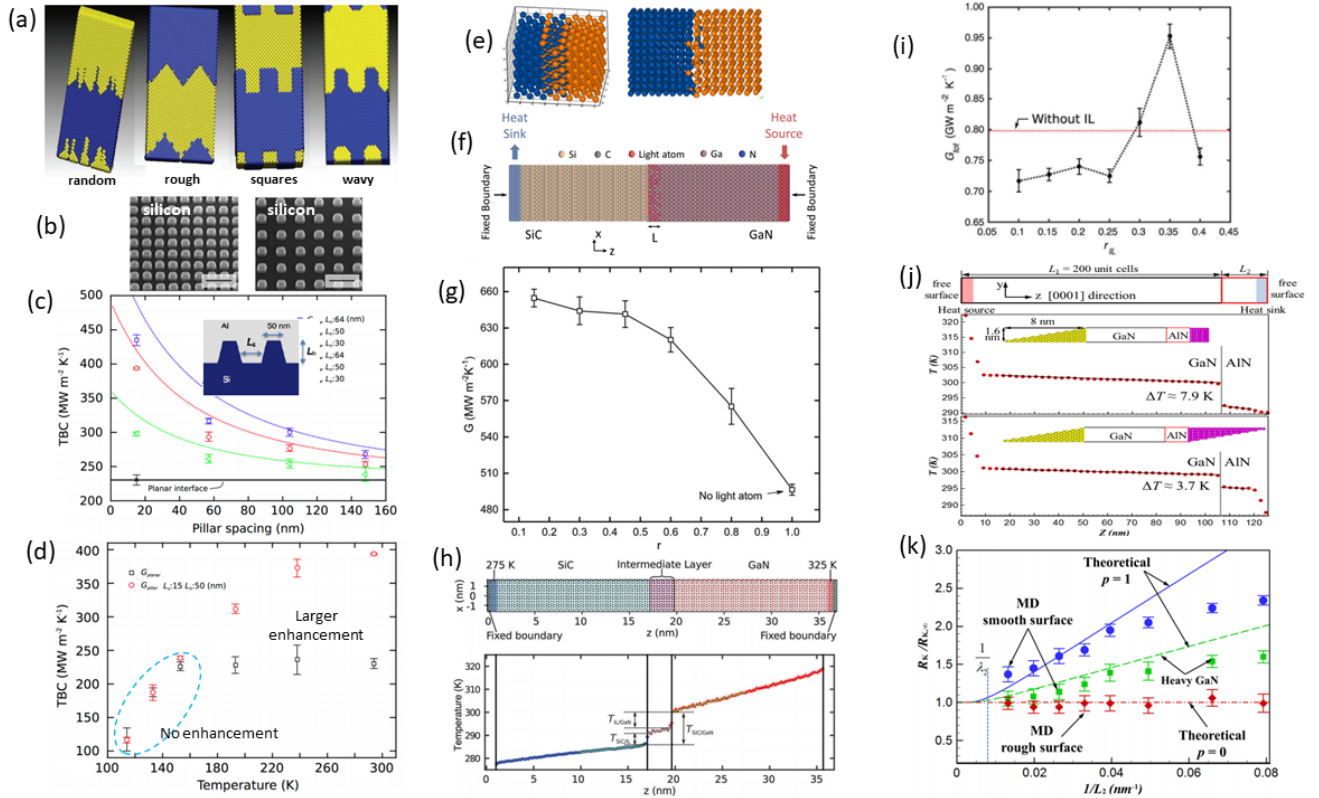

**Figure 12.** (a) MD simulations of rough interfaces (reprinted from Ref [384], Copyright 2014, with permission from the American Physical Society). (b)-(d) Experimental realization and measurement of conductance across nanostructured interfaces (reprinted from Ref [398], Copyright 2016, with permission from the American Chemical Society). MD simulations of interfaces with (e) interatomic mixing (reprinted from Ref [361], Copyright 2007, with permission from Elsevier), (f)-(g) added light atoms in the interface region (reprinted from Ref [401], Copyright 2019, with permission from the Royal Society of Chemistry), and (h)-(i) added intermediate layer between the two structures (reprinted from Ref [402], Copyright 2017, with permission from the Royal Society of Chemistry). Panels (j) and (k) show how surface roughness at the sides of the interface structure can (j) increase the interface conductance and (k) make the interface conductance to become size-independent, (reprinted from Ref [400], Copyright 2014, with permission from the American Physical Society).

### 4.3.2. Interatomic mixing, adding intermediate-layers and mass-grading

The consensus among the existing literature is that by disrupting the distribution of atoms from its perfect form in the interface region an improvement in TIC is observed. One of the seminal studies around this issue is the one by Stevens *et al*.[361], where by using NEMD, they showed that interatomic mixing increases heat transfer across interfaces (Figure 12(e)). Their observations were later confirmed by theoretical analysis based on a modified DMM approach[403] and by experimental measurements.[403,404] Although AGF has been used to further study the effect of interatomic mixing under the harmonic framework,[347] the authors are unaware of any other major MD study this effect. Particularly, detecting the temperature-dependent anharmonic interactions that give rise to the interface heat transfer caused by interatomic mixing, using a MD-based modal analysis approach, seems to be an insightful follow-on study. Isotope effects have also been shown to be an effective approach to increase interfacial heat transfer. Using NEMD, Li *et al*.[401] have recently shown that by introducing light atoms across SiC/GaN interfaces, interfacial heat transfer improves (Figure 12(f)-(g)). In addition, adding an extra layer at the interface has been proposed as a strategy to increase heat transfer as well. Using NEMD Lee and Luo[402] showed that adding an AlN layer to the interface of SiC/GaN increases heat transfer (Figure 12(h)-(i)). The effectiveness of an intermediate layer has also been shown by other NEMD studies[363] and PGM-based techniques such as DMM[405] and AGF[406]. Different MD simulations have also shown that mass grading the interface region is an effective option to increase the TIC,[206,363,407,408] which have also been supported by experiments.[359,409] One of the common qualitative predictive tools to study interfacial heat transfer is evaluation of the degree of overlap of phonon density of states (DOS) across the sides of the interface. Current theory suggests that in the elastic scattering limit, larger overlap results in higher TIC. A systematic study of over 2000 interfaces by Gordiz and Henry further tested this idea and showed that, although predictive in many cases, in general, DOS overlap is not the key descriptor for TIC. Recent theoretical observations by Giri and Hopkin[368] and Gordiz and Henry[207] have also shown that DOS overlap is not applicable in different scenarios, which has also been recently experimentally confirmed by Gaskins *et al.*[364] Thus, further studies, particularly based on the MD-based modal analysis are needed to gain insight into the exact phonon interactions that lead to an increase in interface heat transfer caused by changes in atomic distributions around the interface region.

### 4.3.3. Interfacial interaction/bonding modifications

It can be seen from Equation 19 that interfacial forces directly impact the interfacial heat flow. Thus, by being able to modify them we can have opportunities to control the TIC. Gordiz and Henry[394] explored this idea by systematically changing the mass/mismatch and interfacial forces across more than 2000 Lennard-Jones interfaces and calculating the TIC of each of them using EMD simulations (Figure 13). It should be noted that although the masses and forces constants are changing, there are specific variations of these two parameters that lead to exactly matched phonon dispersion curves for the two sides of the interface (the points on the diagonal of Figure 13), which can result in 100% bulk DOS overlap. Theoretically, in the harmonic limit, complete overlap of DOS should result in the maximum TIC values among interfaces with identical dispersion curves (on the diagonal of Figure 13). However, the calculations by Gordiz and Henry[394] showed that the maximum conductance does not necessarily happen at the interfaces with 100% bulk DOS overlap. In fact, as Figure 13 shows maximum conductance happens on an off-diagonal region that corresponds to some non-obvious interfaces. Gordiz and Henry[394] showed that DOS overlap can to some degree explain these observations, if the DOS overlap is calculated for the interfacial region instead of the bulk region. However, that still seems insufficient, and because of the anharmonic interactions at play (Figure 13(a)) temperature-dependent anaharmonic MD-based modal analysis interactions are needed to shed more light on the exact phonon interactions that cause such observation. For similar Lennard-Jones systems, Giri *et al*.[410] have also shown that TIC increases when the temperature is raised, which shows the role of anharmonic interactions in interfacial heat transfer[342,359,411,412], which can be captured by MD simulations.[76,153,207,208,386,387,389,390,394,413,414]

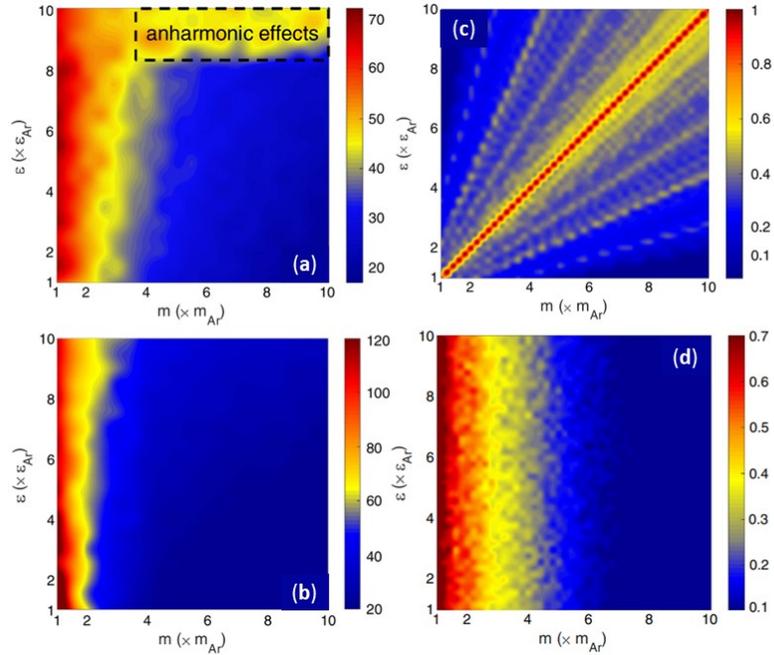

**Figure 13**. (a-b) TIC values across more than 2000 Lennard-Jones interfaces obtained from EMD simulations at (**a**) high temperature (40K) and (**b**) low temperature (5K) simulations.[394] High and low temperature terms utilized here are with respect to argon's melting point of 83K. Phonon DOS degrees of overlap for all interfaces when only (**c**) bulk and (**d**) interfacial vibrations are included in the DOS overlap calculations.

### 4.3.4. Interfaces between amorphous materials

It has been shown that detecting the temperature drop across the interface of amorphous materials is difficult to detect both experimentally and also in NEMD simulations (Figure 14(b)).[368] To overcome this difficulty, instead of analyzing a single amorphous interface, prior work[415] used NEMD across superlattices and then using the thermal circuit model to extract the conductance corresponding to a single interface (Figure 14(a)-(c)).[415] Another method that has been shown to be powerful in resolving this issue is using EMD across single amorphous interfaces.[207,208] Using these approaches Giri and Hopkins[368] and Gordiz and Henry[207] have shown that contrary to the notion that high TC materials lead to higher TIC values, amorphous-amorphous interfaces have a very high TIC despite the fact that both sides can have low TC corresponding to amorphous solids. The high values of TIC across amorphous-amorphous interfaces have also been confirmed by experimental measurements.[407,416] Because of the broken symmetry across amorphous interfaces, application of PGM-based approaches across these interfaces seems questionable. Therefore, further studies across amorphous interfaces, using MD-based modal analysis techniques can be an interesting area of research to provide more insight into the reported high TIC values.

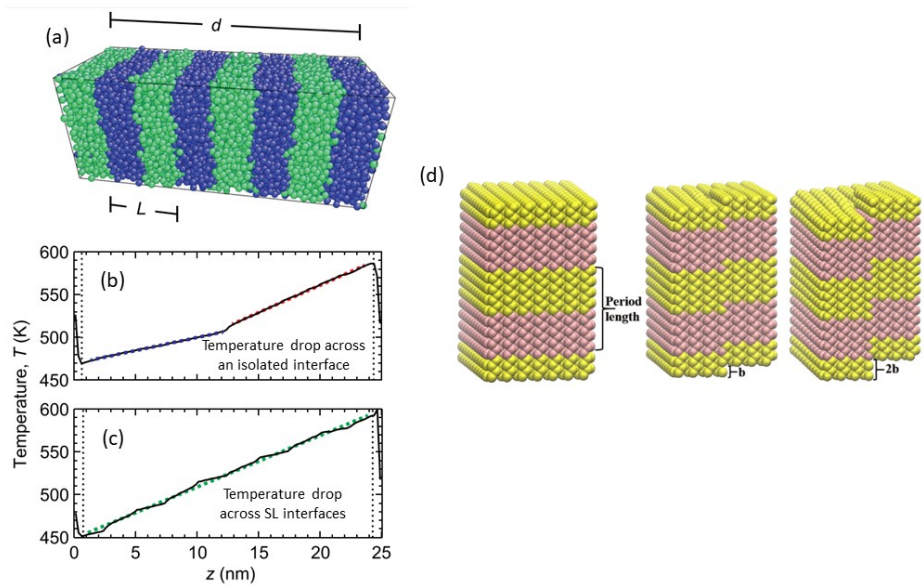

**Figure 14.** (a)-(c) NEMD simulation across the Stillinger-Weber amorphous superlattice (SL) with clearly small temperature drops across interfaces (reprinted from Ref [415], Copyright 2015, with permission from AIP Publishing). (d) Si/Ge SL structures with different degrees of dislocations (no dislocations on the left structure and with dislocations in the middle and on the right one) (reprinted from Ref [252], Copyright 2019, with permission from American Physical Society).

### 4.3.5. Dislocations

During the synthesis of an interface between two lattice-mismatched materials, strain relaxation can lead to dislocations,[366] which have been shown to be important in thermal transport in solid materials.[417] Using AGF,[418] dislocations at an individual Si/Ge interface has been investigated. Although MD simulations have been done to capture the correct atomic orientations at the γ/γ′ phase interface in a Ni-based single-crystal superalloy[419,420] or Ni/Al interface[421], very few MD simulations have been performed to investigate the heat transfer across interfaces with dislocations. Thus, it is an area that remains open, with many unanswered questions. However, one example is the work by Hu *et al.*,[252] where they showed that interfacial dislocations in Si/Ge superlattice nanowires decrease the TC in these structures (Figure 14d). Nonetheless, the main reason there is a lack of MD simulations analyzing thermal transport at dislocated interfaces, is the large number of atoms that are needed in such simulations.

### 5. Comparison between MD-based modal analysis techniques

As explained in previous sections, various methods [76,137,151–154,386,387,389,390] are able to capture the modal contributions to thermal transport in the bulk or through interfaces of solid materials based on MD simulations. All these proposed methods [76,137,151–154,386,387,389,390] can provide the spectral contributions to heat transfer in all phases of solid materials and interfaces. However, when it comes to providing additional information about the eigenvectors of these contributing modes of vibration, these spectral methods [137,151,153,386,390] are restricted to cases when crystalline solids are present in the bulk or at the sides of interface, since they are dependent on a well-defined definition of a group velocity. Studies have extended the applicability of these methods to amorphous materials by defining an effective phonon dispersion curve for these broken symmetry systems and using SED method [137] to identify the modal contributions. [142] However, the only two methods that can provide both spectral and modal contributions to thermal transport in all phases of solids without any simplifying assumptions are GKMA [154] and ICMA.[76,389] Having access to the exact eigenvectors of the contributing modes of vibration and how each mode is spatially and directionally distributed in space is an important knowledge needed for future designs. Here, to understand the differences and similarities between different MD-based modal analysis approaches, we discuss and compare their predictions for the prototypical example of crystalline Si/Ge interface. Several MD studies have calculated the conductance across Si/Ge interface, however only a few have reported anharmonic modal contributions in the system. Using spectral analysis approach, Chalopin and Volz[153] and Murakami *et al.*[422] showed that at a crystalline Si/Ge interface, the modes of vibration around 12-13 THz contribute largely to the interface conductance. Zhou and Hu[387] had a similar observation. In addition, using an extension of the spectral analysis approach they[387] could also determine the harmonic and third-order anharmonic contributions to the conductance (Figure 15(d)-(e)). Using ICMA method, Gordiz and Henry[413] showed three distinct features of these highly contributing modes around 12-13 THz. First, the number of the contributing modes is less than 0.3% of the total number of modes in the system. Second, the eigen-vectors of vibration for these interfacial modes have a large degree of localization around the interface, however, they also have vibrations on the bulk of the Si side (Figure 15(f)-(g)). Third, the eigen-vectors of vibrations have allowed these modes of vibration with frequencies around 12-13 THz to strongly couple to all the other modes of vibration in the system and transfer their energy across the interface, which is reflected on the strong cross-correlation band around the 12-13 THz region (Figure 15 (b)-(c)).

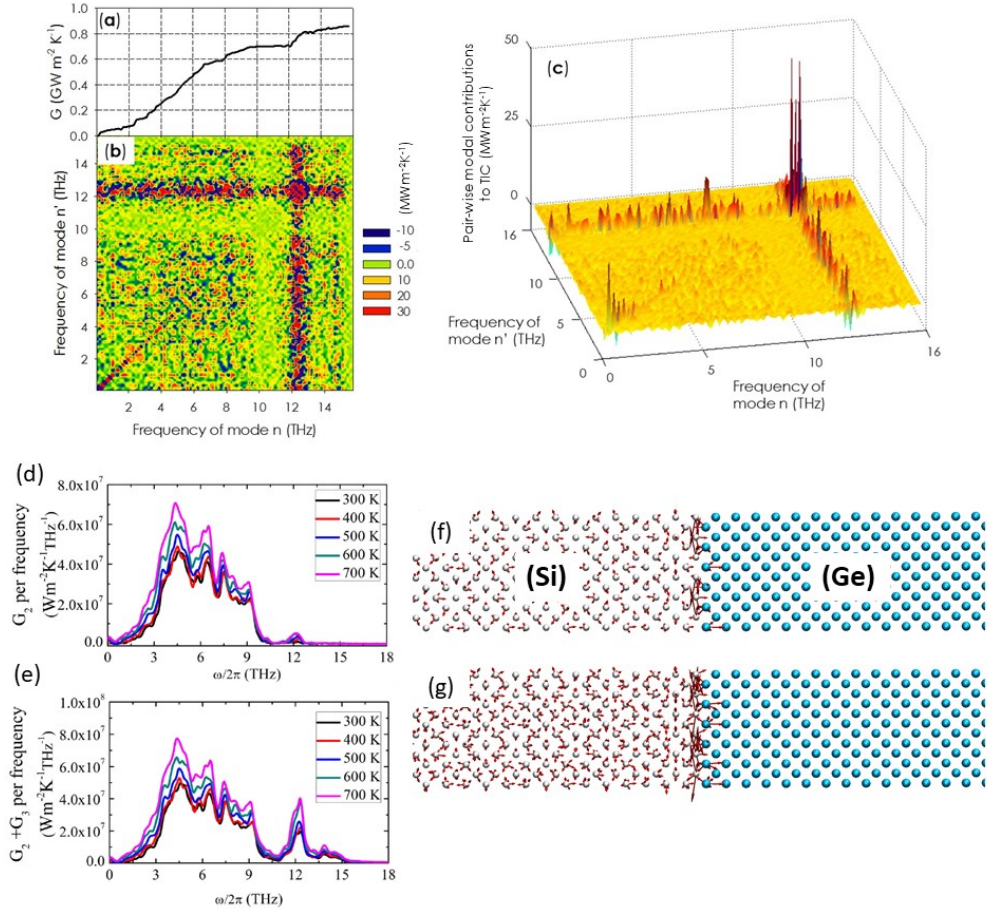

**Figure 15.** (a)-(c) are from the study by Gordiz and Henry.[413] (a) TIC accumulation function, (b) 2D map and (c) 3D perspective depiction of the data in (b) showing the magnitudes of the correlations as elevations above the plane of two frequency axes. (d-e) are from the study by Zhou *et al.* (reprinted from Ref [387], Copyright 2017, with permission from the American Physical Society) and they show the (d) harmonic and (e) third-order anharmonic contribution of different frequencies of vibration in the system at different temperatures. (f-g) are from the study by Gordiz

and Henry[413] and show two examples from the highly contributing modes of vibration around 12-13 THz. The red arrows show the eigenvectors of vibration.

It should be noted however that the implementation of spectral methods is seemingly easier than GKMA and ICMA and their convergence are faster than GKMA and ICMA since almost all spectral methods [151–154,386,387,390] are based on NEMD simulations. A new more general and easier implementation of GKMA in LAMMPS [423] has been recently proposed, [385] but ICMA implementation is still dependent on the details of the structure and the chosen interatomic potential. In addition, the convergence of GKMA and ICMA might require more computational expense since they are based on EMD simulations. Another issue to note here is that the LD calculations required by GKMA and ICMA methods are based on gamma point SCLD. For mid-size MD structures (~10,000 atoms) calculating the vibrational modes based on SCLD is tractable. However, when it comes to arbitrarily large systems (>100,000 atoms), calculating the modes of vibration becomes the bottleneck to modal analysis. This is because although the dynamical matrix corresponding to a large system can be straightforwardly obtained using approaches such as direct displacements,[424] diagonalizing large dynamical matrix is not straightforward even by using the existing state-of-the-art numerical packages such as ScaLAPACK.[425] Therefore, although GKMA and ICMA are promising in the degree of information they can provide about the modal contributions to thermal transport in all phases of solid materials, efforts are needed to rectify their easier implementation and obtain a straightforward approach to calculate the modes of vibration from SCLD calculations of arbitrarily large systems before their potential can be truly realized.

In sections 3 and 4, we reviewed the theory and application of MD-based approaches to study the effects of physicochemical system parameters on TC and TIC respectively. To utilize the full potential of MD, i.e. in designing new materials with targeted structural and compositions features, it is important to thoroughly benchmark the simulation framework against novel experimental data.

## 6. Validation and Benchmarking: Experimental Measurements

While the MD methodology has been able to model experimental data reasonably well on several occasions, new predictions need to be benchmarked against experimental data to ensure general applicability of MD-based approaches. Here we briefly review experimental methods for measuring TC and TIC. The first experimental measurements of thermal conductivity was reported back to 1861 by Angstrom.[426] A large amount of experimentally measured thermal conductivity data are well-documented in the literature and can be used to compare with calculated ones.[427–429] In the past several decades, with the development of ultrafast lasers, nanoscale probing of thermal conductivity with picosecond time resolution and micrometer spatial resolution have become possible.[430,431] The errors of TC measured by commonly-used transient techniques such as 3-omega and time-domain thermoreflectance (TDTR) are approximately ±10% while those by steady-state techniques such as Raman thermometry are about ±20%.[432] Table 1 shows a comparison of several thermal characterization techniques. In some cases, the techniques are able to probe the phonon mean free path (MFP) and provide additional data to the TC. Picosecond acoustic technique can measure the local thicknesses of multilayer thin films. TDTR is the most widely used tool for TIC measurements.

**Table 1. Comparison of several thermal characterization techniques.**

|  | TDTR | FDTR | PA | Raman | 3-Omega | Laser flash |
|---|---|---|---|---|---|---|
| TIC | ✓ | ✓ | X | Limited | Limited | Limited |
| Anisotropy | ✓ | ✓ | X | ✓ | ✓ | Limited |
| Rough surface | X | X | ✓ | ✓ | Limited | ✓ |
| Metal transducer | ✓ | ✓ | ✓ | X | ✓ | ✓ |
| Phonon MFP | ✓ | ✓ | X | X | X | X |

| Picosecond acoustic | ✓ | X | X | X | X | X |
|---|---|---|---|---|---|---|
| Penetration depth | nm-μm | nm-μm | μm | | μm | μm-mm |
| Spatial resolution | μm | μm | mm | mm | μm | mm |
| Mapping | ✓ | ✓ | X | ✓ | X | X |

Unlike TC measurements, experimental measurements of TIC were not reported near room temperature until 1980s and 1990s with the developments of a transient hot-strip technique and a picosecond transient thermoreflectance (PTT).[433,434] Enabled by the development of TDTR, there are about 50 papers (covering about 60 interfaces) in total to date for experimentally measured TIC. Most of them are metal-nonmetal interfaces because TDTR measurements need metal films on the sample surface as transducers: Au/Al$_2$O$_3$[433,435], TiN/Al$_2$O$_3$[436], TiN/MgO[436], Au/Si[437], Al/Ge[438], Al/Si[438], NiSi$_2$/Si[439], Al/MgO[440], Al/diamond[441], Pb/diamond[441], Pt/diamond[441], Au/diamond[441], NiSi/Si[442], CoSi$_2$/Si[442], TiSi$_2$/Si[442], Rh/Al$_2$O$_3$[343], Ti/diamond[443], Al/Al$_2$O$_3$[433], Pb/diamond[433], Al/AlN[444], Pt/Si[444], H$_2$O/Al with SAM[445], Au/GaAs with SAM[446], Bi/diamond[411], Cr/Si[404], Pt/Al$_2$O$_3$[359], Bi/Si[447], Ti/graphite[448], Al/diamond[449], Al/SiC[450], Al/Si[451], Al/graphene[452], Au/quartz with SAM[453], Cu/silica with SAM[454], Al/O/diamond[455], Au/Ti/Si[456], metal carbides/diamond[457], Au/hexadecane[458], AuCu/sapphire[459], Al/Si[460], SiO$_2$/Pt[416], Au/Cu/Al$_2$O$_3$[461]. For these measurements, the metal-substrate TIC usually has high sensitivity which results in small errors (about ±7%).[435] However, for the measurements of nonmetal-nonmetal interfaces, the heating on the metal transducer surface needs to penetrate through the metal transducer layer, the metal-nonmetal interface, and the nonmetal thin layer to reach the nonmetal-nonmetal interface. This leads to low measurement sensitivity of the buried nonmetal-nonmetal interfaces and extra unknown parameters in the data fittings. Due to the experimental difficulties, the measurement errors (±20-30%) are usually much larger than those of the metal-nonmetal interfaces.[364] Much less interfaces have been measured: Si/diamond[462], AlN/GaN[463], GaSb/GaAs[464], organic/clay superlattice[465], GaN/SiC[466], Si/Si van der Waals interface[467], $SiO_2/Al_2O_3$ superlattice[468], ZnO/HQ/ZnO superlattice[469], PS/sapphire with self-assembled monolayer (SAM)[470], GaN/Si[471], $SiO_2$/Si[416], ZnO/GaN[364], SiC:H/SiOC:H superlattice[407], bonded GaN/SiC[472], Si/diamond[473], $Ga_2O_3$/diamond van der Waals interface[474]. Additionally, electron-mediated TIC are orders of magnitude higher than phonon-mediated TIC and only a few metal-metal interfaces have been measured: Al/Cu[475], and Pd/Ir[476].

Among these experimentally measured interfaces, only several interfaces are epitaxially grown which are supposed to have high-quality interfaces.[364,435,436,440,442,463,466] However, the real interfacial structures of most of these epi-interfaces are unknown due to the lack of detailed material characterizations except an Al-sapphire interface reported recently.[435] For metal-nonmetal interface growth, metals may react with the substrates and are affected by contamination, disorder, alloying, mixing, etching, and oxidation.[436,442] The lack of detailed material characterizations of the interfaces in the literature hinders building the relationship of the interface structures and the measured TIC because interfacial reactions, mixing, bonding, crystalline orientations, surface chemistry, pressure, roughness, and interfacial disorder all affect TIC.[435] All these factors make the measured TIC values in the literature inconsistent even for same interfaces.[435]

The large gap between interfacial structures and the measured TIC induced by the lack of material characterizations impedes the development of theoretical understanding of interfacial thermal transport. Theoretical or modeling works are often based on the presumption of perfect interfaces or compared with the measured TIC of interfaces with unknown interfacial structures. This usually results in large deviations between the calculated TIC and the measured TIC. The deviations resulting from structure and imperfections of the real interface being measured are usually compounded with the deviations resulting from the inaccuracy of the theoretical models, which impedes the development of fundamental understanding of TIC. For future experimental works, detailed material characterizations are imperative to understand the real interfacial structures for different growth conditions and techniques, especially for technologically important interfaces and new heterogeneously integrated interfaces.[472,477] High-throughput advanced thermal measurement techniques are also on developing to measure the TIC of buried interfaces with small errors.[478] For future theoretical works, more theoretical modeling or calculations which can include the effects of the interfacial imperfections should be developed to understand the real interfaces.

## 7. Open questions and challenges

Before examining the challenges associated with MD itself, it is worth mentioning that materials with defects, polymers, amorphous alloys and their interfaces have received much less attention in the literature. For materials with defects, to our knowledge, all prior studies have only reported how defects affect TC, but not the changes in modal character and their influence on thermal transport. While modal contributions to polymers, interfaces, and alloys have also been studied, interactions between these modes have not been explored enough. For instance, how do diffusons and locons interact with each other and with other quasiparticles? In this regard, one may also ask how modes which contribute strongly to TC or TIC gain their energy from other modes; how do specific mode interactions contribute to thermal transport? Gordiz *et al.*[208,479] have shown that interface modes have finite contributions to TIC and correlate strongly with other modes, but it is not known why this is true or if it can be exploited to design thermal transport across interfaces. The reason for these materials and interfaces not receiving much attention can be attributed partly to the lack of interatomic potentials and partly to the lack of fast, accurate, and reliable MD based formalisms to calculate the modal contributions. However, recent progress in the development of interatomic potential and the emergence of modal analysis techniques, future studies can reveal the mechanisms of thermal transport in these systems which will be of great importance to numerous technological applications.[480–482]

There are also several hurdles to be overcome to ensure fast, accurate, and reliable MD calculations. MD simulations are well-established with convenient and powerful codes such as LAMMPS[423], GROMACS[483], and NAMD[484], which have been used extensively over the past two decades. Development of faster MD methods is crucial to bringing more practicality and usefulness to thermal transport property prediction in MD, i.e. by providing quick calculations of TC and TIC. One promising method here is the use of graphics processing units (GPUs) to accelerate the sampling of phase space required by thermal transport calculations, which have been shown by Yang *et al.*[485] to result in an order of magnitude speed-up when calculating TC of solid argon. Regarding software development for the use of GPUs in MD, Fan *et al.*[486] recently developed a code called GPUMD which incorporates a unique algorithm for calculating force and heat current on GPUs, and they showed that this resulted in more than an order of magnitude speedup compared to LAMMPS calculations.

While such efforts result in increased speed of MD simulations, the basic theory for calculating TC and TIC remains the same. Since the MD methods to calculate TC and TIC (viz. EMD, NEMD, and AEMD) are well-established, recent and future work regarding simulations of phonon transport involves a deeper understanding into the underlying mechanisms behind thermal transport, and the modal analysis techniques of Section 3.2 highlight the recent work in this area.

Accurately simulating the dynamics and performing modal analysis, however, requires accurate IAPs to ensure that the correct modes of a system are being described (the harmonic part of the potential), as well as the correct interactions between those modes (the anharmonic part of the potential). An accurate representation of the underlying potential energy surface of a system is needed for an accurate sampling of space, one can in turn accurately sample phase space and use statistical mechanical approaches like EMD-GK to extract bulk thermal properties. Thus, a critical challenge for the community is the development of accurate IAPs that can be validated and used in a predictive capacity.

Another significant challenge hindering the broad applicability of MD-based methods is that phonon populations are governed by incorrect statistics. As classical simulations, phonon populations follow the equipartition theorem, whereas in real quantum systems, phonons exhibit a Bose-Einstein distribution in their populations. This can cause appreciable deviations at temperatures well below the Debye limit that necessitate the use of correction schemes. An important study by Turney *et al.* self-consistently assess the validity of a common correction scheme mapping classical thermal conductivity to quantum thermal conductivity and demonstrated rather poor agreement between corrected and actual quantum thermal conductivity.[487] They demonstrated that, while the heat capacities can be easily corrected at a modal level, there is no clear way to account for increased phonon scattering due to elevated mode occupations in low temperature classical simulations relative to that expected from quantum statistics. Indeed, a recent paper by Hu *et al.* confirmed that increased phonon scattering due to elevated mode occupation was the primary reason for the discrepancy in thermal conductivity predictions at low temperature from classical MD systems.[488]

It has been argued that in systems in which phonon-phonon scattering is not the dominant form of scattering, as in the case when boundary effects are strong, the effects of incorrect mode occupations are less relevant (for more details, refer to the tutorial by Seyf *et al.* [385], Section II.C., for a further discussion). Likewise, quantum corrections applied at a frequency-level, rather than a mode-level, can be a more reasonable post processing correction scheme, as demonstrated in the study by in a recent study investigating frequency dependent mean free path in amorphous silicon.[194,489] Nevertheless, developing heuristic correction schemes that can account for the excess scattering due to elevated mode occupations populations at lower temperatures in classical simulations would be a significant step forward. Unfortunately, no clear path exists to do this.

In principle, it is possible to go beyond *post-hoc* quantum correction schemes and directly incorporate nuclear quantum effects. One approach is to use a quantum thermostat, coupling a classical system to a quantum bath, in order to induce the correct Bose-Einstein statistics.[490,491] This is an exciting approach and has been applied to the study of thermal conductivity in the past.[492] Unfortunately, such methods can be problematic for anharmonic systems, suffering from what is known as zero point energy leakage, where energy from high frequency modes incorrectly transfer to low frequency modes.[493] Nonetheless, efforts to address this issue are ongoing.[494]

An even more rigorous approach to incorporating nuclear quantum effects is the use of path integral molecular dynamics, based on the path integral technique of Feynman. Here, quantum dynamics can be replicated by simulating multiple images of a system, harmonically corrected, at every time-step, governed by what is known as a "ring-polymer" Hamiltonian (for more details, refer to the review by Markland and Ceriotti for an overview of the technique and its variants[495]). However, this makes such an approach enormously expensive, and there are further issues with the incorporation of nonlinear operators (like the heat flux).[496,497] Still, the application of methods that directly incorporate nuclear quantum effects into MD simulations of thermal properties is very undeveloped but could be a highly fruitful path of research going forward.

## 8. Summary

We have reviewed the theory and application of classical MD methodology to extract the phonon transport properties (TC and TIC). MD approach has several advantages over ALD based solution of BTE – implicit inclusion of anharmonicity to full order and no restrictive dependence on a plane wave phonon model thereby letting the user conduct numerical experiments on perfect crystals as well as materials over varying degrees of disorder and interfaces, often inaccessible to DFT-BTE and *ab initio* MD. Classical MD also lets one attain large length scales (several nm) and long timescales (several ns) which are cost prohibitive with DFT at the length scales needed ($\geq 1$ nm $\rightarrow$ $10^2$-$10^4$ atoms).

We have also reviewed the theory of LD and how MD can be used in combination with LD to extract phonon modal information, especially modal contribution to TC and TIC. Upon surveying the studies by various research groups to understand and predict phonon modal contribution to TC and TIC, the span of materials considered is broad and the effects of intrinsic and extrinsic parameters considered, namely size effects, structural and compositional disorder, and interfacial properties such as roughness, dislocation, atomic mixing, mass grading, nanostructuring, and bonding modifications constitute a comprehensive understanding of phonon transport properties. Notably, results of these studies converge to PGM on one end of the spectrum for crystalline materials, to the Allen-Feldman theory for strongly disordered materials at the other end of the spectrum. We have also revisited how different researchers have used MD simulations to examine the general applicability of widely used theoretical models such as the VCA approximation with Tamura model and the K-C model developed for materials with compositional and structural defects respectively. In essence, MD can be judiciously applied to design novel materials with desired thermal transport properties for diverse applications in the fields of energy, biomedical, microelectronics, textiles, and much more.

Nevertheless, we have also identified critical deficiencies associated with the MD methodology and suggested solutions to be explored in the coming years. Firstly, accuracy of results relies heavily on the efficacy of the interatomic potentials in capturing phonon properties. This, to an extent, can be addressed by developing vibrationally accurate interatomic potentials from the first principles and benchmarking against independently obtained experimental data. And secondly, being based on classical mechanics, MD does not include quantum effects, leading to reduced predictive accuracy at low temperatures. While semi-empirical quantum correction schemes have been used in the past, it is important to explore frequency level quantum correction schemes and more rigorous nuclear quantum correction schemes to ensure high levels of accuracy in predictions.

## Acknowledgements

The authors would like to thank the United States Office of Naval Research Multidisciplinary University Research Initiative (N00014-18-1-2429) and the National Science Foundation CAREER (1554050) awarded to A. H. for financial support. A. R. was supported by the National Science Foundation Graduate Research Fellowship (1122374). Any opinion, findings, and conclusions or recommendations expressed in this material are those of the authors and do not necessarily reflect the views of the National Science Foundation.

**References**

[1]   Cahill D G, Ford W K, Goodson K E, Mahan G D, Majumdar A, Maris H J, Merlin R and Phillpot S R 2003 Nanoscale thermal transport *J. Appl. Phys.* **93** 793–818

[2]   Cahill D G, Braun P V, Chen G, Clarke D R, Fan S, Goodson K E, Keblinski P, King W P, Mahan G D and Majumdar A 2014 Nanoscale thermal transport. II. 2003–2012 *Appl. Phys. Rev.* **1** 11305

[3]   Moore A L and Shi L 2014 Emerging challenges and materials for thermal management of electronics *Mater. today* **17** 163–74

[4]   Costescu R M, Cahill D G, Fabreguette F H, Sechrist Z A and George S M 2004 Ultra-low thermal conductivity in W/Al2O3 nanolaminates *Science (80-. ).* **303** 989–90

[5]   Williams B S 2008 Terahertz quantum cascade lasers *Asia Optical Fiber Communication and Optoelectronic Exposition and Conference* (Optical Society of America) p SuG3

[6]   Hao M, Li J, Park S, Moura S and Dames C 2018 Efficient thermal management of Li-ion batteries with a passive interfacial thermal regulator based on a shape memory alloy *Nat. Energy* **3** 899–906

[7]   Wang M and Lin S 2016 Anisotropic and ultralow phonon thermal transport in organic–inorganic hybrid perovskites: atomistic insights into solar cell thermal management and thermoelectric energy conversion efficiency *Adv. Funct. Mater.* **26** 5297–306

[8]   Saadah M, Hernandez E and Balandin A A 2017 Thermal management of concentrated multi-junction solar cells with graphene-enhanced thermal interface materials *Appl. Sci.* **7**

589

[9]  Cui H and Overend M 2019 A review of heat transfer characteristics of switchable insulation technologies for thermally adaptive building envelopes *Energy Build.* **199** 427–44

[10]  Henry A, Prasher R and Majumdar A 2020 Five thermal energy grand challenges for decarbonization *Nat. Energy* 1–3

[11]  Wu S, Yan T, Kuai Z and Pan W 2020 Thermal conductivity enhancement on phase change materials for thermal energy storage: A review *Energy Storage Mater.* **25** 251–95

[12]  Li C, Fu L, Ouyang J, Tang A and Yang H 2015 Kaolinite stabilized paraffin composite phase change materials for thermal energy storage *Appl. Clay Sci.* **115** 212–20

[13]  Yifan Z, Wei H, Le Z, Yong M, Lei C, Zongxiang L and Ling D 2020 Power and energy flexibility of district heating system and its application in wide-area power and heat dispatch *Energy* **190** 116426

[14]  Kavvadias K C and Quoilin S 2018 Exploiting waste heat potential by long distance heat transmission: Design considerations and techno-economic assessment *Appl. Energy* **216** 452–65

[15]  Tian Z, Lee S and Chen G 2014 Comprehensive review of heat transfer in thermoelectric materials and devices *Annu. Rev. Heat Transf.* **17**

[16]  Shakouri A 2006 Nanoscale thermal transport and microrefrigerators on a chip *Proc. IEEE* **94** 1613–38

[17]  Ziman J M 2001 *Electrons and phonons: the theory of transport phenomena in solids* (Oxford university press)

[18]  Dove M T and Dove M T 1993 *Introduction to lattice dynamics* vol 4 (Cambridge

university press)

[19] Srivastava G P 2019 *The physics of phonons* (Routledge)

[20] Peierls R 1929 Zur kinetischen theorie der wärmeleitung in kristallen *Ann. Phys.* **395** 1055–101

[21] Klemens P G 1958 Thermal conductivity and lattice vibrational modes *Solid state physics* vol 7 (Elsevier) pp 1–98

[22] Callaway J 1959 Model for {Lattice} {Thermal} {Conductivity} at {Low} {Temperatures} *Phys. Rev.* **113** 1046–51

[23] Holland M G 1963 Analysis of lattice thermal conductivity *Phys. Rev.* **132** 2461

[24] Carruthers P 1961 Theory of thermal conductivity of solids at low temperatures *Rev. Mod. Phys.* **33** 92

[25] Klemens P G 1951 The thermal conductivity of dielectric solids at low temperatures (theoretical) *Proc. R. Soc. London. Ser. A. Math. Phys. Sci.* **208** 108–33

[26] Wingert M C, Zheng J, Kwon S and Chen R 2016 Thermal transport in amorphous materials: a review *Semicond. Sci. Technol.* **31** 113003

[27] DeAngelis F, Muraleedharan M G, Moon J, Seyf H R, Minnich A J, McGaughey A J H and Henry A 2019 Thermal Transport in Disordered Materials *Nanoscale Microscale Thermophys. Eng.* **23** 81–116

[28] Peierls R E 1996 *Quantum theory of solids* (Clarendon Press)

[29] Srivastava G P 1990 *The physics of phonons* (CRC press)

[30] Frenkel D, Smit B, Tobochnik J, McKay S R and Christian W 1997 Understanding Molecular Simulation *Comput. Phys.* **11** 351

[31] Allen M P and Tildesley D J 2017 *Computer simulation of liquids* (Oxford university

[32]   Landauer R 1970 Electrical resistance of disordered one-dimensional lattices *Philos. Mag.* **21** 863–7

[33]   Khalatnikov I M 1952 *Teploobmen Mezhdu Tverdym Telom I Geliem-Ii *Zhurnal Eksp. I Teor. Fiz.* **22** 687–704

[34]   Little W A 1959 The transport of heat between dissimilar solids at low temperatures *Can. J. Phys.* **37** 334–49

[35]   Swartz E T and Pohl R O 1989 Thermal boundary resistance *Rev. Mod. Phys.* **61** 605

[36]   Mingo N and Yang L 2003 Phonon transport in nanowires coated with an amorphous material: An atomistic Green's function approach *Phys. Rev. B* **68** 245406

[37]   Kakodkar R R and Feser J P 2015 A framework for solving atomistic phonon-structure scattering problems in the frequency domain using perfectly matched layer boundaries *J. Appl. Phys.* **118** 94301

[38]   Neeper D A and Dillinger J R 1964 Thermal resistance at indium-sapphire boundaries between 1.1 and 2.1 K *Phys. Rev.* **135** A1028

[39]   Dai J and Tian Z 2020 Rigorous formalism of anharmonic atomistic Green's function for three-dimensional interfaces *Phys. Rev. B* **101** 41301

[40]   Schelling P K, Phillpot S R and Keblinski P 2002 Phonon wave-packet dynamics at semiconductor interfaces by molecular-dynamics simulation *Appl. Phys. Lett.* **80** 2484–6

[41]   Broido D A, Malorny M, Birner G, Mingo N and Stewart D A 2007 Intrinsic lattice thermal conductivity of semiconductors from first principles *Appl. Phys. Lett.* **91** 231922

[42]   Stillinger F H and Weber T A 1985 Computer simulation of local order in condensed phases of silicon *Phys. Rev. B* **31** 5262

[43]     Tersoff J 1988 Empirical interatomic potential for carbon, with applications to amorphous carbon *Phys. Rev. Lett.* **61** 2879–82

[44]     Brenner D 1990 Empirical potential for hydrocarbons for use in simulating the chemical vapor deposition of diamond films *Phys. Rev. B* **42** 9458–71

[45]     Weber W 1977 Adiabatic bond charge model for the phonons in diamond, Si, Ge, and α− Sn *Phys. Rev. B* **15** 4789

[46]     Lindsay L, Broido D A and Reinecke T L 2013 First-principles determination of ultrahigh thermal conductivity of boron arsenide: a competitor for diamond? *Phys. Rev. Lett.* **111** 25901

[47]     Lindsay L, Broido D A and Reinecke T L 2013 Phonon-isotope scattering and thermal conductivity in materials with a large isotope effect: A first-principles study *Phys. Rev. B* **88** 144306

[48]     Feng T, Lindsay L and Ruan X 2017 Four-phonon scattering significantly reduces intrinsic thermal conductivity of solids *Phys. Rev. B* **96** 161201

[49]     Protik N H, Carrete J, Katcho N A, Mingo N and Broido D 2016 Ab initio study of the effect of vacancies on the thermal conductivity of boron arsenide *Phys. Rev. B* **94** 45207

[50]     Wang C Z, Chan C T and Ho K M 1990 Tight-binding molecular-dynamics study of phonon anharmonic effects in silicon and diamond *Phys. Rev. B* **42** 11276

[51]     Lee Y H, Biswas R, Soukoulis C M, Wang C Z, Chan C T and Ho K M 1991 Molecular-dynamics simulation of thermal conductivity in amorphous silicon *Phys. Rev. B* **43** 6573

[52]     Marcolongo A, Umari P and Baroni S 2016 Microscopic theory and quantum simulation of atomic heat transport *Nat. Phys.* **12** 80–4

[53]     Carbogno C, Ramprasad R and Scheffler M 2017 Ab initio Green-Kubo approach for the

thermal conductivity of solids *Phys. Rev. Lett.* **118** 175901

[54] McGaughey A J H and Kaviany M %J P R B 2004 Quantitative validation of the Boltzmann transport equation phonon thermal conductivity model under the single-mode relaxation time approximation **69** 94303

[55] Chen G 2005 *Nanoscale energy transport and conversion: a parallel treatment of electrons, molecules, phonons, and photons* (Oxford University Press)

[56] Nelson E 1966 Derivation of the Schrödinger equation from Newtonian mechanics *Phys. Rev. D* **150** 1079

[57] Müser M H 2001 Simulation of material properties below the debye temperature: a path-integral molecular dynamics case study of quartz *J. Chem. Phys.* **114** 6364–70

[58] Poltavsky I and Tkatchenko A 2016 Modeling quantum nuclei with perturbed path integral molecular dynamics *Chem. Sci.* **7** 1368–72

[59] Batcho P F and Schlick T 2001 Special stability advantages of position-Verlet over velocity-Verlet in multiple-time step integration *J. Chem. Phys.* **115** 4019–29

[60] Harrison R W 1993 Stiffness and energy conservation in molecular dynamics: An improved integrator *J. Comput. Chem.* **14** 1112–22

[61] Skeel R D, Zhang G and Schlick T 1997 A family of symplectic integrators: stability, accuracy, and molecular dynamics applications *SIAM J. Sci. Comput.* **18** 203–22

[62] Hoover W G 1985 Canonical dynamics: Equilibrium phase-space distributions *Phys. Rev. A* **31** 1695

[63] Grønbech-Jensen N 2019 Complete set of stochastic Verlet-type thermostats for correct Langevin simulations *Mol. Phys.* 1–25

[64] Ping Y, Correa A A, Ogitsu T, Draeger E, Schwegler E, Ao T, Widmann K, Price D F,

Lee E, Tam H, Springer P T, Hanson D, Koslow I, Prendergast D, Collins G and Ng A 2010 Warm dense matter created by isochoric laser heating *High Energy Density Phys.*

[65] Zarkadoula E, Samolyuk G, Xue H, Bei H and Weber W J 2016 Effects of two-temperature model on cascade evolution in Ni and NiFe *Scr. Mater.*

[66] Allen P B 1987 Theory of thermal relaxation of electrons in metals *Phys. Rev. Lett.*

[67] Duffy D M and Rutherford A M 2007 Including the effects of electronic stopping and electron-ion interactions in radiation damage simulations *J. Phys. Condens. Matter*

[68] Tamm A, Samolyuk G, Correa A A, Klintenberg M, Aabloo A and Caro A 2016 Electron-phonon interaction within classical molecular dynamics *Phys. Rev. B*

[69] Tamm A, Caro M, Caro A, Samolyuk G, Klintenberg M and Correa A A 2018 Langevin Dynamics with Spatial Correlations as a Model for Electron-Phonon Coupling *Phys. Rev. Lett.*

[70] Chandrasekaran A, Kamal D, Batra R, Kim C, Chen L and Ramprasad R 2019 Solving the electronic structure problem with machine learning *npj Comput. Mater.*

[71] Dove M T 1993 *Introduction to lattice dynamics* vol 4 (Cambridge university press)

[72] Lv W and Henry A 2016 Examining the Validity of the Phonon Gas Model in Amorphous Materials *Sci. Rep.*

[73] Lv W 2017 *A correlation based theory for phonon transport* (Georgia Institute of Technology)

[74] Seyf H R, Yates L, Bougher T L, Graham S, Cola B A, Detchprohm T, Ji M-H, Kim J, Dupuis R and Lv W %J npj C M 2017 Rethinking phonons: The issue of disorder **3** 1–8

[75] Seyf H R and Henry A 2016 A method for distinguishing between propagons, diffusions, and locons *J. Appl. Phys.*

[76]    Gordiz K and Henry A 2015 A formalism for calculating the modal contributions to thermal interface conductance *New J. Phys.* **17** 103002

[77]    Mahajan S and Sanejouand Y H 2017 Jumping between protein conformers using normal modes *J. Comput. Chem.*

[78]    Go N, Noguti T and Nishikawa T 1983 Dynamics of a small globular protein in terms of low-frequency vibrational modes. *Proc. Natl. Acad. Sci. U. S. A.*

[79]    Scaramozzino D, Lacidogna G and Carpinteri A 2019 Protein Conformational Changes and Low-Frequency Vibrational Modes: A Similarity Analysis *Mech. Biol. Syst. Mater. Micro-and Nanomechanics* **4** 7–10

[80]    Tadano T, Gohda Y and Tsuneyuki S 2014 Anharmonic force constants extracted from first-principles molecular dynamics: applications to heat transfer simulations *J. Phys. Condens. Matter* **26** 225402

[81]    Hellman O, Steneteg P, Abrikosov I A and Simak S I %J P R B 2013 Temperature dependent effective potential method for accurate free energy calculations of solids **87** 104111

[82]    Togo A and Tanaka I 2015 First principles phonon calculations in materials science *Scr. Mater.* **108** 1–5

[83]    Murakami T, Shiga T, Hori T, Esfarjani K and Shiomi J %J E P L 2013 Importance of local force fields on lattice thermal conductivity reduction in PbTe1− xSex alloys **102** 46002

[84]    Tian Z, Garg J, Esfarjani K, Shiga T, Shiomi J and Chen G %J P R B 2012 Phonon conduction in PbSe, PbTe, and PbTe 1− x Se x from first-principles calculations **85** 184303

[85]   Zhou F, Nielson W, Xia Y and Ozoliņš V 2019 Compressive sensing lattice dynamics. I. General formalism *Phys. Rev. B* **100** 184308

[86]   Eriksson F, Fransson E and Erhart P 2018 The Hiphive Package for the Extraction of High-Order Force Constants by Machine Learning *Adv. Theory Simulations* 1800184

[87]   Rohskopf A, Wyant S, Gordiz K, Reza Seyf H, Gopal Muraleedharan M and Henry A 2020 Fast & accurate interatomic potentials for describing thermal vibrations *Comput. Mater. Sci.*

[88]   da Cruz C, Termentzidis K, Chantrenne P and Kleber X 2011 Molecular dynamics simulations for the prediction of thermal conductivity of bulk silicon and silicon nanowires: {Influence} of interatomic potentials and boundary conditions *J. Appl. Phys.* **110** 34309

[89]   Howell P C 2012 Comparison of molecular dynamics methods and interatomic potentials for calculating the thermal conductivity of silicon *J. Chem. Phys.* **137** 224111

[90]   Nejat Pishkenari H, Mohagheghian E and Rasouli A 2016 Molecular dynamics study of the thermal expansion coefficient of silicon *Phys. Lett. A* **380** 4039–43

[91]   Zou J-H (邹济杭), Ye Z-Q (叶振强) and Cao B-Y (曹炳阳) 2016 Phonon thermal properties of graphene from molecular dynamics using different potentials *J. Chem. Phys.* **145** 134705

[92]   Khan A I, Navid I A, Noshin M, Uddin H M A, Hossain F F and Subrina S 2015 Equilibrium Molecular Dynamics (MD) Simulation Study of Thermal Conductivity of Graphene Nanoribbon: A Comparative Study on MD Potentials *Electronics* **4** 1109–24

[93]   Xu K, Gabourie A J, Hashemi A, Fan Z, Wei N, Farimani A B, Komsa H-P, Krasheninnikov A V, Pop E and Ala-Nissila T 2019 Thermal transport in

$\{{\textbackslash}mathrm\{{MoS}}\}_{2}$ from molecular dynamics using different empirical potentials *Phys. Rev. B* **99** 54303

[94]   Muraleedharan M G, Rohskopf A, Yang V and Henry A 2017 Phonon optimized interatomic potential for aluminum *AIP Adv.* **7**

[95]   Gonzalez-Valle C U, Hahn S H, Muraleedharan M G, Zhang Q, van Duin A C T and Ramos-Alvarado B 2020 Investigation into the Atomistic Scale Mechanisms Responsible for the Enhanced Dielectric Response in the Interfacial Region of Polymer Nanocomposites *J. Phys. Chem. C*

[96]   Bartók A P, Kondor R and Csányi G 2013 On representing chemical environments *Phys. Rev. B - Condens. Matter Mater. Phys.*

[97]   Bartók A P, Payne M C, Kondor R and review letters Csányi G %J P 2010 Gaussian approximation potentials: The accuracy of quantum mechanics, without the electrons **104** 136403

[98]   Li R, Liu Z, Rohskopf A, Gordiz K, Henry A, Lee E and Luo T 2020 A deep neural network interatomic potential for studying thermal conductivity of β-Ga2O3 *Appl. Phys. Lett.* **117** 152102

[99]   Zhang C and Sun Q 2019 Gaussian approximation potential for studying the thermal conductivity of silicene *J. Appl. Phys.* **126** 105103

[100]  Qian X, Peng S, Li X, Wei Y and Yang R 2019 Thermal conductivity modeling using machine learning potentials: application to crystalline and amorphous silicon *Mater. Today Phys.* **10** 100140

[101]  Gu X and Zhao C Y 2019 Thermal conductivity of single-layer MoS2(1−x)Se2x alloys from molecular dynamics simulations with a machine-learning-based interatomic potential

*Comput. Mater. Sci.* **165** 74–81

[102] Babaei H, Guo R, Hashemi A and Lee S 2019 Machine-learning-based interatomic potential for phonon transport in perfect crystalline Si and crystalline Si with vacancies *Phys. Rev. Mater.* **3** 074603

[103] Li R, Lee E and Luo T 2020 A unified deep neural network potential capable of predicting thermal conductivity of silicon in different phases *Mater. Today Phys.* **12** 100181

[104] Korotaev P, Novoselov I, Yanilkin A and Shapeev A 2019 Accessing thermal conductivity of complex compounds by machine learning interatomic potentials *Phys. Rev. B* **100** 144308

[105] Sosso G C, Donadio D, Caravati S, Behler J and Bernasconi M 2012 Thermal transport in phase-change materials from atomistic simulations *Phys. Rev. B* **86** 104301

[106] Campi D, Donadio D, Sosso G C, Behler J and Bernasconi M 2015 Electron-phonon interaction and thermal boundary resistance at the crystal-amorphous interface of the phase change compound GeTe *J. Appl. Phys.* **117** 015304

[107] J Evans D and P Morriss G 2007 *Statistical Mechanics of Nonequilbrium Liquids* (ANU Press)

[108] Müller-Plathe F 1997 A simple nonequilibrium molecular dynamics method for calculating the thermal conductivity *J. Chem. Phys.* **106** 6082–5

[109] Lampin E, Palla P L, Francioso P A and Cleri F 2013 Thermal conductivity from approach-to-equilibrium molecular dynamics *J. Appl. Phys.*

[110] Hahn K R, Melis C and Colombo L 2016 Thermal transport in nanocrystalline graphene investigated by approach-to-equilibrium molecular dynamics simulations *Carbon N. Y.* **96**

429–38

[111] Frenkel D and Smit B 2002 Understanding molecular simulation: From algorithms to applications **1** 1–638

[112] Fan Z, Pereira L F C, Wang H Q, Zheng J C, Donadio D and Harju A 2015 Force and heat current formulas for many-body potentials in molecular dynamics simulations with applications to thermal conductivity calculations *Phys. Rev. B - Condens. Matter Mater. Phys.*

[113] Boone P, Babaei H and Wilmer C E 2019 Heat flux for many-body interactions: Corrections to LAMMPS *J. Chem. Theory Comput.* **15** 5579–87

[114] Surblys D, Matsubara H, Kikugawa G and Ohara T %J P R E 2019 Application of atomic stress to compute heat flux via molecular dynamics for systems with many-body interactions **99** 51301

[115] Fan Z, Pereira L F C, Wang H-Q, Zheng J-C, Donadio D and Harju A 2015 Force and heat current formulas for many-body potentials in molecular dynamics simulations with applications to thermal conductivity calculations *Phys. Rev. B* **92** 94301

[116] Khadem M H and Wemhoff A P 2013 Comparison of Green–Kubo and NEMD heat flux formulations for thermal conductivity prediction using the Tersoff potential *Comput. Mater. Sci.* **69** 428–34

[117] Gordiz K, Singh D J and Henry A %J J of A P 2015 Ensemble averaging vs. time averaging in molecular dynamics simulations of thermal conductivity **117** 45104

[118] Travis K P, Daivis P J and Evans D J 1995 Computer simulation algorithms for molecules undergoing planar Couette flow: A nonequilibrium molecular dynamics study *J. Chem. Phys.*

[119] Hoover W G and Hoover C G 2005 Nonequilibrium molecular dynamics *Condens. Matter Phys.*

[120] Xu X, Pereira L F C, Wang Y, Wu J, Zhang K, Zhao X, Bae S, Bui C T, Xie R and Thong J T L 2014 Length-dependent thermal conductivity in suspended single-layer graphene *Nat. Commun.* **5** 3689

[121] Kuang S and Gezelter J D 2012 Velocity shearing and scaling RNEMD: a minimally perturbing method for simulating temperature and momentum gradients *Mol. Phys.* **110** 691–701

[122] Jund P and Jullien R 1999 Molecular-dynamics calculation of the thermal conductivity of vitreous silica *Phys. Rev. B* **59** 13707

[123] Kwon Y K and Kim P 2006 Unusually high thermal conductivity in carbon nanotubes *High Thermal Conductivity Materials* vol 84 pp 227–65

[124] Chiritescu C, Cahill D G, Nguyen N, Johnson D, Bodapati A, Keblinski P and Zschack P %J S 2007 Ultralow thermal conductivity in disordered, layered WSe2 crystals **315** 351–3

[125] Dong H, Fan Z, Shi L, Harju A and Ala-Nissila T %J P R B 2018 Equivalence of the equilibrium and the nonequilibrium molecular dynamics methods for thermal conductivity calculations: From bulk to nanowire silicon **97** 94305

[126] Parker W J, Jenkins R J, Butler C P and Abbott G L 1961 Flash method of determining thermal diffusivity, heat capacity, and thermal conductivity *J. Appl. Phys.* **32** 1679–84

[127] Melis C, Dettori R, Vandermeulen S and Colombo L %J T E P J B 2014 Calculating thermal conductivity in a transient conduction regime: theory and implementation **87** 96

[128] Feng T, Yao W, Wang Z, Shi J, Li C, Cao B and Ruan X %J P R B 2017 Spectral analysis of nonequilibrium molecular dynamics: Spectral phonon temperature and local

nonequilibrium in thin films and across interfaces **95** 195202

[129] Henry A S and Chen G 2008 Spectral phonon transport properties of silicon based on molecular dynamics simulations and lattice dynamics *J. Comput. Theor. Nanosci.* **5** 141–52

[130] Chalise D, Feng T and Ruan X 2017 Spectral Phonon Relaxation Time Calculation Tool Based on Molecular Dynamics

[131] Ladd A J C, Moran B and Hoover W G 1986 Lattice thermal conductivity: A comparison of molecular dynamics and anharmonic lattice dynamics *Phys. Rev. B* **34** 5058

[132] McGaughey A J H and Kaviany M 2004 Thermal conductivity decomposition and analysis using molecular dynamics simulations. Part I. Lennard-Jones argon *Int. J. Heat Mass Transf.* **47** 1783–98

[133] McGaughey A J H and Kaviany M 2004 Quantitative validation of the Boltzmann transport equation phonon thermal conductivity model under the single-mode relaxation time approximation *Phys. Rev. B* **69** 94303

[134] Larkin J M and McGaughey A J H 2013 Predicting alloy vibrational mode properties using lattice dynamics calculations, molecular dynamics simulations, and the virtual crystal approximation *J. Appl. Phys.* **114** 23507

[135] Hori T, Shiga T and Shiomi J 2013 Phonon transport analysis of silicon germanium alloys using molecular dynamics simulations *J. Appl. Phys.* **113** 203514

[136] Goicochea J V, Madrid M and Amon C 2010 Thermal properties for bulk silicon based on the determination of relaxation times using molecular dynamics *J. Heat Transfer* **132** 12401

[137] Thomas J A, Turney J E, Iutzi R M, Amon C H and McGaughey A J H 2010 Predicting

phonon dispersion relations and lifetimes from the spectral energy density *Phys. Rev. B* **81** 81411

[138] Larkin J M, Turney J E, Massicotte A D, Amon C H and McGaughey A J H 2014 Comparison and evaluation of spectral energy methods for predicting phonon properties *J. Comput. Theor. Nanosci.* **11** 249–56

[139] Qiu B, Bao H, Zhang G, Wu Y and Ruan X 2012 Molecular dynamics simulations of lattice thermal conductivity and spectral phonon mean free path of {PbTe}: {Bulk} and nanostructures *Comput. Mater. Sci.* **53** 278–85

[140] Qiu B and Ruan X 2012 Reduction of spectral phonon relaxation times from suspended to supported graphene *Appl. Phys. Lett.* **100** 193101

[141] McGaughey A J H and Kaviany M 2004 Thermal conductivity decomposition and analysis using molecular dynamics simulations: Part II. Complex silica structures *Int. J. Heat Mass Transf.* **47** 1799–816

[142] Larkin J M and McGaughey A J H 2014 Thermal conductivity accumulation in amorphous silica and amorphous silicon *Phys. Rev. B* **89** 144303

[143] Baldi G, Giordano V M, Monaco G, Sette F, Fabiani E, Fontana A and Ruocco G 2008 Thermal conductivity and terahertz vibrational dynamics of vitreous silica *Phys. Rev. B* **77** 214309

[144] Baldi G, Giordano V M, Monaco G and Ruta B 2010 Sound attenuation at terahertz frequencies and the boson peak of vitreous silica *Phys. Rev. Lett.* **104** 195501

[145] Baldi G, Zanatta M, Gilioli E, Milman V, Refson K, Wehinger B, Winkler B, Fontana A and Monaco G 2013 Emergence of crystal-like atomic dynamics in glasses at the nanometer scale *Phys. Rev. Lett.* **110** 185503

[146]  Ruzicka B, Scopigno T, Caponi S, Fontana A, Pilla O, Giura P, Monaco G, Pontecorvo E, Ruocco G and Sette F 2004 Evidence of anomalous dispersion of the generalized sound velocity in glasses *Phys. Rev. B* **69** 100201

[147]  Horbach J, Kob W and Binder K 2001 High frequency sound and the boson peak in amorphous silica *Eur. Phys. J. B-Condensed Matter Complex Syst.* **19** 531–43

[148]  Feldman J L 2002 Calculations of the generalized dynamic structure factor for amorphous silicon *J. Non. Cryst. Solids* **307** 128–34

[149]  Taraskin S N and Elliott S R 1999 Determination of the Ioffe-Regel limit for vibrational excitations in disordered materials *Philos. Mag. B* **79** 1747–54

[150]  Beltukov Y M, Kozub V I and Parshin D A 2013 Ioffe-Regel criterion and diffusion of vibrations in random lattices *Phys. Rev. B* **87** 134203

[151]  Zhou Y, Zhang X and Hu M %J P R B 2015 Quantitatively analyzing phonon spectral contribution of thermal conductivity based on nonequilibrium molecular dynamics simulations. I. From space Fourier transform **92** 195204

[152]  Zhou Y and Hu M 2016 Erratum: Quantitatively analyzing phonon spectral contribution of thermal conductivity based on nonequilibrium molecular dynamics simulations. II. From time Fourier transform [Phys. Rev. B 92, 195205 (2015)] *Phys. Rev. B* **93** 39901

[153]  Chalopin Y and Volz S 2013 A microscopic formulation of the phonon transmission at the nanoscale *Appl. Phys. Lett.* **103** 51602

[154]  Lv W and Henry A 2016 Direct calculation of modal contributions to thermal conductivity via Green–Kubo modal analysis *New J. Phys.* **18** 13028

[155]  Sun T and Allen P B 2010 Lattice thermal conductivity: Computations and theory of the high-temperature breakdown of the phonon-gas model *Phys. Rev. B* **82** 224305

[156] Lv W and Henry A 2016 Non-negligible Contributions to Thermal Conductivity from Localized Modes in Amorphous Silicon Dioxide *Sci. Rep.* **6** 1–8

[157] Li D, Wu Y, Kim P, Shi L, Yang P and Majumdar A 2003 Thermal conductivity of individual silicon nanowires *Appl. Phys. Lett.* **83** 2934–6

[158] Che J, Cagin T and Goddard III W A 2000 Thermal conductivity of carbon nanotubes *Nanotechnology* **11** 65

[159] Volz S G and Chen G 2000 Molecular-dynamics simulation of thermal conductivity of silicon crystals *Phys. Rev. B* **61** 2651

[160] Schelling P K, Phillpot S R and Keblinski P 2002 Comparison of atomic-level simulation methods for computing thermal conductivity *Phys. Rev. B* **65** 144306

[161] Sun L and Murthy J Y 2006 Domain size effects in molecular dynamics simulation of phonon transport in silicon *Appl. Phys. Lett.* **89** 171919

[162] Sellan D P, Landry E S, Turney J E, McGaughey A J H and Amon C H 2010 Size effects in molecular dynamics thermal conductivity predictions *Phys. Rev. B* **81** 214305

[163] Wei Z, Yang F, Bi K, Yang J and Chen Y 2017 Thermal transport properties of all-sp2 three-dimensional graphene: Anisotropy, size and pressure effects *Carbon N. Y.* **113** 212–8

[164] Evans W J, Hu L and Keblinski P 2010 Thermal conductivity of graphene ribbons from equilibrium molecular dynamics: Effect of ribbon width, edge roughness, and hydrogen termination *Appl. Phys. Lett.* **96** 203112

[165] Chantrenne P and Barrat J-L 2004 Finite size effects in determination of thermal conductivities: comparing molecular dynamics results with simple models *J. Heat Transfer* **126** 577–85

[166] Muraleedharan M G, Sundaram D S, Henry A and Yang V 2017 Thermal conductivity calculation of nano-suspensions using Green-Kubo relations with reduced artificial correlations *J. Phys. Condens. Matter* **29**

[167] Oligschleger C and Schön J C 1999 Simulation of thermal conductivity and heat transport in solids *Phys. Rev. B* **59** 4125

[168] Yoon Y-G, Car R, Srolovitz D J and Scandolo S 2004 Thermal conductivity of crystalline quartz from classical simulations *Phys. Rev. B* **70** 12302

[169] Feng X-L 2003 Molecular dynamics simulation of thermal conductivity of nanoscale thin silicon films *Microscale Thermophys. Eng.* **7** 153–61

[170] Sellan D P, Turney J E, McGaughey A J H and Amon C H 2010 Cross-plane phonon transport in thin films *J. Appl. Phys.* **108** 113524

[171] Frachioni A and White Jr B E 2012 Simulated thermal conductivity of silicon-based random multilayer thin films *J. Appl. Phys.* **112** 14320

[172] McGaughey A J H, Landry E S, Sellan D P and Amon C H 2011 Size-dependent model for thin film and nanowire thermal conductivity *Appl. Phys. Lett.* **99** 131904

[173] Zhang Y-Y, Pei Q-X, Jiang J-W, Wei N and Zhang Y-W 2016 Thermal conductivities of single-and multi-layer phosphorene: a molecular dynamics study *Nanoscale* **8** 483–91

[174] Bi K, Chen Y, Yang J, Wang Y and Chen M 2006 Molecular dynamics simulation of thermal conductivity of single-wall carbon nanotubes *Phys. Lett. A* **350** 150–3

[175] Wang Z, Zu X, Gao F, Weber W J and Crocombette J-P 2007 Atomistic simulation of the size and orientation dependences of thermal conductivity in GaN nanowires *Appl. Phys. Lett.* **90** 161923

[176] Guo Z, Zhang D and Gong X-G 2009 Thermal conductivity of graphene nanoribbons

*Appl. Phys. Lett.* **95** 163103

[177] Hu J, Schiffli S, Vallabhaneni A, Ruan X and Chen Y P 2010 Tuning the thermal conductivity of graphene nanoribbons by edge passivation and isotope engineering: A molecular dynamics study *Appl. Phys. Lett.* **97** 133107

[178] Zhang D H, Yao K L and Gao G Y 2011 The peculiar transport properties in pn junctions of doped graphene nanoribbons *J. Appl. Phys.* **110** 13718

[179] Lu S and McGaughey A J H 2015 Thermal conductance of superlattice junctions *AIP Adv.* **5** 53205

[180] Samaraweera N, Chan K L and Mithraratne K 2018 Reduced thermal conductivity of nanotwinned random layer structures: a promising nanostructuring towards efficient {Si} and {Si}/{Ge} thermoelectric materials *J. Phys. D. Appl. Phys.* **51** 204006

[181] Landry E S, Hussein M I and McGaughey A J H 2008 Complex superlattice unit cell designs for reduced thermal conductivity *Phys. Rev. B* **77** 184302

[182] Termentzidis K, Chantrenne P and Keblinski P 2009 Nonequilibrium molecular dynamics simulation of the in-plane thermal conductivity of superlattices with rough interfaces *Phys. Rev. B* **79** 214307

[183] Cao H-Y, Guo Z-X, Xiang H and Gong X-G 2012 Layer and size dependence of thermal conductivity in multilayer graphene nanoribbons *Phys. Lett. A* **376** 525–8

[184] Mortazavi B, Pötschke M and Cuniberti G 2014 Multiscale modeling of thermal conductivity of polycrystalline graphene sheets *Nanoscale* **6** 3344–52

[185] Wei Z, Ni Z, Bi K, Chen M and Chen Y 2011 In-plane lattice thermal conductivities of multilayer graphene films *Carbon N. Y.* **49** 2653–8

[186] Jiang J-W, Park H S and Rabczuk T 2013 Molecular dynamics simulations of single-layer

molybdenum disulphide (MoS2): Stillinger-Weber parametrization, mechanical properties, and thermal conductivity *J. Appl. Phys.* **114** 64307

[187] Liu X, Zhang G, Pei Q-X and Zhang Y-W 2013 Phonon thermal conductivity of monolayer MoS2 sheet and nanoribbons *Appl. Phys. Lett.* **103** 133113

[188] Li S, Ren H, Zhang Y, Xie X, Cai K, Li C and Wei N 2019 Thermal Conductivity of Two Types of 2D Carbon Allotropes: a Molecular Dynamics Study *Nanoscale Res. Lett.* **14** 7

[189] Henry A, Chen G, Plimpton S J and Thompson A 2010 1D-to-3D transition of phonon heat conduction in polyethylene using molecular dynamics simulations *Phys. Rev. B* **82** 144308

[190] Wang J-S and Li B 2004 Mode-coupling theory and molecular dynamics simulation for heat conduction in a chain with transverse motions *Phys. Rev. E* **70** 21204

[191] Allen P B, Feldman J L, Fabian J and Wooten F 1999 Diffusons, locons and propagons: Character of atomie yibrations in amorphous Si *Philos. Mag. B* **79** 1715–31

[192] Allen P B and Feldman J L 1993 Thermal conductivity of disordered harmonic solids *Phys. Rev. B* **48** 12581

[193] Kommandur S and Yee S K 2017 An empirical model to predict temperature-dependent thermal conductivity of amorphous polymers *J. Polym. Sci. Part B Polym. Phys.* **55** 1160–70

[194] Lv W and Henry A 2016 Phonon transport in amorphous carbon using Green–Kubo modal analysis *Appl. Phys. Lett.* **108** 181905

[195] He Y, Donadio D and Galli G 2011 Heat transport in amorphous silicon: Interplay between morphology and disorder *Appl. Phys. Lett.* **98** 144101

[196] Terao T, Lussetti E and Müller-Plathe F 2007 Nonequilibrium molecular dynamics

methods for computing the thermal conductivity: application to amorphous polymers *Phys. Rev. E* **75** 57701

[197]  Lussetti E, Terao T and Müller-Plathe F 2007 Nonequilibrium molecular dynamics calculation of the thermal conductivity of amorphous polyamide-6, 6 *J. Phys. Chem. B* **111** 11516–23

[198]  Monk J D, Haskins J B, Bauschlicher Jr C W and Lawson J W 2015 Molecular dynamics simulations of phenolic resin: construction of atomistic models *Polymer (Guildf).* **62** 39–49

[199]  Monk J D, Bucholz E W, Boghozian T, Deshpande S, Schieber J, Bauschlicher Jr C W and Lawson J W 2015 Computational and experimental study of phenolic resins: thermal–mechanical properties and the role of hydrogen bonding *Macromolecules* **48** 7670–80

[200]  Luo T, Esfarjani K, Shiomi J, Henry A and Chen G 2011 Molecular dynamics simulation of thermal energy transport in polydimethylsiloxane *J. Appl. Phys.* **109** 74321

[201]  Shenogin S, Bodapati A, Keblinski P and McGaughey A J H 2009 Predicting the thermal conductivity of inorganic and polymeric glasses: The role of anharmonicity *J. Appl. Phys.* **105** 34906

[202]  Cevallos J G, Bergles A E, Bar-Cohen A, Rodgers P and Gupta S K 2012 Polymer heat exchangers—history, opportunities, and challenges *Heat Transf. Eng.* **33** 1075–93

[203]  Kikugawa G, Desai T G, Keblinski P and Ohara T 2013 Effect of crosslink formation on heat conduction in amorphous polymers *J. Appl. Phys.* **114** 34302

[204]  Xiong X, Yang M, Liu C, Li X and Tang D 2017 Thermal conductivity of cross-linked polyethylene from molecular dynamics simulation *J. Appl. Phys.* **122** 35104

[205]  Hsieh W-P, Losego M D, Braun P V, Shenogin S, Keblinski P and Cahill D G 2011

Testing the minimum thermal conductivity model for amorphous polymers using high pressure *Phys. Rev. B* **83** 174205

[206] Giri A, Braun J L and Hopkins P E 2016 Effect of crystalline/amorphous interfaces on thermal transport across confined thin films and superlattices *J. Appl. Phys.* **119** 235305

[207] Gordiz K and Henry A 2017 Phonon transport at interfaces between different phases of silicon and germanium *J. Appl. Phys.* **121** 25102

[208] Gordiz K, Muraleedharan M G and Henry A 2019 Interface conductance modal analysis of a crystalline Si-amorphous SiO2 interface *J. Appl. Phys.* **125** 135102

[209] Callaway J 1959 Model for Lattice Thermal Conductivity at Low Temperatures *Phys. Rev.* **113** 1046–51

[210] Gurunathan R, Hanus R, Dylla M, Katre A and Snyder G J 2020 Analytical Models of Phonon--Point-Defect Scattering *Phys. Rev. Appl.* **13** 034011

[211] Klemens P G 1960 Thermal Resistance due to Point Defects at High Temperatures *Phys. Rev.* **119** 507–9

[212] Klemens P G 1955 The Scattering of Low-Frequency Lattice Waves by Static Imperfections *Proc. Phys. Soc. Sect. A* **68** 1113–1128

[213] Li W, Carrete J, A. Katcho N and Mingo N 2014 ShengBTE: A solver of the Boltzmann transport equation for phonons *Comput. Phys. Commun.* **185** 1747–58

[214] Katcho N A, Carrete J, Li W and Mingo N 2014 Effect of nitrogen and vacancy defects on the thermal conductivity of diamond: An ab initio Green's function approach *Phys. Rev. B* **90** 094117

[215] Polanco C A and Lindsay L 2018 Thermal conductivity of InN with point defects from first principles *Phys. Rev. B* **98** 014306

[216]  Polanco C A 2018 *Ab initio* phonon point defect scattering and thermal transport in graphene *Phys. Rev. B* **97**

[217]  Wang T, Carrete J, van Roekeghem A, Mingo N and Madsen G K H 2017 Ab initio phonon scattering by dislocations *Phys. Rev. B* **95** 245304

[218]  Wang T, Carrete J, Mingo N and Madsen G K H 2019 Phonon Scattering by Dislocations in GaN *ACS Appl. Mater. Interfaces* **11** 8175–81

[219]  Wang T, Madsen G K H and Hartmaier A 2014 Atomistic study of the influence of lattice defects on the thermal conductivity of silicon *Model. Simul. Mater. Sci. Eng.* **22** 35011

[220]  Feng T, Qiu B and Ruan X 2015 Coupling between phonon-phonon and phonon-impurity scattering: A critical revisit of the spectral Matthiessen's rule *Phys. Rev. B* **92** 235206

[221]  Dongre B, Wang T and Madsen G K H 2017 Comparison of the {Green}–{Kubo} and homogeneous non-equilibrium molecular dynamics methods for calculating thermal conductivity *Model. Simul. Mater. Sci. Eng.* **25** 54001

[222]  Crocombette J-P and Proville L 2011 Thermal conductivity degradation induced by point defects in irradiated silicon carbide *Appl. Phys. Lett.* **98** 191905

[223]  Samolyuk G D, Golubov S I, Osetsky Y N and Stoller R E 2011 Molecular dynamics study of influence of vacancy types defects on thermal conductivity of β-{SiC} *J. Nucl. Mater.* **418** 174–81

[224]  Mao Y, Li Y, Xiong Y and Xiao W 2018 Point defect effects on the thermal conductivity of β-{SiC} by molecular dynamics simulations *Comput. Mater. Sci.* **152** 300–7

[225]  Liu X-Y, Cooper M W D, McClellan K J, Lashley J C, Byler D D, Bell B D C, Grimes R W, Stanek C R and Andersson D A 2016 Molecular Dynamics Simulation of Thermal Transport in ${\mathrm{UO}}_{2}$ Containing Uranium, Oxygen, and Fission-product

Defects *Phys. Rev. Appl.* **6** 044015

[226] Troncoso J F, Aguado-Puente P and Kohanoff J 2019 Effect of intrinsic defects on the thermal conductivity of {PbTe} from classical molecular dynamics simulations *J. Phys. Condens. Matter* **32** 45701

[227] Bedoya-Martínez O N, Hashibon A and Elsässer C 2016 Influence of point defects on the phonon thermal conductivity and phonon density of states of {Bi2Te3} *Phys. status solidi* **213** 684–93

[228] Ji H S, Kim H, Lee C, Rhyee J-S, Kim M H, Kaviany M and Shim J H 2013 Vacancy-suppressed lattice conductivity of high-$\mathit{ZT}$ In${}_{4}$Se${}_{3\ensuremath{-}x}$ *Phys. Rev. B* **87** 125111

[229] Liu Z, Yang X, Chen G and Zhai P 2016 Molecular dynamics study of the influence of Sb-vacancy defects on the lattice thermal conductivity of crystalline CoSb3 *Comput. Mater. Sci.* **124** 403–10

[230] Chen H and McGaughey A J H 2011 Thermal Conductivity of Carbon Nanotubes With Defects (American Society of Mechanical Engineers Digital Collection)

[231] Chien S-K, Yang Y-T and Chen C-K 2011 The effects of vacancy defects and nitrogen doping on the thermal conductivity of armchair (10, 10) single-wall carbon nanotubes *Solid State Commun.* **151** 1004–8

[232] Zhang H, Lee G and Cho K 2011 Thermal transport in graphene and effects of vacancy defects *Phys. Rev. B* **84** 115460

[233] Hu S, Chen J, Yang N and Li B 2017 Thermal transport in graphene with defect and doping: Phonon modes analysis *Carbon N. Y.* **116** 139–44

[234] Yang H, Tang Y, Gong J, Liu Y, Wang X, Zhao Y, Yang P and Wang S 2013 Influence

of doped nitrogen and vacancy defects on the thermal conductivity of graphene nanoribbons *J. Mol. Model.* **19** 4781–8

[235] Noshin M, Khan A I, Navid I A, Uddin H M A and Subrina S 2017 Impact of vacancies on the thermal conductivity of graphene nanoribbons: {A} molecular dynamics simulation study *AIP Adv.* **7** 15112

[236] Cui L, Zhang Y, Du X and Wei G 2018 Computational study on thermal conductivity of defective carbon nanomaterials: carbon nanotubes versus graphene nanoribbons *J. Mater. Sci.* **53** 4242–51

[237] Ding Z, Pei Q-X, Jiang J-W and Zhang Y-W 2015 Manipulating the Thermal Conductivity of Monolayer MoS2 via Lattice Defect and Strain Engineering *J. Phys. Chem. C* **119** 16358–65

[238] Wang Y, Zhang K and Xie G 2016 Remarkable suppression of thermal conductivity by point defects in MoS2 nanoribbons *Appl. Surf. Sci.* **360** 107–12

[239] Li H and Zhang R 2012 Vacancy-defect–induced diminution of thermal conductivity in silicene *EPL (Europhysics Lett.* **99** 36001

[240] Meem A U H, Chowdhury O and Morshed A M 2018 Effects of vacancy defects location on thermal conductivity of silicon nanowire: a molecular dynamics study *Micro & Nano Lett.* **13** 1146–50

[241] Sun Y, Zhou Y, Han J, Hu M, Xu B and Liu W 2020 Molecular dynamics simulations of the effect of dislocations on the thermal conductivity of iron *J. Appl. Phys.* **127** 45106

[242] Deng B, Chernatynskiy A, Shukla P, Sinnott S B and Phillpot S R 2013 Effects of edge dislocations on thermal transport in {UO2} *J. Nucl. Mater.* **434** 203–9

[243] Sun Y, Zhou Y, Han J, Liu W, Nan C, Lin Y, Hu M and Xu B 2019 Strong phonon

localization in PbTe with dislocations and large deviation to Matthiessen's rule *arXiv:1901.10250 [physics]*

[244] Spiteri D, Pomeroy J W and Kuball M 2013 Influence of microstructural defects on the thermal conductivity of GaN: A molecular dynamics study *Phys. status solidi* **250** 1541–5

[245] Al-Ghalith J, Ni Y and Dumitrică T 2016 Nanowires with dislocations for ultralow lattice thermal conductivity *Phys. Chem. Chem. Phys.* **18** 9888–92

[246] Ni Y, Xiong S, Volz S and Dumitrică T 2014 Thermal Transport Along the Dislocation Line in Silicon Carbide *Phys. Rev. Lett.* **113** 124301

[247] Termentzidis K, Isaiev M, Salnikova A, Belabbas I, Lacroix D and Kioseoglou J 2018 Impact of screw and edge dislocations on the thermal conductivity of individual nanowires and bulk {GaN}: a molecular dynamics study *Phys. Chem. Chem. Phys.* **20** 5159–72

[248] Chen Y, Lukes J R, Li D, Yang J and Wu Y 2004 Thermal expansion and impurity effects on lattice thermal conductivity of solid argon *J. Chem. Phys.* **120** 3841–6

[249] Yao M, Watanabe T, Schelling P K, Keblinski P, Cahill D G and Phillpot S R 2008 Phonon-defect scattering in doped silicon by molecular dynamics simulation *J. Appl. Phys.* **104** 24905

[250] Li M, Zheng B, Duan K, Zhang Y, Huang Z and Zhou H 2018 Effect of Defects on the Thermal Transport across the Graphene/Hexagonal Boron Nitride Interface *J. Phys. Chem. C* **122** 14945–53

[251] Giaremis S, Kioseoglou J, Desmarchelier P, Tanguy A, Isaiev M, Belabbas I, Komninou P and Termentzidis K 2020 Decorated Dislocations against Phonon Propagation for Thermal Management *ACS Appl. Energy Mater.*

[252] Hu S, Zhang H, Xiong S, Zhang H, Wang H, Chen Y, Volz S and Ni Y 2019 Screw
122

dislocation induced phonon transport suppression in SiGe superlattices *Phys. Rev. B* **100** 75432

[253]   Jones R E and Ward D K 2018 Influence of defects on the thermal conductivity of compressed {LiF} *Phys. Rev. B* **97** 54103

[254]   Cui L, Shi S, Li Z, Wei G and Du X 2018 Reduction of thermal conductivity in silicene nanomesh: insights from coherent and incoherent phonon transport *Phys. Chem. Chem. Phys.* **20** 27169–75

[255]   de Sousa Oliveira L and Neophytou N 2019 Large-scale molecular dynamics investigation of geometrical features in nanoporous {Si} *Phys. Rev. B* **100** 35409

[256]   Chen W, Cooper M W D, Xiao Z, Andersson D A and Bai X-M 2019 Effect of Xe bubble size and pressure on the thermal conductivity of UO2—A molecular dynamics study *J. Mater. Res.* **34** 2295–305

[257]   Lee C-W, Chernatynskiy A, Shukla P, Stoller R E, Sinnott S B and Phillpot S R 2015 Effect of pores and He bubbles on the thermal transport properties of UO2 by molecular dynamics simulation *J. Nucl. Mater.* **456** 253–9

[258]   Tonks M R, Liu X-Y, Andersson D, Perez D, Chernatynskiy A, Pastore G, Stanek C R and Williamson R 2016 Development of a multiscale thermal conductivity model for fission gas in {UO2} *J. Nucl. Mater.* **469** 89–98

[259]   Yousefi F, Khoeini F and Rajabpour A 2020 Thermal conductivity and thermal rectification of nanoporous graphene: {A} molecular dynamics simulation *Int. J. Heat Mass Transf.* **146** 118884

[260]   Zhou X W and Jones R E 2012 Effects of nano-void density, size and spatial population on thermal conductivity: a case study of {GaN} crystal *J. Phys. Condens. Matter* **24**

325804

[261] Yang N, Ni X, Jiang J-W and Li B 2012 How does folding modulate thermal conductivity of graphene? *Appl. Phys. Lett.* **100** 93107

[262] Anon Wrinkling and thermal conductivity of one graphene sheet under shear: Molecular Simulation: Vol 41, No 4

[263] Wang C, Liu Y, Li L and Tan H 2014 Anisotropic thermal conductivity of graphene wrinkles *Nanoscale* **6** 5703–7

[264] Cui L, Du X, Wei G and Feng Y 2016 Thermal Conductivity of Graphene Wrinkles: A Molecular Dynamics Simulation *J. Phys. Chem. C* **120** 23807–12

[265] Zhang C, Hao X-L, Wang C-X, Wei N and Rabczuk T 2017 Thermal conductivity of graphene nanoribbons under shear deformation: {A} molecular dynamics simulation *Sci. Rep.* **7** 1–8

[266] Jiang J-W, Yang N, Wang B-S and Rabczuk T 2013 Modulation of Thermal Conductivity in Kinked Silicon Nanowires: Phonon Interchanging and Pinching Effects *Nano Lett.* **13** 1670–4

[267] Zhao Y, Yang L, Liu C, Zhang Q, Chen Y, Yang J and Li D 2019 Kink effects on thermal transport in silicon nanowires *Int. J. Heat Mass Transf.* **137** 573–8

[268] Termentzidis K, Barreteau T, Ni Y, Merabia S, Zianni X, Chalopin Y, Chantrenne P and Volz S 2013 Modulated SiC nanowires: Molecular dynamics study of their thermal properties *Phys. Rev. B* **87** 125410

[269] Bodapati A, Schelling P K, Phillpot S R and Keblinski P 2006 Vibrations and thermal transport in nanocrystalline silicon *Phys. Rev. B* **74** 245207

[270] Ju S and Liang X 2012 Thermal conductivity of nanocrystalline silicon by direct

molecular dynamics simulation *J. Appl. Phys.* **112** 64305

[271] Ju S, Liang X and Xu X 2011 Out-of-plane thermal conductivity of polycrystalline silicon nanofilm by molecular dynamics simulation *J. Appl. Phys.* **110** 54318

[272] da Cruz C, Katcho N A, Mingo N and Veiga R G A 2013 Thermal conductivity of nanocrystalline {SiGe} alloys using molecular dynamics simulations *J. Appl. Phys.* **114** 164310

[273] Dong H, Wen B, Zhang Y and Melnik R 2017 Thermal Conductivity of Diamond/SiC Nano-Polycrystalline Composites and Phonon Scattering at Interfaces *ACS Omega* **2** 2344–50

[274] Liu H K, Lin Y and Luo S N 2014 Grain Boundary Energy and Grain Size Dependences of Thermal Conductivity of Polycrystalline Graphene *J. Phys. Chem. C* **118** 24797–802

[275] Fan Z, Hirvonen P, Pereira L F C, Ervasti M M, Elder K R, Donadio D, Harju A and Ala-Nissila T 2017 Bimodal Grain-Size Scaling of Thermal Transport in Polycrystalline Graphene from Large-Scale Molecular Dynamics Simulations *Nano Lett.* **17** 5919–24

[276] Wang Y, Song Z and Xu Z 2014 Characterizing phonon thermal conduction in polycrystalline graphene *J. Mater. Res.* **29** 362–72

[277] Wu P H, Quek S S, Sha Z D, Dong Z L, Liu X J, Zhang G, Pei Q X and Zhang Y W 2014 Thermal transport behavior of polycrystalline graphene: {A} molecular dynamics study *J. Appl. Phys.* **116** 204303

[278] Mortazavi B, Pereira L F C, Jiang J-W and Rabczuk T 2015 Modelling heat conduction in polycrystalline hexagonal boron-nitride films *Sci. Rep.* **5** 13228

[279] Proshchenko V S, Dholabhai P P, Sterling T C and Neogi S 2019 Heat and charge transport in bulk semiconductors with interstitial defects *Phys. Rev. B* **99** 14207

[280] Estreicher S K, Gibbons T M and Bebek M B 2015 Thermal phonons and defects in semiconductors: {The} physical reason why defects reduce heat flow, and how to control it *J. Appl. Phys.* **117** 112801

[281] Bebek M B, Stanley C M, Gibbons T M and Estreicher S K 2016 Temperature dependence of phonon-defect interactions: phonon scattering vs. phonon trapping *Sci. Rep.* **6** 1–10

[282] Li S, Ding X, Ren J, Moya X, Li J, Sun J and Salje E K H 2014 Strain-controlled thermal conductivity in ferroic twinned films *Sci. Rep.* **4** 1–7

[283] Kuryliuk V, Nepochatyi O, Chantrenne P, Lacroix D and Isaiev M 2019 Thermal conductivity of strained silicon: {Molecular} dynamics insight and kinetic theory approach *J. Appl. Phys.* **126** 55109

[284] Nakajima K 2010 Production method of thermoelectric semiconductor alloy, thermoelectric conversion module and thermoelectric power generating device

[285] Lin Y-M, Rabin O, Cronin S B, Ying J Y and Dresselhaus M S 2002 Semimetal–semiconductor transition in Bi$_{1-x}$Sb$_x$ alloy nanowires and their thermoelectric properties *Appl. Phys. Lett.* **81** 2403–5

[286] Baker C H and Norris P M 2015 Effect of long-and short-range order on SiGe alloy thermal conductivity: molecular dynamics simulation *Phys. Rev. B* **91** 180302

[287] Tamura S 1983 Isotope scattering of dispersive phonons in Ge *Phys. Rev. B* **27** 858

[288] Shiga T, Hori T and Shiomi J 2014 Influence of mass contrast in alloy phonon scattering *Jpn. J. Appl. Phys.* **53** 21802

[289] Mei S and Knezevic I 2018 Thermal conductivity of ternary III-V semiconductor alloys: The role of mass difference and long-range order *J. Appl. Phys.* **123** 125103

[290]   Zhang H, Han H, Xiong S, Wang H, Volz S and Ni Y 2017 Impeded thermal transport in composition graded SiGe nanowires *Appl. Phys. Lett.* **111** 121907

[291]   Giri A, Braun J L, Tomko J A and Hopkins P E 2017 Reducing the thermal conductivity of chemically ordered binary alloys below the alloy limit via the alteration of phonon dispersion relations *Appl. Phys. Lett.* **110** 233112

[292]   Xiong S, Sääskilahti K, Kosevich Y A, Han H, Donadio D and Volz S 2016 Blocking phonon transport by structural resonances in alloy-based nanophononic metamaterials leads to ultralow thermal conductivity *Phys. Rev. Lett.* **117** 25503

[293]   Melis C and Colombo L 2014 Lattice thermal conductivity of Si 1− x Ge x nanocomposites *Phys. Rev. Lett.* **112** 65901

[294]   Norouzzadeh P, Nozariasbmarz A, Krasinski J S and Vashaee D 2015 Thermal conductivity of nanostructured {SixGe1}−x in amorphous limit by molecular dynamics simulation *J. Appl. Phys.* **117** 214303

[295]   Gasanly N M 2005 Compositional dependence of the Raman lineshapes in GaSxSe1− x layered mixed crystals *J. Raman Spectrosc.* **36** 879–83

[296]   Lyver IV J W and Blaisten-Barojas E 2006 Computational study of heat transport in compositionally disordered binary crystals *Acta Mater.* **54** 4633–9

[297]   Skye A and Schelling P K 2008 Thermal resistivity of Si–Ge alloys by molecular-dynamics simulation *J. Appl. Phys.* **103** 113524

[298]   Somayajulu P S, Ghosh P S, Arya A, Devi K V V, Sathe D B, Banerjee J, Khan K B, Dey G K, Dutta B K %J J of A and Compounds 2016 Thermal expansion and thermal conductivity of (Th, Pu) O2 mixed oxides: A molecular dynamics and experimental study **664** 291–303

[299]   Dettori R, Melis C, Rurali R and Colombo L %J J of A P 2016 Thermal rectification in silicon by a graded distribution of defects **119** 215102

[300]   Xu R-F, Han K and Li H-P 2018 Effect of isotope doping on phonon thermal conductivity of silicene nanoribbons: A molecular dynamics study *Chinese Phys. B* **27** 26801

[301]   Zhang X, Zhang J and Yang M %J S S C 2020 Molecular dynamics study on the thermal conductivity of bilayer graphene with nitrogen doping **309** 113845

[302]   Yue C, Feng J, Feng J and Jiang Y %J I J of T S 2018 Study on the solid thermal insulation mechanisms of nitrogen-doped graphene aerogels by molecular dynamics simulations and experiments **133** 162–9

[303]   Navid I A and advances Subrina S %J R S C 2018 Thermal transport characterization of carbon and silicon doped stanene nanoribbon: an equilibrium molecular dynamics study **8** 31690–9

[304]   Wei J, Liu H J, Cheng L, Zhang J, Jiang P H, Liang J H, Fan D D and Shi J %J P L A 2017 Molecular dynamics simulations of the lattice thermal conductivity of thermoelectric material CuInTe2 **381** 1611–4

[305]   Takaki H, Kobayashi K, Shimono M, Kobayashi N, Hirose K, Tsujii N and Mori T %J M T P 2017 Thermoelectric properties of a magnetic semiconductor CuFeS2 **3** 85–92

[306]   Wu X, Yang N and Luo T 2015 Unusual isotope effect on thermal transport of single layer molybdenum disulphide *Appl. Phys. Lett.* **107** 191907

[307]   Kang J S, Li M, Wu H, Nguyen H and Hu Y 2018 Experimental observation of high thermal conductivity in boron arsenide *Science (80-. ).* **361** 575–8

[308]   Tian F, Song B, Chen X, Ravichandran N K, Lv Y, Chen K, Sullivan S, Kim J, Zhou Y

and Liu T-H 2018 Unusual high thermal conductivity in boron arsenide bulk crystals *Science (80-. ).* **361** 582–5

[309] Zheng Q, Li S, Li C, Lv Y, Liu X, Huang P Y, Broido D A, Lv B and Cahill D G 2018 High Thermal Conductivity in Isotopically Enriched Cubic Boron Phosphide *Adv. Funct. Mater.* **28** 579–81

[310] Park M, Lee I-H and Kim Y-S 2014 Lattice thermal conductivity of crystalline and amorphous silicon with and without isotopic effects from the ballistic to diffusive thermal transport regime *J. Appl. Phys.* **116** 43514

[311] Zhang H, Lee G, Fonseca A F, Borders T L and Cho K 2010 Isotope effect on the thermal conductivity of graphene *J. Nanomater.* **2010**

[312] MARUYAMA S, IGARASHI Y, TANIGUCHI Y and SHIOMI J 2006 Anisotropic Heat Transfer of Single-Walled Carbon Nanotubes *J. Therm. Sci. Technol.* **1** 138–48

[313] Li X, Chen J, Yu C and Zhang G 2013 Comparison of isotope effects on thermal conductivity of graphene nanoribbons and carbon nanotubes *Appl. Phys. Lett.* **103** 13111

[314] Pei Q-X, Zhang Y-W, Sha Z-D and Shenoy V B 2013 Tuning the thermal conductivity of silicene with tensile strain and isotopic doping: A molecular dynamics study *J. Appl. Phys.* **114** 33526

[315] Lindsay L, Broido D A and Reinecke T L 2012 Thermal conductivity and large isotope effect in GaN from first principles *Phys. Rev. Lett.* **109** 95901

[316] Olson J R, Pohl R O, Vandersande J W, Zoltan A, Anthony T R and Banholzer W F 1993 Thermal conductivity of diamond between 170 and 1200 K and the isotope effect *Phys. Rev. B* **47** 14850

[317] Chang C W, Fennimore A M, Afanasiev A, Okawa D, Ikuno T, Garcia H, Li D,

Majumdar A and Zettl A 2006 Isotope effect on the thermal conductivity of boron nitride nanotubes *Phys. Rev. Lett.* **97** 85901

[318] Tamura S 1984 Isotope scattering of large-wave-vector phonons in GaAs and InSb: Deformation-dipole and overlap-shell models *Phys. Rev. B* **30** 849

[319] Wang Y, Huang H and Ruan X 2014 Decomposition of coherent and incoherent phonon conduction in superlattices and random multilayers *Phys. Rev. B* **90** 165406

[320] Lax M, Hu P and Narayanamurti V 1981 Spontaneous phonon decay selection rule: N and U processes *Phys. Rev. B* **23** 3095

[321] Tian F, Song B, Lv B, Sun J, Huyan S, Wu Q, Mao J, Ni Y, Ding Z and Huberman S 2018 Seeded growth of boron arsenide single crystals with high thermal conductivity *Appl. Phys. Lett.* **112** 31903

[322] Yee S K, Malen J A, Majumdar A and Segalman R A 2011 Thermoelectricity in fullerene–metal heterojunctions *Nano Lett.* **11** 4089–94

[323] Kim P, Shi L, Majumdar A and McEuen P L 2001 Thermal transport measurements of individual multiwalled nanotubes *Phys. Rev. Lett.* **87** 215502

[324] Catalan G, Seidel J, Ramesh R and Scott J F 2012 Domain wall nanoelectronics *Rev. Mod. Phys.* **84** 119

[325] Pop E 2010 Energy dissipation and transport in nanoscale devices *Nano Res.* **3** 147–69

[326] Tien C-L and Chen G 1994 Challenges in microscale conductive and radiative heat transfer *J. Heat Transfer* **116** 799–807

[327] Muraleedharan M G, Unnikrishnan U, Henry A and Yang V 2019 Flame propagation in nano-aluminum–water (nAl–H2O) mixtures: The role of thermal interface resistance *Combust. Flame*

[328]  Min C, Nuofu C, Xiaoli Y, Yu W, Yiming B and Xingwang Z 2009 Thermal analysis and test for single concentrator solar cells *J. Semicond.* **30** 44011

[329]  Katz E A, Gordon J M and Feuermann D 2006 Effects of ultra-high flux and intensity distribution in multi-junction solar cells *Prog. photovoltaics* **14** 297–304

[330]  Ortiz B R, Gordiz K, Gomes L C, Braden T, Adamczyk J M, Qu J, Ertekin E and Toberer E S 2019 Carrier density control in Cu 2 HgGeTe 4 and discovery of Hg 2 GeTe 4 via phase boundary mapping *J. Mater. Chem. A* **7** 621–31

[331]  Ortiz B R, Adamczyk J M, Gordiz K, Braden T and Toberer E S 2019 Towards the high-throughput synthesis of bulk materials: thermoelectric PbTe–PbSe–SnTe–SnSe alloys *Mol. Syst. Des. Eng.* **4** 407–20

[332]  Zhao L-D, Lo S-H, Zhang Y, Sun H, Tan G, Uher C, Wolverton C, Dravid V P and Kanatzidis M G 2014 Ultralow thermal conductivity and high thermoelectric figure of merit in SnSe crystals *Nature* **508** 373–7

[333]  Toprak M S, Stiewe C, Platzek D, Williams S, Bertini L, Müller E, Gatti C, Zhang Y, Rowe M and Muhammed M 2004 The impact of nanostructuring on the thermal conductivity of thermoelectric CoSb3 *Adv. Funct. Mater.* **14** 1189–96

[334]  He J, Girard S N, Kanatzidis M G and Dravid V P 2010 Microstructure-lattice thermal conductivity correlation in nanostructured PbTe0. 7S0. 3 thermoelectric Materials *Adv. Funct. Mater.* **20** 764–72

[335]  Medlin D L and Snyder G J 2009 Interfaces in bulk thermoelectric materials: a review for current opinion in colloid and interface science *Curr. Opin. Colloid Interface Sci.* **14** 226–35

[336]  Gordiz K, Menon A K and Yee S K 2017 Interconnect patterns for printed organic

thermoelectric devices with large fill factors *J. Appl. Phys.* **122** 124507

[337] Tang J J, Bai Y, Zhang J C, Liu K, Liu X Y, Zhang P, Wang Y, Zhang L, Liang G Y and Gao Y 2016 Microstructural design and oxidation resistance of CoNiCrAlY alloy coatings in thermal barrier coating system *J. Alloys Compd.* **688** 729–41

[338] Padture N P, Gell M and Jordan E H 2002 Thermal barrier coatings for gas-turbine engine applications *Science (80-. ).* **296** 280–4

[339] Cao X Q, Vassen R and Stöver D 2004 Ceramic materials for thermal barrier coatings *J. Eur. Ceram. Soc.* **24** 1–10

[340] Minnich Aj, Dresselhaus M S, Ren Z F and Chen G 2009 Bulk nanostructured thermoelectric materials: current research and future prospects *Energy Environ. Sci.* **2** 466–79

[341] Kapitza P L 1941 The study of heat transfer in helium II *J. Phys.(USSR)* **4** 181–210

[342] Hopkins P E, Duda J C and Norris P M 2011 Anharmonic phonon interactions at interfaces and contributions to thermal boundary conductance *J. Heat Transfer* **133**

[343] Swartz E T and Pohl R O 1987 Thermal resistance at interfaces *Appl. Phys. Lett.* **51** 2200–2

[344] Zhang W, Fisher T S and Mingo N 2007 The atomistic Green's function method: An efficient simulation approach for nanoscale phonon transport *Numer. Heat Transf. Part B Fundam.* **51** 333–49

[345] Zhang W, Mingo N and Fisher T S 2007 Simulation of interfacial phonon transport in Si–Ge heterostructures using an atomistic Green's function method *J. Heat Transfer* **129** 483–91

[346] Mingo N 2009 Green's Function Methods for Phonon Transport Through Nano-Contacts

*Thermal nanosystems and nanomaterials* (Springer) pp 63–94

[347] Tian Z, Esfarjani K and Chen G 2012 Enhancing phonon transmission across a Si/Ge interface by atomic roughness: First-principles study with the Green's function method *Phys. Rev. B* **86** 235304

[348] Young D A and Maris H J 1989 Lattice-dynamical calculation of the Kapitza resistance between fcc lattices *Phys. Rev. B* **40** 3685

[349] Sun H and Pipe K P 2012 Perturbation analysis of acoustic wave scattering at rough solid-solid interfaces *J. Appl. Phys.* **111** 23510

[350] Zhao H and Freund J B 2009 Phonon scattering at a rough interface between two fcc lattices *J. Appl. Phys.* **105** 13515

[351] Kakodkar R R and Feser J P 2016 Probing the validity of the diffuse mismatch model for phonons using atomistic simulations *arXiv Prepr. arXiv1607.08572*

[352] Zhang Z M 2007 *Nano/microscale heat transfer* (McGraw-Hill New York)

[353] Landauer R 1957 Spatial variation of currents and fields due to localized scatterers in metallic conduction *IBM J. Res. Dev.* **1** 223–31

[354] Hopkins P E 2009 Multiple phonon processes contributing to inelastic scattering during thermal boundary conductance at solid interfaces *J. Appl. Phys.* **106** 13528

[355] Hopkins P E and Norris P M 2007 Effects of joint vibrational states on thermal boundary conductance *Nanoscale Microscale Thermophys. Eng.* **11** 247–57

[356] Prasher R 2009 Acoustic mismatch model for thermal contact resistance of van der Waals contacts *Appl. Phys. Lett.* **94** 41903–5

[357] Reddy P, Castelino K and Majumdar A 2005 Diffuse mismatch model of thermal boundary conductance using exact phonon dispersion *Appl. Phys. Lett.* **87** 211908

[358] Shin S, Kaviany M, Desai T and Bonner R 2010 Roles of atomic restructuring in interfacial phonon transport *Phys. Rev. B* **82** 81302

[359] Hopkins P E, Norris P M and Stevens R J 2008 Influence of inelastic scattering at metal-dielectric interfaces *J. Heat Transfer* **130**

[360] Ong Z-Y and Pop E 2010 Frequency and polarization dependence of thermal coupling between carbon nanotubes and SiO2 *J. Appl. Phys.* **108** 103502

[361] Stevens R J, Zhigilei L V and Norris P M 2007 Effects of temperature and disorder on thermal boundary conductance at solid–solid interfaces: Nonequilibrium molecular dynamics simulations *Int. J. Heat Mass Transf.* **50** 3977–89

[362] Chalopin Y, Mingo N, Diao J, Srivastava D and Volz S 2012 Large effects of pressure induced inelastic channels on interface thermal conductance *Appl. Phys. Lett.* **101** 221903

[363] English T S, Duda J C, Smoyer J L, Jordan D A, Norris P M and Zhigilei L V 2012 Enhancing and tuning phonon transport at vibrationally mismatched solid-solid interfaces *Phys. Rev. B* **85** 35438

[364] Gaskins J T, Kotsonis G, Giri A, Ju S, Rohskopf A, Wang Y, Bai T, Sachet E, Shelton C T and Liu Z 2018 Thermal boundary conductance across heteroepitaxial ZnO/GaN interfaces: Assessment of the phonon gas model *Nano Lett.* **18** 7469–77

[365] Minnich A J 2014 Towards a microscopic understanding of phonon heat conduction *arXiv Prepr. arXiv1405.0532*

[366] Hopkins P E 2013 Thermal transport across solid interfaces with nanoscale imperfections: effects of roughness, disorder, dislocations, and bonding on thermal boundary conductance *ISRN Mech. Eng.* **2013**

[367] Monachon C, Weber L and Dames C 2016 Thermal boundary conductance: A materials

science perspective *Annu. Rev. Mater. Res.* **46** 433–63

[368] Giri A and Hopkins P E 2019 A Review of Experimental and Computational Advances in Thermal Boundary Conductance and Nanoscale Thermal Transport across Solid Interfaces *Adv. Funct. Mater.* 1903857

[369] Gordiz K and Allaei S M V 2014 Thermal rectification in pristine-hydrogenated carbon nanotube junction: A molecular dynamics study *J. Appl. Phys.* **115** 163512

[370] Wu X and Luo T 2014 The importance of anharmonicity in thermal transport across solid-solid interfaces *J. Appl. Phys.* **115** 14901

[371] Rajabpour A, Allaei S M V and Kowsary F 2011 Interface thermal resistance and thermal rectification in hybrid graphene-graphane nanoribbons: A nonequilibrium molecular dynamics study *Appl. Phys. Lett.* **99** 51917

[372] Hu M, Keblinski P and Schelling P K 2009 Kapitza conductance of silicon–amorphous polyethylene interfaces by molecular dynamics simulations *Phys. Rev. B* **79** 104305

[373] Kubo R 1966 The fluctuation-dissipation theorem *Reports Prog. Phys.* **29** 255

[374] Puech L, Bonfait G and Castaing B 1986 Mobility of the 3He solid-liquid interface: Experiment and theory *J. Low Temp. Phys.* **62** 315–27

[375] Domingues G, Volz S, Joulain K and Greffet J-J 2005 Heat transfer between two nanoparticles through near field interaction *Phys. Rev. Lett.* **94** 85901

[376] Barrat J-L and Chiaruttini F 2003 Kapitza resistance at the liquid—solid interface *Mol. Phys.* **101** 1605–10

[377] Chalopin Y, Esfarjani K, Henry A, Volz S and Chen G 2012 Thermal interface conductance in Si/Ge superlattices by equilibrium molecular dynamics *Phys. Rev. B* **85** 195302

[378] Rajabpour A and Volz S 2010 Thermal boundary resistance from mode energy relaxation times: Case study of argon-like crystals by molecular dynamics *J. Appl. Phys.* **108** 94324

[379] Landry E S and McGaughey A J H 2009 Thermal boundary resistance predictions from molecular dynamics simulations and theoretical calculations *Phys. Rev. B* **80** 165304

[380] Giri A, Donovan B F and Hopkins P E 2018 Localization of vibrational modes leads to reduced thermal conductivity of amorphous heterostructures *Phys. Rev. Mater.* **2** 56002

[381] Shenogin S, Xue L, Ozisik R, Keblinski P and Cahill D G 2004 Role of thermal boundary resistance on the heat flow in carbon-nanotube composites *J. Appl. Phys.* **95** 8136–44

[382] Hu L, Desai T and Keblinski P 2011 Determination of interfacial thermal resistance at the nanoscale *Phys. Rev. B* **83** 195423

[383] Merabia S, Keblinski P, Joly L, Lewis L J and Barrat J-L 2009 Critical heat flux around strongly heated nanoparticles *Phys. Rev. E* **79** 21404

[384] Merabia S and Termentzidis K 2014 Thermal boundary conductance across rough interfaces probed by molecular dynamics *Phys. Rev. B* **89** 54309

[385] Seyf H R, Gordiz K, DeAngelis F and Henry A 2019 Using Green-Kubo modal analysis (GKMA) and interface conductance modal analysis (ICMA) to study phonon transport with molecular dynamics *J. Appl. Phys.* **125** 81101

[386] Feng T, Zhong Y, Shi J and Ruan X 2019 Unexpected high inelastic phonon transport across solid-solid interface: Modal nonequilibrium molecular dynamics simulations and Landauer analysis *Phys. Rev. B* **99** 45301

[387] Zhou Y and Hu M 2017 Full quantification of frequency-dependent interfacial thermal conductance contributed by two-and three-phonon scattering processes from nonequilibrium molecular dynamics simulations *Phys. Rev. B* **95** 115313

[388]  Hardy R J 1963 Energy-flux operator for a lattice *Phys. Rev.* **132** 168

[389]  Gordiz K and Henry A 2016 Phonon transport at interfaces: Determining the correct modes of vibration *J. Appl. Phys.* **119** 15101

[390]  Sääskilahti K, Oksanen J, Tulkki J and Volz S 2014 Role of anharmonic phonon scattering in the spectrally decomposed thermal conductance at planar interfaces *Phys. Rev. B* **90** 134312

[391]  Dresselhaus M S, Chen G, Ren Z F, Dresselhaus G, Henry A and Fleurial J-P 2009 New composite thermoelectric materials for energy harvesting applications *Jom* **61** 86–90

[392]  Roberts N A and Walker D G 2010 Phonon wave-packet simulations of Ar/Kr interfaces for thermal rectification *J. Appl. Phys.* **108** 123515

[393]  Gordiz K 2017 *Modal decomposition of thermal conductance* (Georgia Institute of Technology)

[394]  Gordiz K and Henry A 2015 Examining the effects of stiffness and mass difference on the thermal interface conductance between Lennard-Jones solids *Sci. Rep.* **5** 18361

[395]  Lee E, Zhang T, Hu M and Luo T 2016 Thermal boundary conductance enhancement using experimentally achievable nanostructured interfaces–analytical study combined with molecular dynamics simulation *Phys. Chem. Chem. Phys.* **18** 16794–801

[396]  Zhou X W, Jones R E, Kimmer C J, Duda J C and Hopkins P E 2013 Relationship of thermal boundary conductance to structure from an analytical model plus molecular dynamics simulations *Phys. Rev. B* **87** 94303

[397]  Hu M, Zhang X, Poulikakos D and Grigoropoulos C P 2011 Large "near junction" thermal resistance reduction in electronics by interface nanoengineering *Int. J. Heat Mass Transf.* **54** 5183–91

[398]  Lee E, Zhang T, Yoo T, Guo Z and Luo T 2016 Nanostructures significantly enhance thermal transport across solid interfaces *ACS Appl. Mater. Interfaces* **8** 35505–12

[399]  Lee E and Luo T 2017 Investigation of thermal transport across solid interfaces with randomly distributed nanostructures *2017 16th IEEE Intersociety Conference on Thermal and Thermomechanical Phenomena in Electronic Systems (ITherm)* (IEEE) pp 363–7

[400]  Liang Z, Sasikumar K and Keblinski P 2014 Thermal Transport across a Substrate–Thin-Film Interface: Effects of Film Thickness and Surface Roughness *Phys. Rev. Lett.* **113** 65901

[401]  Li R, Gordiz K, Henry A, Hopkins P E, Lee E and Luo T 2019 Effect of light atoms on thermal transport across solid–solid interfaces *Phys. Chem. Chem. Phys.* **21** 17029–35

[402]  Lee E and Luo T 2017 The role of optical phonons in intermediate layer-mediated thermal transport across solid interfaces *Phys. Chem. Chem. Phys.* **19** 18407–15

[403]  Norris P M and Hopkins P E 2009 Examining interfacial diffuse phonon scattering through transient thermoreflectance measurements of thermal boundary conductance *J. Heat Transfer* **131**

[404]  Hopkins P E, Norris P M, Stevens R J, Beechem T E and Graham S 2008 Influence of interfacial mixing on thermal boundary conductance across a chromium/silicon interface *J. Heat Transfer* **130** 62402

[405]  Le N Q, Duda J C, English T S, Hopkins P E, Beechem T E and Norris P M 2012 Strategies for tuning phonon transport in multilayered structures using a mismatch-based particle model *J. Appl. Phys.* **111** 84310

[406]  Polanco C A, Rastgarkafshgarkolaei R, Zhang J, Le N Q, Norris P M and Ghosh A W 2017 Design rules for interfacial thermal conductance: Building better bridges *Phys. Rev.*

*B* **95** 195303

[407] Giri A, King S W, Lanford W A, Mei A B, Merrill D, Li L, Oviedo R, Richards J, Olson D H and Braun J L 2018 Interfacial defect vibrations enhance thermal transport in amorphous multilayers with ultrahigh thermal boundary conductance *Adv. Mater.* **30** 1804097

[408] Hahn K R, Puligheddu M and Colombo L 2015 Thermal boundary resistance at Si/Ge interfaces determined by approach-to-equilibrium molecular dynamics simulations *Phys. Rev. B* **91** 195313

[409] Gorham C S, Hattar K, Cheaito R, Duda J C, Gaskins J T, Beechem T E, Ihlefeld J F, Biedermann L B, Piekos E S and Medlin D L 2014 Ion irradiation of the native oxide/silicon surface increases the thermal boundary conductance across aluminum/silicon interfaces *Phys. Rev. B* **90** 24301

[410] Giri A, Braun J L and Hopkins P E 2016 Implications of interfacial bond strength on the spectral contributions to thermal boundary conductance across solid, liquid, and gas interfaces: A molecular dynamics study *J. Phys. Chem. C* **120** 24847–56

[411] Lyeo H-K and Cahill D G 2006 Thermal conductance of interfaces between highly dissimilar materials *Phys. Rev. B* **73** 144301

[412] Duda J C, Smoyer J L, Norris P M and Hopkins P E 2009 Extension of the diffuse mismatch model for thermal boundary conductance between isotropic and anisotropic materials *Appl. Phys. Lett.* **95** 31912

[413] Gordiz K and Henry A 2016 Phonon transport at crystalline Si/Ge interfaces: the role of interfacial modes of vibration *Sci. Rep.* **6** 1–9

[414] Gordiz K and Henry A 2016 Interface conductance modal analysis of lattice matched

InGaAs/InP *Appl. Phys. Lett.* **108** 181606

[415] Giri A, Hopkins P E, Wessel J G and Duda J C 2015 Kapitza resistance and the thermal conductivity of amorphous superlattices *J. Appl. Phys.* **118** 165303

[416] Kimling J, Philippi-Kobs A, Jacobsohn J, Oepen H P and Cahill D G 2017 Thermal conductance of interfaces with amorphous SiO 2 measured by time-resolved magneto-optic Kerr-effect thermometry *Phys. Rev. B* **95** 184305

[417] Lee S-M, Cahill D G and Venkatasubramanian R 1997 Thermal conductivity of Si–Ge superlattices *Appl. Phys. Lett.* **70** 2957–9

[418] Li X and Yang R 2012 Effect of lattice mismatch on phonon transmission and interface thermal conductance across dissimilar material interfaces *Phys. Rev. B* **86** 54305

[419] Zhu T and Wang C 2005 Misfit dislocation networks in the γ∕γ′ phase interface of a Ni-based single-crystal superalloy: Molecular dynamics simulations *Phys. Rev. B* **72** 14111

[420] Li N-L, Wu W-P and Nie K 2018 Molecular dynamics study on the evolution of interfacial dislocation network and mechanical properties of Ni-based single crystal superalloys *Phys. Lett. A* **382** 1361–7

[421] Zheng D L, Chen S D, Soh A K and Ma Y 2010 Molecular dynamics simulations of glide dislocations induced by misfit dislocations at the Ni/Al interface *Comput. Mater. Sci.* **48** 551–5

[422] Murakami T, Hori T, Shiga T and Shiomi J 2014 Probing and tuning inelastic phonon conductance across finite-thickness interface *Appl. Phys. Express* **7** 121801

[423] Plimpton S 1993 *Fast parallel algorithms for short-range molecular dynamics* (Sandia National Labs., Albuquerque, NM (United States))

[424] Parlinski K 1999 Calculation of phonon dispersion curves by the direct method *AIP*

*Conference Proceedings* vol 479 (American Institute of Physics) pp 121–6

[425] Choi J, Dongarra J J, Pozo R and Walker D W 1992 ScaLAPACK: A scalable linear algebra library for distributed memory concurrent computers *The Fourth Symposium on the Frontiers of Massively Parallel Computation* (IEEE Computer Society) pp 120–1

[426] Angstrom A J 1861 A new method of determining the thermal conductivity of solids *Annln. Phys* **64** 513–30

[427] Ho C Y, Powell R W and Liley P E 1972 Thermal Conductivity of the Elements *J. Phys. Chem. Ref. Data*

[428] Childs G E, Ericks L J and Powell R L 1973 Thermal Conductivity of Solids at Room Temperature and Below - A Review and Compilation of the Literature *NBS Monogr. 131*

[429] Ho C Y, Powell R W and Liley P E 1968 Thermal Conductivity of Selected Materials, Part 2 *Nsrds-Nbs 16*

[430] Schmidt A J 2013 Pump-probe thermoreflectance *Annu. Rev. Heat Transf.* **16**

[431] Sood A, Cheaito R, Bai T, Kwon H, Wang Y, Li C, Yates L, Bougher T, Graham S and Asheghi M 2018 Direct Visualization of Thermal Conductivity Suppression Due to Enhanced Phonon Scattering Near Individual Grain Boundaries *Nano Lett.*

[432] Beechem T, Yates L and Graham S 2015 Invited Review Article: Error and uncertainty in Raman thermal conductivity measurements *Rev. Sci. Instrum.* **86** 41101

[433] Stoner R J and Maris H J 1993 Kapitza conductance and heat flow between solids at temperatures from 50 to 300 K *Phys. Rev. B* **48** 16373

[434] Paddock C A and Eesley G L 1986 Transient thermoreflectance from thin metal films *J. Appl. Phys.* **60** 285–90

[435] Cheng Z, Koh Y R, Ahmad H, Hu R, Shi J, Liao M E, Wang Y, Bai T, Li R, Lee E,

Clinton E A, Matthews C M, Engel Z, Yates L, Luo T, Goorsky M S, Doolittle W A, Tian Z, Hopkins P E and Graham S 2020 Thermal conductance across harmonic-matched epitaxial Al-sapphire heterointerfaces *Commun. Phys.*

[436]  Costescu R M, Wall M A and Cahill D G 2003 Thermal conductance of epitaxial interfaces *Phys. Rev. B* **67** 54302

[437]  Oh D, Kim S, Rogers J A, Cahill D G and Sinha S 2011 Interfacial Thermal Conductance of Transfer-Printed Metal Films *Adv. Mater.* **23** 5028–33

[438]  Wilson R B and Cahill D G 2014 Anisotropic failure of Fourier theory in time-domain thermoreflectance experiments *Nat. Commun.* **5** 5075

[439]  Liu D, Xie R, Yang N, Li B and Thong J T L 2014 Profiling nanowire thermal resistance with a spatial resolution of nanometers *Nano Lett.* **14** 806–12

[440]  Wilson R B, Apgar B A, Hsieh W-P, Martin L W and Cahill D G 2015 Thermal conductance of strongly bonded metal-oxide interfaces *Phys. Rev. B* **91** 115414

[441]  Hohensee G T, Wilson R B and Cahill D G 2015 Thermal conductance of metal–diamond interfaces at high pressure *Nat. Commun.* **6** 6578

[442]  Ye N, Feser J P, Sadasivam S, Fisher T S, Wang T, Ni C and Janotti A 2017 Thermal transport across metal silicide-silicon interfaces: An experimental comparison between epitaxial and nonepitaxial interfaces *Phys. Rev. B* **95** 85430

[443]  Stoner R J, Maris H J, Anthony T R and Banholzer W F 1992 Measurements of the Kapitza conductance between diamond and several metals *Phys. Rev. Lett.* **68** 1563

[444]  Stevens R J, Smith A N and Norris P M 2005 Measurement of thermal boundary conductance of a series of metal-dielectric interfaces by the transient thermoreflectance technique *J. Heat Transfer* **127** 315–22

[445] Ge Z, Cahill D G and Braun P V 2006 Thermal conductance of hydrophilic and hydrophobic interfaces *Phys. Rev. Lett.* **96** 186101

[446] Wang R Y, Segalman R A and Majumdar A 2006 Room temperature thermal conductance of alkanedithiol self-assembled monolayers *Appl. Phys. Lett.* **89** 173113

[447] Hanisch A, Krenzer B, Pelka T, Möllenbeck S and Horn-von Hoegen M 2008 Thermal response of epitaxial thin Bi films on Si (001) upon femtosecond laser excitation studied by ultrafast electron diffraction *Phys. Rev. B* **77** 125410

[448] Schmidt A J, Collins K C, Minnich A J and Chen G 2010 Thermal conductance and phonon transmissivity of metal–graphite interfaces *J. Appl. Phys.* **107** 104907

[449] Collins K C, Chen S and Chen G 2010 Effects of surface chemistry on thermal conductance at aluminum–diamond interfaces *Appl. Phys. Lett.* **97** 83102

[450] Hsieh W-P, Lyons A S, Pop E, Keblinski P and Cahill D G 2011 Pressure tuning of the thermal conductance of weak interfaces *Phys. Rev. B* **84** 184107

[451] Duda J C and Hopkins P E 2012 Systematically controlling Kapitza conductance via chemical etching *Appl. Phys. Lett.* **100** 111602

[452] Hopkins P E, Baraket M, Barnat E V, Beechem T E, Kearney S P, Duda J C, Robinson J T and Walton S G 2012 Manipulating thermal conductance at metal–graphene contacts via chemical functionalization *Nano Lett.* **12** 590–5

[453] Losego M D, Grady M E, Sottos N R, Cahill D G and Braun P V 2012 Effects of chemical bonding on heat transport across interfaces *Nat. Mater.* **11** 502–6

[454] O'Brien P J, Shenogin S, Liu J, Chow P K, Laurencin D, Mutin P H, Yamaguchi M, Keblinski P and Ramanath G 2013 Bonding-induced thermal conductance enhancement at inorganic heterointerfaces using nanomolecular monolayers *Nat. Mater.* **12** 118–22

[455] Monachon C and Weber L 2013 Influence of diamond surface termination on thermal boundary conductance between Al and diamond *J. Appl. Phys.* **113** 183504

[456] Duda J C, Yang C-Y, Foley B M, Cheaito R, Medlin D L, Jones R E and Hopkins P E 2013 Influence of interfacial properties on thermal transport at gold: silicon contacts *Appl. Phys. Lett.* **102** 81902

[457] Monachon C and Weber L 2014 Thermal boundary conductance between refractory metal carbides and diamond *Acta Mater.* **73** 337–46

[458] Sun F, Zhang T, Jobbins M M, Guo Z, Zhang X, Zheng Z, Tang D, Ptasinska S and Luo T 2014 Molecular Bridge Enables Anomalous Enhancement in Thermal Transport across Hard-Soft Material Interfaces *Adv. Mater.* **26** 6093–9

[459] Freedman J P, Yu X, Davis R F, Gellman A J and Malen J A 2016 Thermal interface conductance across metal alloy–dielectric interfaces *Phys. Rev. B* **93** 35309

[460] Hua C, Chen X, Ravichandran N K and Minnich A J 2017 Experimental metrology to obtain thermal phonon transmission coefficients at solid interfaces *Phys. Rev. B* **95** 205423

[461] Saha D, Yu X, Jeong M, Darwish M, Weldon J, Gellman A J and Malen J A 2019 Impact of metal adhesion layer diffusion on thermal interface conductance *Phys. Rev. B* **99** 115418

[462] Goodson K E, Käding O W, Rösler M and Zachai R 1995 Experimental investigation of thermal conduction normal to diamond-silicon boundaries *J. Appl. Phys.* **77** 1385–92

[463] Koh Y K, Cao Y, Cahill D G and Jena D 2009 Heat-Transport Mechanisms in Superlattices *Adv. Funct. Mater.* **19** 610–5

[464] Hopkins P E, Duda J C, Clark S P, Hains C P, Rotter T J, Phinney L M and Balakrishnan

G 2011 Effect of dislocation density on thermal boundary conductance across GaSb/GaAs interfaces *Appl. Phys. Lett.* **98** 161913

[465] Losego M D, Blitz I P, Vaia R A, Cahill D G and Braun P V 2013 Ultralow thermal conductivity in organoclay nanolaminates synthesized via simple self-assembly *Nano Lett.* **13** 2215–9

[466] Ziade E, Yang J, Brummer G, Nothern D, Moustakas T and Schmidt A J 2015 Thermal transport through GaN–SiC interfaces from 300 to 600 K *Appl. Phys. Lett.* **107** 91605

[467] Schroeder D P, Aksamija Z, Rath A, Voyles P M, Lagally M G and Eriksson M A 2015 Thermal resistance of transferred-silicon-nanomembrane interfaces *Phys. Rev. Lett.* **115** 256101

[468] Fong S W, Sood A, Chen L, Kumari N, Asheghi M, Goodson K E, Gibson G A and Wong H-S 2016 Thermal conductivity measurement of amorphous dielectric multilayers for phase-change memory power reduction *J. Appl. Phys.* **120** 15103

[469] Giri A, Niemelä J-P, Tynell T, Gaskins J T, Donovan B F, Karppinen M and Hopkins P E 2016 Heat-transport mechanisms in molecular building blocks of inorganic/organic hybrid superlattices *Phys. Rev. B* **93** 115310

[470] Zheng K, Sun F, Zhu J, Ma Y, Li X, Tang D, Wang F and Wang X 2016 Enhancing the thermal conductance of polymer and sapphire interface via self-assembled monolayer *ACS Nano* **10** 7792–8

[471] Bougher T L, Yates L, Lo C-F, Johnson W, Graham S and Cola B A 2016 Thermal boundary resistance in GaN films measured by time domain thermoreflectance with robust Monte Carlo uncertainty estimation *Nanoscale Microscale Thermophys. Eng.* **20** 22–32

[472] Mu F, Cheng Z, Shi J, Shin S, Xu B, Shiomi J, Graham S and Suga T 2019 High Thermal

Boundary Conductance across Bonded Heterogeneous GaN-SiC Interfaces *ACS Appl. Mater. Interfaces* **11** 7

[473] Cheng Z, Bai T, Shi J, Feng T, Wang Y, Mecklenburg M, Li C, Hobart K D, Feygelson T and Tadjer M J 2019 Tunable Thermal Energy Transport across Diamond Membranes and Diamond-Si Interfaces by Nanoscale Graphoepitaxy *ACS Appl. Mater. Interfaces*

[474] Cheng Z, Yates L, Shi J, Tadjer M J, Hobart K D and Graham S 2019 Thermal conductance across β-Ga2O3-diamond van der Waals heterogeneous interfaces *APL Mater.* **7** 31118

[475] Gundrum B C, Cahill D G and Averback R S 2005 Thermal conductance of metal-metal interfaces *Phys. Rev. B* **72** 245426

[476] Wilson R B and Cahill D G 2012 Experimental validation of the interfacial form of the Wiedemann-Franz law *Phys. Rev. Lett.* **108** 255901

[477] Cheng Z, Mu F, Yates L, Suga T and Graham S 2020 Interfacial Thermal Conductance across Room-Temperature-Bonded GaN/Diamond Interfaces for GaN-on-Diamond Devices *ACS Appl. Mater. Interfaces* **12** 8376–84

[478] Yuan C, Pomeroy J W and Kuball M 2018 Above bandgap thermoreflectance for non-invasive thermal characterization of GaN-based wafers *Appl. Phys. Lett.* **113** 102101

[479] Gordiz K and reports Henry A %J S 2016 Phonon transport at crystalline Si/Ge interfaces: the role of interfacial modes of vibration **6** 1–9

[480] Hummel P, Lechner A M, Herrmann K, Biehl P, Rössel C, Wiedenhöft L, Schacher F H and Retsch M 2020 Thermal Transport in Ampholytic Polymers: The Role of Hydrogen Bonding and Water Uptake *Macromolecules* **53** 5528–37

[481] Henry A 2014 Thermal transport in polymers *Annu. Rev. heat Transf.* **17**

[482]  Buschow K H J 1983 Stability and electrical transport properties of amorphous Ti1-xNix alloys *J. Phys. F Met. Phys.* **13** 563

[483]  Berendsen H J C, van der Spoel D and van Drunen R 1995 GROMACS: A message-passing parallel molecular dynamics implementation *Comput. Phys. Commun.*

[484]  Phillips J C, Braun R, Wang W, Gumbart J, Tajkhorshid E, Villa E, Chipot C, Skeel R D, Kalé L and Schulten K 2005 Scalable molecular dynamics with NAMD *J. Comput. Chem.*

[485]  Yang J, Wang Y and Chen Y 2007 GPU accelerated molecular dynamics simulation of thermal conductivities *J. Comput. Phys.*

[486]  Fan Z, Chen W, Vierimaa V and Harju A 2017 Efficient molecular dynamics simulations with many-body potentials on graphics processing units *Comput. Phys. Commun.*

[487]  Turney J E, McGaughey A J H and Amon C H 2009 Assessing the applicability of quantum corrections to classical thermal conductivity predictions *Phys. Rev. B* **79** 224305

[488]  Puligheddu M, Xia Y, Chan M and Galli G 2019 Computational prediction of lattice thermal conductivity: A comparison of molecular dynamics and Boltzmann transport approaches *Phys. Rev. Mater.* **3** 85401

[489]  Sääskilahti K, Oksanen J, Tulkki J, McGaughey A J H and Volz S 2016 Vibrational mean free paths and thermal conductivity of amorphous silicon from non-equilibrium molecular dynamics simulations *AIP Adv.* **6** 121904

[490]  Wang J-S 2007 Quantum Thermal Transport from Classical Molecular Dynamics *Phys. Rev. Lett.* **99** 160601

[491]  Ceriotti M, Bussi G and Parrinello M 2009 Nuclear Quantum Effects in Solids Using a Colored-Noise Thermostat *Phys. Rev. Lett.* **103** 030603

[492]  Bedoya-Martínez O N, Barrat J-L and Rodney D 2014 Computation of the thermal

conductivity using methods based on classical and quantum molecular dynamics *Phys. Rev. B* **89** 014303

[493] Brieuc F, Bronstein Y, Dammak H, Depondt P, Finocchi F and Hayoun M 2016 Zero-Point Energy Leakage in Quantum Thermal Bath Molecular Dynamics Simulations *J. Chem. Theory Comput.* **12** 5688–97

[494] Mangaud E, Huppert S, Plé T, Depondt P, Bonella S and Finocchi F 2019 The Fluctuation–Dissipation Theorem as a Diagnosis and Cure for Zero-Point Energy Leakage in Quantum Thermal Bath Simulations *J. Chem. Theory Comput.* **15** 2863–80

[495] Markland T E and Ceriotti M 2018 Nuclear quantum effects enter the mainstream *Nat. Rev. Chem.* **2** 1–14

[496] Jang S and Voth G A 1999 A derivation of centroid molecular dynamics and other approximate time evolution methods for path integral centroid variables *J. Chem. Phys.* **111** 2371–84

[497] Paesani F and Voth G A 2008 Nonlinear quantum time correlation functions from centroid molecular dynamics and the maximum entropy method *J. Chem. Phys.* **129** 194113